\documentclass{aa}  

\usepackage{graphicx}
\usepackage[varg]{txfonts}
\usepackage{multirow}
\usepackage{amsmath}
\usepackage{lscape}
\usepackage{bm}
\usepackage{ulem}

\usepackage{color}

\usepackage{natbib}
\newcommand{\myemail}{hanae.inami@univ-lyon1.fr}

\newcommand{\Ang}{${\rm \AA}$\xspace}

\newcommand{\kms}{km\,s$^{-1}$\xspace}

\newcommand{\lya}{Ly$\alpha$\xspace}

\newcommand{\ciii}{C\,\textsc{iii}]\xspace}
\newcommand{\oii}{[O\,\textsc{ii}]\xspace}
\newcommand{\oiii}{[O\,\textsc{iii}]\xspace}

\newcommand{\feii}{Fe\,\textsc{ii}\xspace}
\newcommand{\aliii}{Al\,\textsc{iii}\xspace}

\newcommand{\mgii}{Mg\,\textsc{ii}\xspace}
\newcommand{\neiii}{[Ne\,\textsc{iii}]\xspace}
\newcommand{\ha}{H$\alpha$\xspace}
\newcommand{\hb}{H$\beta$\xspace}
\newcommand{\hg}{H$\gamma$\xspace}
\newcommand{\hd}{H$\delta$\xspace}

\newcommand{\udft}{\textsf{udf-10}\xspace}
\newcommand{\mosaic}{\textsf{mosaic}\xspace}
 
\newcommand{\UDForgHSTpri}{1095}    
\newcommand{\UDFHSTpri}{854}        
\newcommand{\MOSorgHSTpri}{7904}    
\newcommand{\MOSHSTpri}{6288}       
\newcommand{\MOSHSTpriMAGcut}{1147} 
\newcommand{\COMBHSTpri}{\MOSorgHSTpri} 

\newcommand{\UDFORIGIN}{306}     
\newcommand{\MOSORIGIN}{1251}    

\newcommand{\UDFzHSTCone}{282}     
\newcommand{\UDFzHSTCtwo}{223}     
\newcommand{\UDFzHSTCthree}{87}    
\newcommand{\UDFzORGCone}{160}     
\newcommand{\UDFzORGCtwo}{154}     
\newcommand{\UDFzORGCthree}{46}    
\newcommand{\UDFzORGonlyCone}{31}  
\newcommand{\UDFzORGonlyCtwo}{30}  
\newcommand{\UDFzORGonlyCthree}{2} 
\newcommand{\UDFzALLCone}{313}     
\newcommand{\UDFzALLCtwo}{253}     
\newcommand{\UDFzALLCthree}{89}    

\newcommand{\MOSzHSTCone}{714}      
\newcommand{\MOSzHSTCtwo}{636}      
\newcommand{\MOSzHSTCthree}{515}    
\newcommand{\MOSzHSTallCone}{1295}  
\newcommand{\MOSzHSTallCtwo}{1133}  
\newcommand{\MOSzHSTallCthree}{547} 
\newcommand{\MOSzORGCone}{1163}     
\newcommand{\MOSzORGCtwo}{1057}     
\newcommand{\MOSzORGCthree}{450}    
\newcommand{\MOSzORGonlyCone}{144}  
\newcommand{\MOSzORGonlyCtwo}{114}  
\newcommand{\MOSzORGonlyCthree}{15} 
\newcommand{\MOSzALLCone}{1439}     
\newcommand{\MOSzALLCtwo}{1247}     
\newcommand{\MOSzALLCthree}{562}    

\newcommand{\COMBzHSTCone}{1414}     
\newcommand{\COMBzHSTCtwo}{1206}     
\newcommand{\COMBzHSTCthree}{570}    
\newcommand{\COMBzORGCone}{1214}     
\newcommand{\COMBzORGCtwo}{1112}     
\newcommand{\COMBzORGCthree}{467}    
\newcommand{\COMBzORGonlyCone}{160}  
\newcommand{\COMBzORGonlyCtwo}{132}  
\newcommand{\COMBzORGonlyCthree}{15} 
\newcommand{\COMBzALLCone}{1574}     
\newcommand{\COMBzALLCtwo}{1338}     
\newcommand{\COMBzALLCthree}{585}    

\newcommand{\COMBprevzUVUDF}{150}    
\newcommand{\COMBprevzVUDS}{11}      
\newcommand{\COMBprevz}{161}         

\newcommand{\mkcatupnum}{200}
\newcommand{\numnotinraf}{\COMBzORGonlyCone} 
\newcommand{\mkcatupnsigma}{3} 
\newcommand{\maxmatchdist}{0.30}
\newcommand{\rafprobinnc}{76}    
\newcommand{\aperturediameter}{0.50}

\newcommand{\UDFver}{udf10_c042_e031_withz_iter6}
\newcommand{\MOSver}{mosaic_c042_e030_withz_iter8}
\newcommand{\COMBver}{combined_udf10_c042_e031_withz_iter6_mosaic_c042_e030_withz_iter8}

\begin{document}

\title{The MUSE Hubble Ultra Deep Field Survey:}

\subtitle{II. Spectroscopic redshifts and \\
  comparisons to color selections of high-redshift galaxies
  \thanks{Based on observations made with ESO telescopes at the La
    Silla Paranal Observatory under program IDs 094.A-0289(B),
    095.A-0010(A), 096.A-0045(A) and 096.A-0045(B). },\thanks{MUSE
    Ultra Deep Field redshift catalogs (Full Table A.1) is only
    available at the CDS via anonymous ftp to
    \texttt{cdsarc.u-strasbg.fr} (\texttt{130.79.128.5}) or via
    \texttt{http://cdsarc.u-strasbg.fr/viz-bin/qcat?J/A+A/vol/page} }
}

\author{
      H. Inami\inst{1} \thanks{\email{\myemail}}, 
      R. Bacon\inst{1}, 
      J. Brinchmann\inst{2,3},
      J. Richard\inst{1},
      T. Contini\inst{4},
      S. Conseil\inst{1}, 
      S. Hamer\inst{1}, 
      M. Akhlaghi\inst{1}, 
      N. Bouch\'e\inst{4}, 
      B. Cl\'ement\inst{1}, 
      G. Desprez\inst{1}, 
      A. B. Drake\inst{1}, 
      T. Hashimoto\inst{1}, 
      F. Leclercq\inst{1}, 
      M. Maseda\inst{2},
      L. Michel-Dansac\inst{1}, 
      M. Paalvast\inst{2}, 
      L. Tresse\inst{1}, 
      E. Ventou\inst{4},
      W. Kollatschny\inst{5}, 
      L. A. Boogaard\inst{2},
      H. Finley\inst{4},
      R. A. Marino\inst{6},
      J. Schaye\inst{2},
      L. Wisotzki\inst{7}
}

\institute{ 
  Univ Lyon, Univ Lyon1, Ens de Lyon, CNRS, Centre de Recherche
  Astrophysique de Lyon (CRAL) UMR5574, F-69230, Saint-Genis-Laval,
  France
  \and 
  Leiden Observatory, Leiden University, P.O. Box 9513, 2300 RA
  Leiden, The Netherlands
  \and
  Instituto de Astrof\'isica e Ci\^encias do Espa\c{c}o, Universidade
  do Porto, CAUP, Rua das Estrelas, PT4150-762 Porto, Portugal
  \and 
  IRAP, Institut de Recherche en Astrophysique et Plan\'etologie,
  CNRS, Universit\'e de Toulouse, 14 avenue Edouard Belin, 31400
  Toulouse, France
  \and 
  Institut f\"ur Astrophysik, Universit\"at G\"ottingen,
  Friedrich-Hund-Platz 1, 37077 G\"ottingen, Germany
  \and
  ETH Zurich, Institute of Astronomy, Wolfgang-Pauli-Str. 27, CH-8093
  Zurich, Switzerland
  \and
  Leibniz-Institut f\"ur Astrophysik Potsdam (AIP), An der Sternwarte
  16, D-14482 Potsdam, Germany
}

\date{Accepted August 2, 2017}

\abstract {We have conducted a two-layered spectroscopic survey
  ($1\arcmin \times 1\arcmin$ ultra deep and
  $3\arcmin \times 3\arcmin$ deep regions) in the {\it Hubble} Ultra
  Deep Field (HUDF) with the Multi Unit Spectroscopic Explorer
  (MUSE). The combination of a large field of view, high sensitivity,
  and wide wavelength coverage provides an order of magnitude
  improvement in spectroscopically confirmed redshifts in the HUDF;
  i.e., $\COMBzHSTCtwo$ secure spectroscopic redshifts for {\it HST}
  continuum selected objects, which corresponds to $15\%$ of the total
  ($\COMBHSTpri$). The redshift distribution extends well beyond
  $z > 3$ and to {\it HST}/F775W magnitudes as faint as
  $\approx 30$~mag (AB, $1\sigma$).  In addition, $\COMBzORGonlyCtwo$
  secure redshifts were obtained for sources with no\ Hubble Space
  Telescope ({\it HST)} counterparts that were discovered in the MUSE
  data cubes by a blind search for emission-line features.  In total,
  we present $\COMBzALLCtwo$ high quality redshifts, which is a factor
  of eight increase compared with the previously known spectroscopic
  redshifts in the same field. We assessed redshifts mainly with the
  spectral features \oii at $z < 1.5$ (473 objects) and \lya at
  $2.9 < z < 6.7$ (692 objects).  With respect to F775W magnitude, a
  $50\%$ completeness is reached at $26.5$~mag for ultra deep and
  $25.5$~mag for deep fields, and the completeness remains
  $\gtrsim 20\%$ up to $28 - 29$~mag and $\approx 27$~mag,
  respectively.  We used the determined redshifts to test continuum
  color selection (dropout) diagrams of high-$z$ galaxies. The
  selection condition for F336W dropouts successfully captures
  $\approx 80\%$ of the targeted $z \sim 2.7$ galaxies. However, for
  higher redshift selections (F435W, F606W, and F775W dropouts), the
  success rates decrease to $\approx 20-40\%$. We empirically redefine
  the selection boundaries to make an attempt to improve them to
  $\approx 60\%$. The revised boundaries allow bluer colors that
  capture \lya emitters with high \lya equivalent widths falling in
  the broadbands used for the color-color selection. Along with this
  paper, we release the redshift and line flux catalog. }


\titlerunning{MUSE Ultra Deep Field Redshift Survey}
\authorrunning{Inami et al.}

\maketitle


\section{Introduction}\label{sec:intro}

The distance from an observer to a galaxy is nearly a prerequisite for
any scientific study. Fundamental physical properties of galaxies
(e.g., luminosity, star formation rate, and stellar masses) always include
uncertainties from distance measurements. Ideally, precise redshifts
are desired for all galaxies detected by imaging.  Although the
current technology cannot yet meet this demand, large efforts have
been made to obtain spectroscopic data as efficiently as possible.

Large spectroscopic galaxy surveys have greatly enhanced our ability
to explore the formation and evolution of galaxies by gathering a
wealth of data sets of spectral features (emission lines, absorptions
lines, and breaks), which in turn provide redshifts
\citep[e.g.,][]{York00, Lill07, Newm13, LeFe15}. Multi-object
spectroscopy (MOS) is the most widely used technique for
simultaneously obtaining a large number of spectra over a wide area,
but this technique requires preselection of the targets from imaging
data and has difficulty with dense fields because of potential slit or
fiber collisions (when only passing the field once).  A magnitude cut
is a typical selection method for a target sample. This process can
exclude faint galaxies, including quiescent absorption line galaxies
or dust attenuated red galaxies.  Also, a relatively bright magnitude
cut on the target sample and restrictions on slit positioning lead to
sparse sky sampling and introduce low spectroscopic completeness, in
particular toward higher density regions.

Another approach is photometric redshift (photo-$z$), which is derived
by fitting multiple galaxy templates as a function of redshift to the
observed photometry of a source.  Although photo-$z$ may be an
alternative to redshifts from MOS observations, its error is typically
$\Delta z/(1+z) \sim 0.05$ \citep[e.g.,][]{Bonn16,Beck17}. This
precision is inadequate for tasks such as finding (close) galaxy
pairs, identifying overdensity regions, and studying (local)
environmental effects.

An integral field unit (IFU) with a wide field of view (FoV), high
sensitivity, wide wavelength coverage, and high spectral resolution
can remedy some of these issues. The IFU yields spectroscopy for each
individual pixel over the entire FoV, which means that every detected
object in the field has a spectrum.  However, earlier generation IFUs
had FoV sizes that were too restrictive, insufficient spectral resolution, or
spatial sampling that was too low for spectral surveys.

The Multi Unit Spectroscopic Explorer \citep[MUSE;][]{Baco10}, is an
IFU on the Very Large Telescope (VLT) Yepun (UT4) of the European
Southern Observatory (ESO).  It is unique in having a large FoV
($1\arcmin \times 1\arcmin$), wide simultaneous wavelength coverage
($4650 - 9300$\Ang), relatively high spectral resolution
($R \sim 3000$), and high throughput (35\% end-to-end).  The MUSE
instrument has a wide range of applications for observing objects such
as H\,\textsc{ii} regions, globular clusters, nearby galaxies, lensing
cluster fields, quasars, and cosmological deep fields.  Despite its
universality, MUSE was originally designed and optimized for deep
spectroscopic observations.  One of its advantages is that it can deal
with dense fields without making any target preselection, achieving a
spatially homogeneous spectroscopic completeness and reducing the
uncertainty in assigning a measured redshift to a photometric object
\citep[cf.][]{Baco15,subm_Brin17}.

We have conducted deep cosmological surveys in the {\it Hubble} Ultra
Deep Field \citep[HUDF;][]{Beck06} with MUSE. This field has some of
the deepest multiwavelength observations on the sky, from X-ray to
radio. However, the previously known secure spectroscopic redshifts
(spec-$z$) in the HUDF are limited to about only $2\%$ of the detected
galaxies \citep[169 out of 9927 galaxies;][]{Rafe15}.  The
three-dimensional data of MUSE also facilitate blind identifications
of previously unknown sources by searching for emission lines directly
in the data cube \citep[cf.][]{Baco15}. In addition to increasing the
spectroscopic completeness in the HUDF with MUSE, the unique
capabilities of MUSE have permitted some of the first detailed
investigations of photo-$z$ calibration of faint objects up to 30th
magnitude \citep[][hereafter Paper~III]{subm_Brin17}, properties of a
statistical sample of \ciii emitters \citep{subm_Mase17}, spatially
resolved stellar kinematics of intermediate redshift galaxies
\citep{subm_Guer17}, constraints on the faint-end slope of
Lyman-$\alpha$ (\lya) luminosity functions and its evolution
\citep{subm_Drak17}, spatial extents of a large number of \lya haloes
\citep{subm_Lecl17}, redshift dependence of the \lya equivalent width
and the UV continuum slope \citep{subm_Hash17}, galactic winds in
star-forming galaxies \citep{subm_Finl17}, evolution of the galaxy
merger rate \citep{subm_Vent17}, and analyses of the cosmic web
\citep{subm_Gall17}.  In this paper, we report on the first set of
redshift determinations in the two layered MUSE UDF fields with
$\sim 10$~hour and $\sim 30$~hour integration times used for all of
the studies above.

The paper is organized as follows: In \S\ref{sec:obs}, we explain our
deep surveys, observations, and data reduction in brief.  Then we
describe the source and spectral extraction methods, procedure of
the redshift determination, detected emission line flux measurements,
and continuum flux measurements in \S\ref{sec:analysis}.  The
resulting redshift distributions and global properties are shown in
\S\ref{sec:result}, followed by \S\ref{sec:discussion} in which we
discuss success rates of the redshift measurements, emission line
detected objects, comparisons with previous spec-$z$ and photo-$z$,
and color selections of high-$z$ galaxies.  Finally, the summary and
conclusion are given in \S\ref{sec:summary}. Along with this paper, we
release the redshift catalogs (see Appendix~\ref{app:cat}).  The
magnitudes are given in the AB system throughout the paper.

\section{Observations and data reduction}\label{sec:obs}

The detailed survey strategy, data reduction, and data quality
assessments are presented in \citet[hereafter
Paper~I]{subm_Baco17}. Here we provide a brief outline.  As part of
the MUSE Guaranteed Time Observing (GTO) program, we carried out deep
surveys in the HUDF.  There are two layers of different depths in
overlapping areas: the $3\arcmin \times 3\arcmin$ medium deep and
$1\arcmin \times 1\arcmin$ ultra deep fields. The medium deep field
was observed at a position angle (PA) of $-42 \deg$ with a
$3\arcmin \times 3\arcmin$ mosaic (\textsf{udf-01} to
\textsf{udf-09}), and thus it is named the \mosaic. The ultra deep
region (named \udft) is located inside the \mosaic with a PA of
$0 \deg$.  We selected this field to maximize the overlap region with
the ASPECS\,\footnote{The ALMA SPECtroscopic Survey in the Hubble
  Ultra Deep Field} project \citep{Walt16}.

The observations were conducted between September 2014 and February
2016. during eight GTO runs under clear nights, good seeing, and
photometric conditions.  The average seeing measured with the obtained
data has a full width at the half maximum (FWHM) of $\approx 0.6''$ at
$7750$\Ang. The total integration time for the \mosaic is
$\approx 10$~hours for the entire field ($9.92$ sq. arcmin). The \udft
was observed for $\approx 21$~hours (at a PA of $0 \deg$), but the final
product was added together with the overlap region of the \mosaic to
achieve $\approx 31$~hours.

The data reduction begins with the MUSE standard pipeline
\citep{Weil12}. For each exposure, it uses the corresponding
calibration files (flats, bias, arc lamps) to generate a {\it
  pixtable} (pixel table), which stores information of wavelength,
photon count, and its variance at each pixel location. After
performing astrometric and flux calibrations on the pixtable, a data
cube is built. Then we implemented the following calibrations beyond
the standard pipeline.  The remaining low-level flat fielding was
removed by self-calibration, artifacts were masked, and the sky
background was subtracted. Finally, we combined the 227 individual
data cubes to create the final data cube for the \mosaic. For \udft,
51 of the PA $0 \deg$ data cubes and 105 overlapping \mosaic data
cubes were used to make the final product.

The MUSE data cubes used for this work are version 0.42 for both the
\udft and \mosaic.  The estimated $1\sigma$ surface brightness limits
are $2.8$ and
$5.5 \times 10^{-20} \, {\rm
  erg\,s^{-1}\,cm^{-2\,}\,\AA^{-1}\,arcsec^{-2}}$
in the wavelength range of $7000-8500$\Ang (excluding regions of OH
sky emission) for \udft and the \mosaic, respectively. The $3\sigma$
emission line flux limits for a point-like source are $1.5$ and
$3.1 \times 10^{-19} \, {\rm erg\,s^{-1}\,cm^{-2\,}}$, respectively,
at $\sim 7000$\Ang, where has no OH sky emission line.

\section{Analysis}\label{sec:analysis}

\subsection{Source and spectral extractions}

We performed two different source extractions in the MUSE Ultra Deep
Field (UDF).  One uses the {\it HST} detected sources from the UVUDF
catalog\,\footnote{The ultraviolet UDF (UVUDF) catalog includes the
  photometries of 11 {\it HST} broadbands from F225W to F160W.}
\citep{Rafe15} as the priors to extract continuum selected
objects. The other searches for emission lines blindly (without prior
information) in the cube directly. The details of the source
detections in the MUSE UDF are discussed in Paper~I. Here we briefly
summarize the methods of the source detection and spectral extraction
of the detected sources.

\subsubsection{Continuum selected objects}\label{sec:analysis_cont}

Among all of the 9969 galaxies in the UVUDF catalog, $\UDForgHSTpri$
and $\MOSorgHSTpri$ of these serve as the priors to extract objects in
the \udft and \mosaic survey fields, respectively. We performed the
extraction at all of the {\it HST} source positions regardless of
whether they are detected in the MUSE data cube.  When the {\it
  HST}-detected galaxies are located within $0.6''$, which cannot be
spatially resolved by our observations, they were merged and treated
as a single object.  We computed the new coordinates for a merged
object via the {\it HST} F775W flux-weighted center of all objects
composing the new merged source.  We neither used their photometric
measurements in this work nor reported the photometric measurements in
the catalogs (see note (e) of Table~\ref{tbl:cat}).  In total, we
acquired $\UDFHSTpri$ and $\MOSHSTpri$ MUSE objects\,\footnote{These
  numbers increase after the redshift analysis because we manage to
  ``split'' some of these objects based on their emission line
  narrowband images (see \S\ref{sec:result_cont}).} for the \udft and
\mosaic fields, respectively.  Then we made a
$5\arcsec \times 5\arcsec$ subcube (or larger for extended objects)
for each of the extracted objects so that they are easy to handle.
For each object, we convolved its {\it HST} segmentation map (provided
by the UVUDF catalog) with the MUSE point spread function (PSF) to
construct a mask image to mask out emission from nearby objects.

For the purpose of redshift determination, while we used all of the
extracted objects in \udft , we only inspected the objects with
$\rm F775W \leq 27$~mag ($\MOSHSTpriMAGcut$ MUSE objects) in the
\mosaic. This magnitude cut was selected after we completed the
redshift analysis in \udft. We discuss in detail how we decided
to make the cut at this magnitude in Section~\ref{subsec:mag_cut}.

\subsubsection{Emission line selected objects}\label{subsec:analysis_origin}

We also searched for emission lines directly in the data cubes to
identify objects.  We adopted the software {\tt ORIGIN} \citep[detectiOn
and extRactIon of Galaxy emIssion liNes;][]{prep_Mary17} to perform
blind detections.

{\tt The ORIGIN} software uses a matched filter in three-dimensional
data to detect signals by correlating with a set of spectral
templates.  It first uses a principal component analysis (PCA) to
remove continuum emission.  A matched filter is then applied to the
continuum removed data cube and the $P$-value test is used to assign a
detection probability to each emission line candidate.  When the test
gives a high probability, a narrowband image is created using the raw
data cube to further test the significance of the line. The line is
only considered to be a real detected line when it survives the
narrowband image test.  Finally, the spatial position of each detected
line is estimated by spatial deconvolution. For more details on the
procedures and parameters used for the detections in \udft and the
\mosaic, we refer to Paper~I.  In \udft and the \mosaic, $\UDFORIGIN$
and $\MOSORIGIN$ emission line objects were detected with {\tt
  ORIGIN}, respectively.

In addition to {\tt ORIGIN}, we also used the {\tt MUSELET}
software\,\footnote{{\tt MUSELET} is publicly available as a part of
  the {\tt MPDAF} software \citep{subm_Piqu17}. More details at
  \url{http://mpdaf.readthedocs.io/en/latest/muselet.html}} to test
our blind search for line emitters in the Mosaic field. {\tt The MUSELET}
software is a {\tt SExtractor}-based detection tool, which identifies line
emission sources in continuum-subtracted narrowband images of a
cube. The narrowband images are produced by collapsing (weighted
  average) the closest 5 wavelength planes (6.25\Ang) and subtracting
a median continuum from the closest 20 wavelength planes (25\Ang) on
each of the blue and red sides of the narrowband regions. By
comparing the UVUDF sources and {\tt ORIGIN} detections with the raw
{\tt MUSELET} catalogs of emission line sources in the \mosaic, 
  {\tt MUSELET} only detects 60\% of the objects which {\tt ORIGIN}
  detects, but it finds 16 additional sources. The lower detection
  rate and small number of additional detections of {\tt MUSELET}
  strengthens our choice of {\tt ORIGIN} as the primary line
detection software for the analyses of these fields. All of the
sources detected only by {\tt MUSELET} are also included in the
analyses of this work and the released catalogs.

\subsubsection{Spectral extractions}

We adopted three different methods for spatially integrating the
data cube to create one-dimensional spectral extractions: unweighted
sum, white-light weighted, and PSF weighted. The unweighted sum is a
simple summation of all of the flux in each wavelength slice over the
mask region (the MUSE PSF convolved {\it HST} segmentation map).
The white-light weighted spectrum is computed by giving a weight
  based on the flux in each spatial element of the MUSE white-light
  image, which is made by collapsing the MUSE data in the wavelength
  direction.  For the PSF-weighted spectrum, the estimated PSF (as a
function of wavelength) is used as the weight for the spectral
extraction.  Local residual background emission is then subtracted
from the extracted spectrum. The local background region is defined by
the area outside of the combined resampled {\it HST} segmentation maps
of all of the prior sources in the $5\arcsec \times 5\arcsec$
subcube.  The global sky background emission has already
been removed during the data reduction.

The weighted optimal extractions offer the advantage of reducing
contamination from neighboring objects that are not covered by the
mask. Although in many cases the weighted extractions provide a higher
signal-to-noise ratio (S/N) in the extracted spectrum, they can be
disadvantageous for objects whose emission line features are spatially
extended or offset from the position of the white-light emission or
the PSF. This characteristic is often seen in \lya emitters
\citep[e.g.,][]{Wiso16, subm_Lecl17}.

As the default for the redshift determination process, because we
preferentially require spectra with a good S/N, we primarily use
white-light and PSF weighted spectra, depending on the galaxy size
found by {\tt SExtractor} \citep{Bert96} in the {\it HST} F775W image
reported in the UVUDF catalog.  When the FWHM of an object is extended
more than the average seeing, 0.7\arcsec, the primary spectrum to
inspect is white-light weighted.  For the galaxies with FWHM
$< 0.7$\arcsec, the PSF-weighted spectrum is used. However, when
necessary, redshift investigators can use any of the extracted spectra
or a user-defined spectrum to investigate spectral features in
depth. For some cases, a simple extraction in a small circular
aperture ($\approx 0.8''$ in diameter) is useful to confirm
observed features.

\subsection{Redshift determination}\label{subsec:z_meas}

We adopted a semi-automatic method for the redshift identification. For
the continuum selected galaxies, we used the redshift fitting software
{\tt MARZ} \citep{Hint16}. The redshift determination of {\tt MARZ} is
based on a modified version of the {\tt AUTOZ} cross-correlation
algorithm \citep{Bald14}.  From a user-defined input list of spectral
templates, it automatically finds the best-fit template to the input
spectrum and determines the redshift. If the resulting redshift is not
ideal, then the user can also update it interactively in the same
window. We used the peak position of the most significant detected
feature to define redshift, including \lya.  The \lya peak
emission may not represent the systemic redshift due to radiation
transfer effects of the resonant \lya line in bulk motion of neutral
hydrogen gas \citep{Hash13}.

\begin{figure*}
  \begin{center}
    \includegraphics[width=\textwidth]{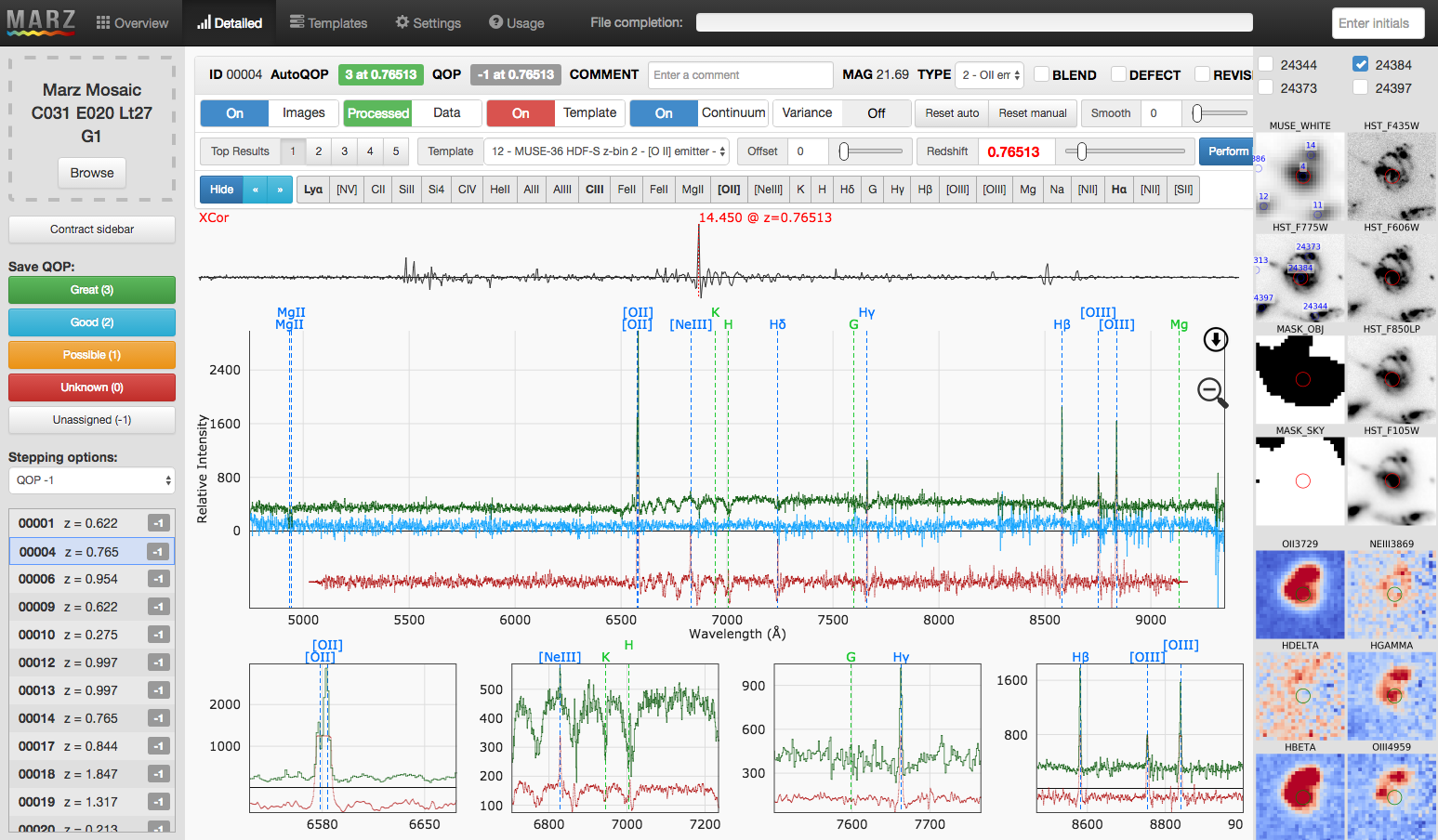}
    \caption{Redshift tool {\tt MARZ} with some modifications for
      our redshift analysis. This tool reports the most likely redshift and
      the quality (QOP) by cross-correlating with input spectral
      templates. The redshift identifiers verify the suggested
      redshift or manually find a better redshift and provide its
      quality (indicated as CONFID in the catalogs). We also implement
      a feature to show images in the panel on the right.  }
    \label{fig:marz}
  \end{center} 
\end{figure*}

In addition to its original capabilities, we customized {\tt MARZ} to
match our needs. We added a panel to display {\it HST} images along
with the UVUDF ID numbers and MUSE white-light images.  The original
{\tt MARZ} already has a button for confidence level implemented
(called the quality operator, QOP, in {\tt MARZ}) and, in addition, we
integrated TYPE and DEFECT buttons (see below for the description of
these parameters).  Instead of only inputting a one-dimensional
spectrum, we can feed the subcube such that it can create narrowband
images simultaneously when a redshift is selected or when certain
emission/absorption features are selected in the spectrum.  A
screenshot of the customized {\tt MARZ} is presented in
Figure~\ref{fig:marz}.

The input spectral templates for {\tt MARZ} are listed and explained
in Appendix~\ref{app:temples}. Some of the templates ($\rm No.\,6-19$)
are made using the MUSE data. During the redshift identification
process, no photometric redshift information is provided to the
inspectors to avoid biases when we compare MUSE determined redshift
against photometric redshift (Paper~III). Instead, we display all of
the existing {\it HST} UV to near-infrared images as supplemental
information to the MUSE spectra. Although it is not yet
  implemented in our modified {\tt MARZ}, we found that {\it HST} color
  images also help constrain spectroscopic redshifts and identify
  the corresponding object that is associated with the determined
  redshift.

For each of the determined redshifts, a confidence level (CONFID) from
3 to 1 is assigned:

\begin{itemize}
\item[{\bf 3}:] Secure redshift, determined by multiple features
\item[{\bf 2}:] Secure redshift, determined by a single feature
\item[{\bf 1}:] Possible redshift, determined by a single feature
  whose spectral identification remains uncertain
\end{itemize}

Although we identified emission and absorption features in the
one-dimensional spectra, we employed narrowband images of the features
to confirm the significance of the detections. The CONFID is assigned
only when the identified features are seen in its narrowband
images. The redshifts with CONFID of 2 or 3 are reliable, but if one
is interested in using the redshifts with CONFID of 1, it is highly
recommended to double check the spectrum and use it with care.

While measuring redshifts, we kept records of which feature is
mainly used for the identification. This parameter is saved as the
integer, TYPE, corresponding to

\begin{itemize}
\item[{\bf 0}:] Star
\item[{\bf 1}:] Nearby emission line object
\item[{\bf 2}:] \oii emitter
\item[{\bf 3}:] Absorption line galaxy
\item[{\bf 4}:] \ciii emitter
\item[{\bf 5}:] \oiii emitter
\item[{\bf 6}:] \lya emitter
\item[{\bf 7}:] Other types
\end{itemize}

In general, the most prominent feature is used to define the type.
We distinguished nearby galaxies (1) and \oii emitters (2) by a
detection of \ha. Thanks to the relatively high spectral resolution of
MUSE, it is possible to distinguish the \oii$\lambda\lambda3726,3729$
and \ciii$\lambda\lambda1907,1909$ doublet features based on their
intrinsic separations at a certain redshift. It also helps to resolve
the distinctive asymmetric profile of \lya.  We discuss this further in
\S\ref{sec:result}, but the majority of redshifts are determined by
identifying the most prominent feature, the \oii and \lya emission at
low redshift ($z < 1.5$), and high redshift ($z > 3$), respectively.
Galaxies are only classified as \oiii emitters when \oiii is more
prominent than \oii in cases both are detected.

We also kept the information of the data quality with the parameter
DEFECT. The default is $\rm DEFECT= 0$, meaning no problem is found in
the data.  When $\rm DEFECT = 1$, it indicates that there are some
issues with the data, but the data are still usable. This flag is
raised often owing to the object lying at the edge of the survey
area. This may indicate that line fluxes of these sources are
underestimated because they are truncated in the data. As long as a
spectral feature can be identified, we are able to determine a
redshift, but we cannot recover the total line fluxes beyond the edge
of the survey area.

For each continuum detected object, at least three investigators
independently assessed the redshift.  For \udft, because it is our
reference field, we had six investigators look at all of the
objects. For the \mosaic, $\MOSHSTpriMAGcut$ objects
($\rm F775W \leq 27$~mag, see Section~\ref{sec:analysis_cont}) are
divided equally for two groups with three investigators each. In the
case of the emission line detected objects, two investigators
individually determined redshift for all of these objects for both
\udft and the \mosaic.  For both continuum and emission line detected
objects, among each of the groups, any disagreements in their
determinations of redshift, CONFID, TYPE, and DEFECT were consolidated
to provide a single set of redshift results. With the consolidated
results, we had at least one more person, who is not in the group,
check all of the determined redshifts and related parameters to
finalize the redshift assessment.

Then we refined the redshift by simultaneously fitting all of the
detected emission lines and measure line fluxes using a spectral
fitting software. This process is discussed in the next subsection.

\subsection{Line flux measurements}\label{subsec:line_flux}

For the redshift determination, the spectra utilized were primarily
extracted based on either the white-light or PSF weighted extraction
in order to optimize S/N. For the purpose of cataloging line fluxes,
we alternatively used the unweighted summed spectra for all of the
sources within each segmentation map\,\footnote{Line fluxes measured
  in the segment may be less than the total flux because of aperture
  effects.}.  This avoids possible biases introduced by the weighting
scheme. It is also better when the spatial distribution of the line
emission is more extended than the white-light emission or PSF. This
is especially relevant for \lya emission \citep[cf.][]{Wiso16,
  subm_Lecl17}.  However, this results in an increased
scatter for faint emission lines.

For the detected emission lines with $\rm S/N > 3$, the line flux and
its uncertainty are provided along with the measured redshift in the
released catalog (Appendix~\ref{app:cat}). We used the spectral line
fitting software {\tt PLATEFIT} \citep{Trem04, Brin08b} to
simultaneously fit all of the emission lines falling within the
spectral range.  This tool first uses a set of model template spectra from
\cite{BC03} with stellar spectra from MILES \citep{Sanc06} to fit the
continuum of the observed spectrum at a predefined redshift with
strong emission lines masked.  After subtracting the fitted continuum
from the observed spectrum, a single Gaussian profile is fitted to
each expected emission line in velocity space.  The velocity offset
(the velocity shift relative to the redshift of the galaxy) and
velocity dispersion of all of the emission lines are tied to be the
same.  See \cite{Trem04} for a more detailed description of the
software.

The velocity offset is limited to vary within
$\pm 300 \, {\rm k m\,s^{-1}}$. For emission line galaxies, we used the
velocity offset output from {\tt PLATEFIT} to refine the original
(input) redshift. The calculation of the velocity offset
includes \lya when it is detected. The typical velocity offset is
about $10$~\kms on average. When the original redshift is obtained
directly from {\tt MARZ}'s cross-correlation, we do not expect the
offset to be significant, but it can be larger when the redshift
identifiers manually estimate the redshift.

As described above, {\tt PLATEFIT} fits single Gaussian functions to
the emission lines. However, this is not suitable for \lya because it
is often observed to have an asymmetric profile.
Thus, we added a feature to the software to fit the complicated profile
of \lya. Instead of fitting \lya with a single Gaussian function, we
used a combination of multiple Gaussian functions (up to 21, although
typically much fewer are used) at fixed relative positions to fit the
\lya profile.  We placed a central component Gaussian function at the
position of \lya as determined by the single Gaussian fitting carried out in
the previous step. We placed an additional 10 components on either side
of this, at a constant spacing of $120\,{\rm km\,s^{-1}}$ to be
consistent with the spectral resolution of the data. As such the
fitting covers a velocity range of $\pm 1200\,{\rm km\,s^{-1}}$ from
the expected position of the line.

We then performed the minimization via a nonlinear least squares
method.  During this process, the fluxes of each Gaussian component
are allowed to vary independently (with most falling to 0 as they lie
in the spectrum where \lya has no emission) until the aggregate of all
components best matches the observed spectrum.  The separation between
each component is kept constant (in velocity) but the wavelength of
the central component is allowed to shift
($\pm 300 \, {\rm km\,s^{-1}}$, consistent with fits for other lines).
This allows for the fitting to correct for small discrepancies in the
central position determined during the multiple Gaussian component
fitting so that it produces a more accurate fit to \lya.  The widths
of individual Gaussian components are forced to be the same but the
width itself is a free parameter. The  {\tt PLATEFIT} tool places a limit of
$< 500 \, {\rm km\,s^{-1}}$ for each line width, but in general the
Gaussian components only vary slightly from the initial guess of
$70 \, {\rm km\,s^{-1}}$ and fall between $60$ and
$120 \, {\rm km\,s^{-1}}$.  Allowing the widths of the components to
vary in this way produces a better fit to the data in a few cases,
compared to running the fitting with a single fixed width for all of
the components. However, the improvement is much less significant than
that obtained by allowing the central velocity of the components to
shift.

\begin{figure}
  \begin{center}
    \includegraphics[width=\textwidth/2, trim=0 0 0 220]{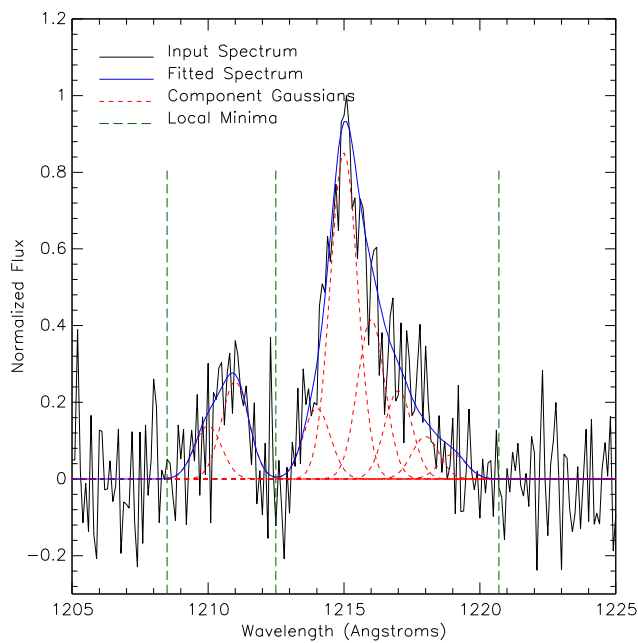}
    \caption{Demonstration of the complex fitting for a \lya
      emission line presenting blue and red bumps. The fitting is performed
      on a simulated spectrum (black line) to better illustrate
      the procedure (the flux is in arbitrarily units). In the real
      data, the final fit is usually comprised of fewer
      components. The dashed red lines are the individual Gaussian
      components and the solid blue line is the total of these
      components (the final fit). The vertical green dashed
      lines indicate the local minima. We
        use the \lya peak position to measure the redshift (see
        \S~\ref{subsec:z_meas}).}
    \label{fig:lya_complex_fit}
  \end{center} 
\end{figure}

After the fitting, we studied the output spectrum of the complex
fitted regions and identified individual complex ``lines'' (a
composite of multiple Gaussian components) by searching the fit
spectrum for local maxima. The routine then scans out from each maxima
on either side of the peak until it identifies a local minimum or the
fitted spectrum returns to 0.  The flux of each complex line is then
calculated along with its associated error and fitted ``lines'' with a
S/N of less than 3 are dropped. This S/N cut successfully avoids
overfitting the noise and removes unphysical fits (such as more than
two components or very broad components) in most cases.  The
properties of the remaining lines are then calculated directly from
the fitted spectrum using a model independent approach. Errors are
determined using a Monte Carlo method by modifying the input spectrum
with random values within the associated errors at each point 100
times. Each line property is then re-extracted for each realization
and the standard deviation of these values is given as the error.  An
example of the complex fitting is presented in
Figure~\ref{fig:lya_complex_fit}.

\subsection{Continuum flux measurements}\label{subsec:NoiseChisel}

Throughout the paper, we used the continuum fluxes from the {\it HST}
observations provided in the UVUDF catalog.  In addition to that, we
measured our own continuum fluxes for the objects that are not in the
UVUDF catalog but are identified by directly detecting their emission
line features with {\tt ORIGIN} and {\tt MUSELET} (see
\S\ref{subsec:analysis_origin}). For these objects, we performed our own
photometric measurements on the {\it HST} images to obtain
segmentation maps and continuum fluxes or upper limits. A full
analysis of why these objects are not in the UVUDF catalog and our
method of extracting their broadband properties are described in
Paper~I. Here we provide a brief summary.

Among $\numnotinraf$ objects that are only found by {\tt
  ORIGIN/MUSELET}, roughly one-quarter have very low or no continuum
emission, and thus they are hardly detectable in any broadband
images.  The rest of the objects can be visually identified in the
images, but they are not detected by {\tt SExtractor}\,\footnote{In
  Section~7.3 of Paper~I, we discuss why they were not detected by
  {\tt SExtractor}.} \citep{Bert96}.  In order to measure continuum
fluxes (or upper limits) of these objects, we used {\tt
  NoiseChisel}\,\footnote{\url{http://www.gnu.org/software/gnuastro/manual/html_node/NoiseChisel.html}},
which is an image analysis software employing a noise-based detection
concept \citep{Akhl15}. It is run independently on each of the {\it
  HST} images and the largest (from all the filters) object closer
than $\maxmatchdist\arcsec$ to the position reported by {\tt
  ORIGIN/MUSELET} is taken as representing the pixels associated with
it to create a segmentation map for each object.  We are able to
associate a {\tt NoiseChisel} segmentation map for $\rafprobinnc\%$ of
the objects only found by {\tt ORIGIN/MUSELET}. However, for the rest
of the objects, no {\tt NoiseChisel} segmentation map (in any of the
{\it HST} broadband images) can be found within
$\maxmatchdist\arcsec$.  For these objects, a
$\aperturediameter\arcsec$ diameter circular aperture was placed
on the position reported by {\tt ORIGIN/MUSELET}.

The broadband magnitude is found by feeding the image of each filter
and the final segmentation map into {\tt MakeCatalog}, which takes
output of {\tt NoiseChisel} as input to directly create a catalog.
The error in magnitude for each segment is derived from the S/N
relation: $\sigma_{M}=2.5/(S/N \times \ln{(10)})$ \citep[see Eq. (3)
in][]{Akhl15}.  {\tt MakeCatalog} produces upper limit magnitudes for
each object, by randomly positioning its segment in $\mkcatupnum$
different blank positions over each broadband image and using 
  $\mkcatupnsigma\sigma$ of the standard deviation of the
  resulting distribution. If the derived magnitude for an object in a
filter is fainter than this upper limit, the upper limit magnitude is
used instead.

\section{Results}\label{sec:result}

\subsection{Redshift determination in the MUSE Ultra Deep Field (\udft)}

In this subsection, first we report the measured redshift and the
parameters associated with it in our reference field, the MUSE Ultra
Deep Field (\udft). In addition to being the deepest spectroscopic
field so far observed with MUSE, all of the extracted MUSE spectra in
this field have been visually inspected. In the \udft field, as
discussed in \S\ref{sec:analysis}, we used two different procedures to
extract the sources. Their redshifts were determined independently
first, and then reconciled. Here we first present the basic
properties of the objects associated with their redshift measurements
for each extraction method independently. Based on our understanding
of the differences and the relationship between the two source
extraction methods and the redshift properties, we tried to find the
best combination of these two methods to efficiently maximize the
detection rate of redshifts in the even larger sample size of the MUSE
Deep Field (the \mosaic).

\subsubsection{Redshifts of the continuum selected objects}\label{sec:result_cont}

Among $\UDFHSTpri$ continuum selected objects in the MUSE Ultra Deep
Field (\udft), we successfully measured the redshifts of $\UDFzHSTCone$
objects in the redshift range from $z=0.21$ to $6.64$. The number of
redshifts with confidence level $\geq 2$ is $\UDFzHSTCtwo$.

\begin{figure*}
  \begin{center}
    \includegraphics[width=\textwidth]{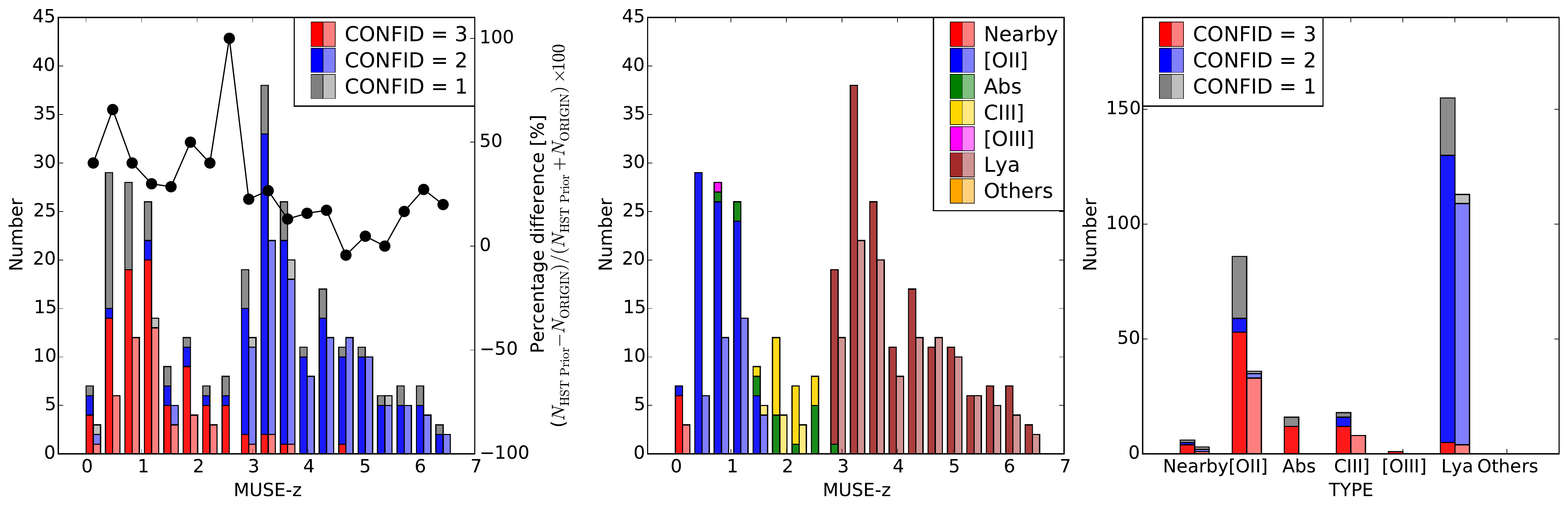}
    \caption{MUSE redshift distributions of the objects extracted
      with the {\it HST} continuum-detected priors (filled bars) and
      the direct line emission searches ({\tt ORIGIN}; filled and
      faded color bars on the right in each bin) in the MUSE Ultra
      Deep Field (\udft). All of the extracted sources are shown
      without removing overlaps between the extraction methods. {\bf
        [Left]} The redshift histogram in bins of $\Delta z = 0.35$.
      The red, blue, and gray represent the confidence levels
      3, 2, and 1, respectively, for the determined redshifts. The
      filled dots show the percentage difference relative to the total
      of {\it HST} prior and {\tt ORIGIN} MUSE-$z$ distributions.
      {\bf [Middle]} The same redshift distribution as the left
      panel, but color coded by the classified type of the
      objects. {\bf [Right]} The histogram of the classified type of
      the objects color coded by the redshift confidence levels. }
    \label{fig:u10_musez_hstpri_org}
  \end{center} 
\end{figure*}

In the left two panels of Figure~\ref{fig:u10_musez_hstpri_org}, we
show the distributions of redshifts found with the MUSE data.  
Because most of the objects at $0 < z < 1.5$ are usually identified by
the \oii doublet, their CONFID is mostly 3 by definition when the
double peaks are resolved and clearly seen.  There are six cases in which
CONFID is 2 in this redshift range, whose spectra in general have
lower S/Ns but the features are obviously detected in narrowband
images.

On the other hand, galaxies at $z > 3$ are often found by a single
feature \lya, which gives CONFID of 2. For the case of $\rm CONFID=3$
at $z > 3$, the \ciii emission or UV absorption features, and in some
rare cases He\,\textsc{ii}, are also discerned. The number of
determined redshifts drops significantly at
$1.5 \lesssim z \lesssim 3.0$. This is a well-known ``redshift
desert'', where \oii shifts out from the red end of the spectral
coverage, although \lya is still too blue to be detected. With the
deep data, we are able to recover some of the redshifts in this range
by detecting \ciii emission and some absorption features (e.g.,
Fe\,\textsc{ii}).

There are also 59 redshifts with $\rm CONFID=1$.  Their redshifts are
difficult to determine with confidence, because in most cases, it is
ambiguous to use their line profile to specify the feature (e.g., \oii
versus \lya). For \oii emitter at $0.25 < z < 0.85$, \hb and \oiii
  are in general also available, but they lie in a sky line crowded
  region, which can sometimes prevent further constraint on the
  measured redshift to give a higher CONFID.

\subsubsection{Redshifts of the emission line selected objects}

Of the $\UDFORIGIN$ emission line selected objects (objects detected
with {\tt ORIGIN} regardless of whether they have UVUDF counterparts),
$\UDFzORGCtwo$ objects have secure redshift measurements with CONFID
of 2 or 3 in the redshift range of $0.28-6.64$. When we include the
$\rm CONFID=1$ redshifts, the resulting number of measured redshifts
is $\UDFzORGCone$.  Since we have to actually detect the emission
features to select this sample, it is expected that the number
fraction of $\rm CONFID=1$ redshifts is much smaller for the emission
line selected objects compared with the continuum selected objects.  A
crucial difference between the continuum and line emission selected
objects is that the former are extracted regardless of whether they
are detected with MUSE or not, but the latter requires actual line
detections.

The redshift distributions of the emission line objects are shown as
the faded color bars (on the right in each bin) in
Figure~\ref{fig:u10_musez_hstpri_org}. Similar to the continuum
selected objects, the majority of the redshifts at $z < 3$ have
$\rm CONFID=3$, while mostly $\rm CONFID=2$ at $z > 3$, because of the
detectable spectral features. The main features that {\tt ORIGIN}
detects to facilitate the redshift identifications are \oii, \lya, and
\ciii. The lack of redshifts in this range is more significant for the
emission line selected objects because by nature {\tt ORIGIN} does not
have the ability to find any absorption galaxies.

\subsubsection{Redshift comparisons between the continuum and emission
  line selected objects}\label{sec:HSTpri_vs_ORIGIN}

When we only consider the secure redshifts ($\rm CONFID \geq 2$), the
derived redshifts of both the continuum and emission line selected
galaxies cover a similar range, from $z=0.2$ to $6.7$. A large
difference is the numbers of identified redshifts between these
two data sets. The continuum selected galaxies have $\UDFzHSTCtwo$
secure redshifts, while the emission line selected galaxies have
$\UDFzORGCtwo$. If we consider that of these $\UDFzORGCtwo$ objects,
$\UDFzORGonlyCtwo$ do not have a continuum detection (not in the prior
list), the difference is even larger.

\begin{figure}
  \resizebox{\hsize}{!}
  {\includegraphics{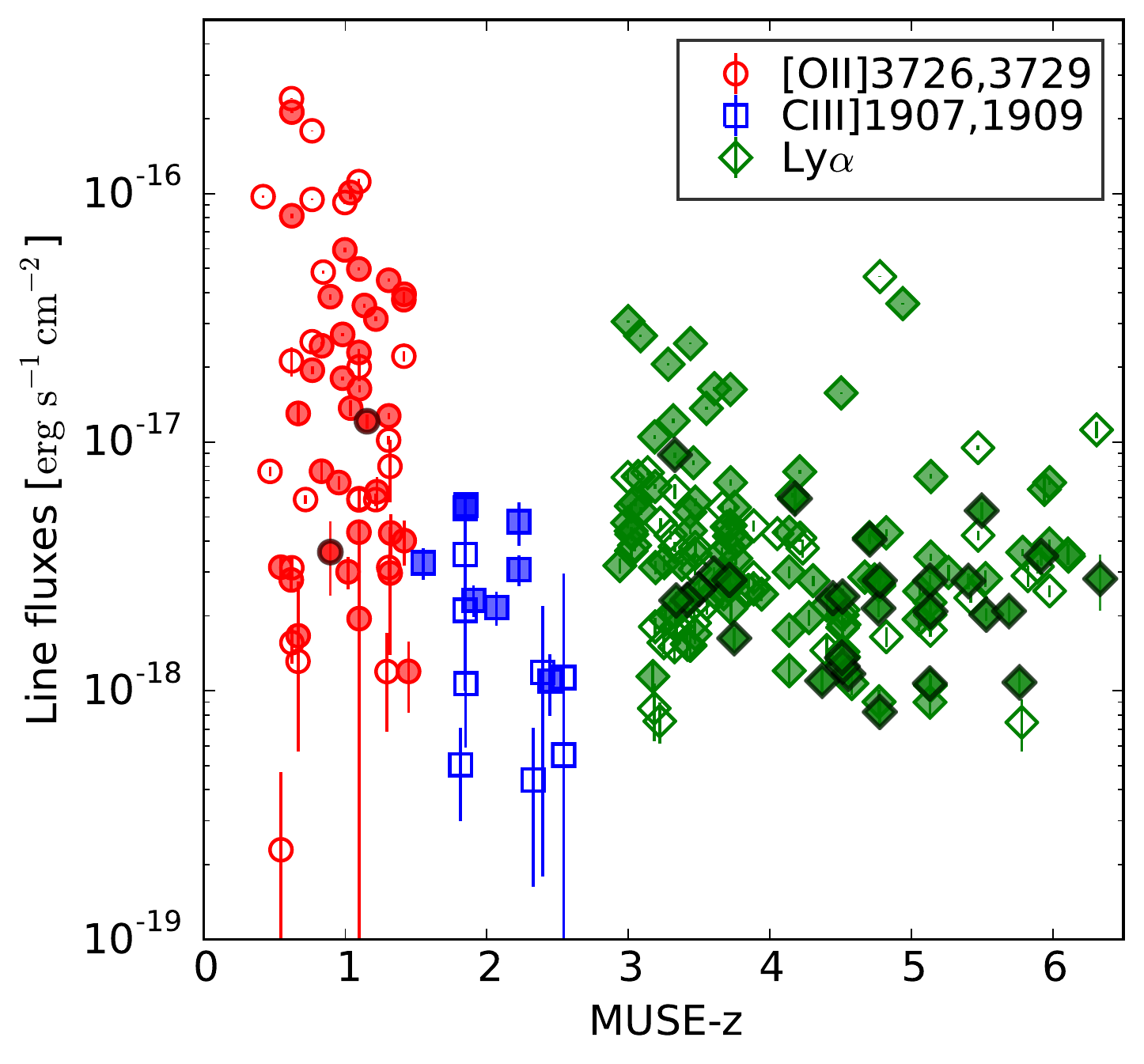}}
  \caption{ Line fluxes of \oii$\lambda\lambda3726,3729$ (the red
    circles), \ciii$\lambda\lambda1907,1909$ (the blue squares), and
    \lya (the green diamonds) of the redshift-identified galaxies in
    \udft ~(CONFID $\geq 2$). The open and filled symbols are
    the {\it HST} continuum and emission line selected objects,
    respectively. The filled symbols with the black edge indicate that
    they are detected only by {\tt ORIGIN}. }
  \label{fig:u10_Fline_hst_org}
\end{figure}

In the left panel of Figure~\ref{fig:u10_musez_hstpri_org}, the
percentage difference of the numbers of determined MUSE redshifts
between the two extraction methods ({\it HST} prior or {\tt ORIGIN})
is shown with the filled dots.  {\tt The ORIGIN} method is optimized to detect
compact emission line objects with faint continuum, such that it is more
sensitive to the detection of high-$z$ emission lines. This method misses
$44\%$ of the \oii emitters identified in the continuum selected
galaxies. For the \lya emitters, $38\%$ of the continuum detected
sample are not found by {\tt ORIGIN}.  There is no obvious trend of
the fraction of missed \oii or \lya with redshift
(Figure~\ref{fig:u10_Fline_hst_org}). There is no trend found with
line surface brightness either.

At $1.5 < z < 3$, the number of confirmed redshifts is smaller for
the emission line selected galaxies because in this redshift range we
mostly rely on galaxies with absorption features to find their
redshifts in addition to \ciii.  {\tt The ORIGIN} method is not designed to
detect absorption features, and thus it is not able to detect any of
these objects. While the other feature, \ciii, is in emission, it is
fairly weak, which makes it harder for {\tt ORIGIN} to detect.  As
shown in Figure~\ref{fig:u10_Fline_hst_org}, line fluxes of the \ciii
emission not detected by {\tt ORIGIN} are at the low end of the line
fluxes.  Although it is possible to tune {\tt ORIGIN} to detect
fainter emission features in the cube, it also dramatically increases
the number of spurious detections.

\subsubsection{Final redshifts for the MUSE Ultra Deep Field (\udft)}\label{sec:udf10_finalz}

\begin{figure}
  \includegraphics[width=\textwidth/2]{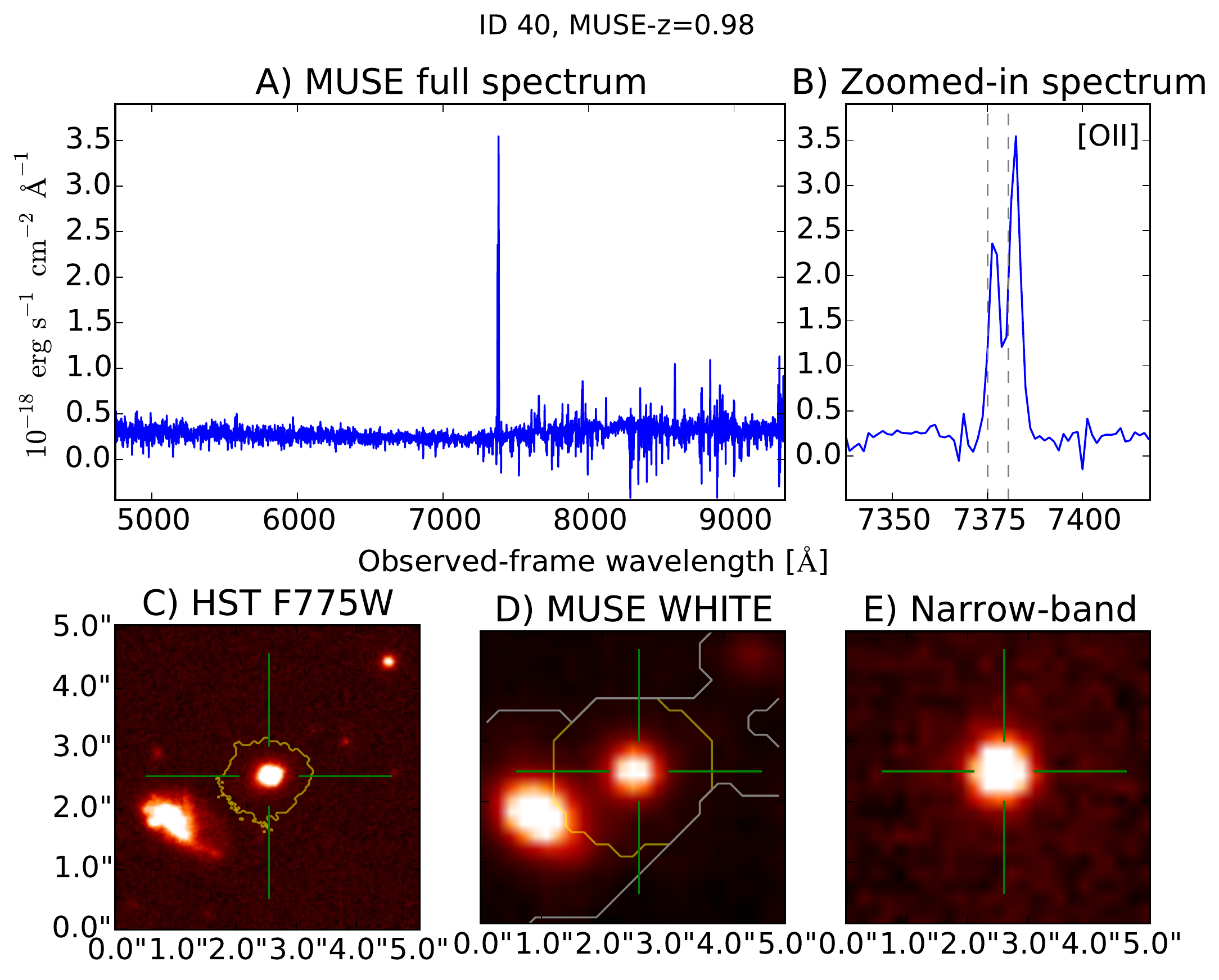}
  \includegraphics[width=\textwidth/2]{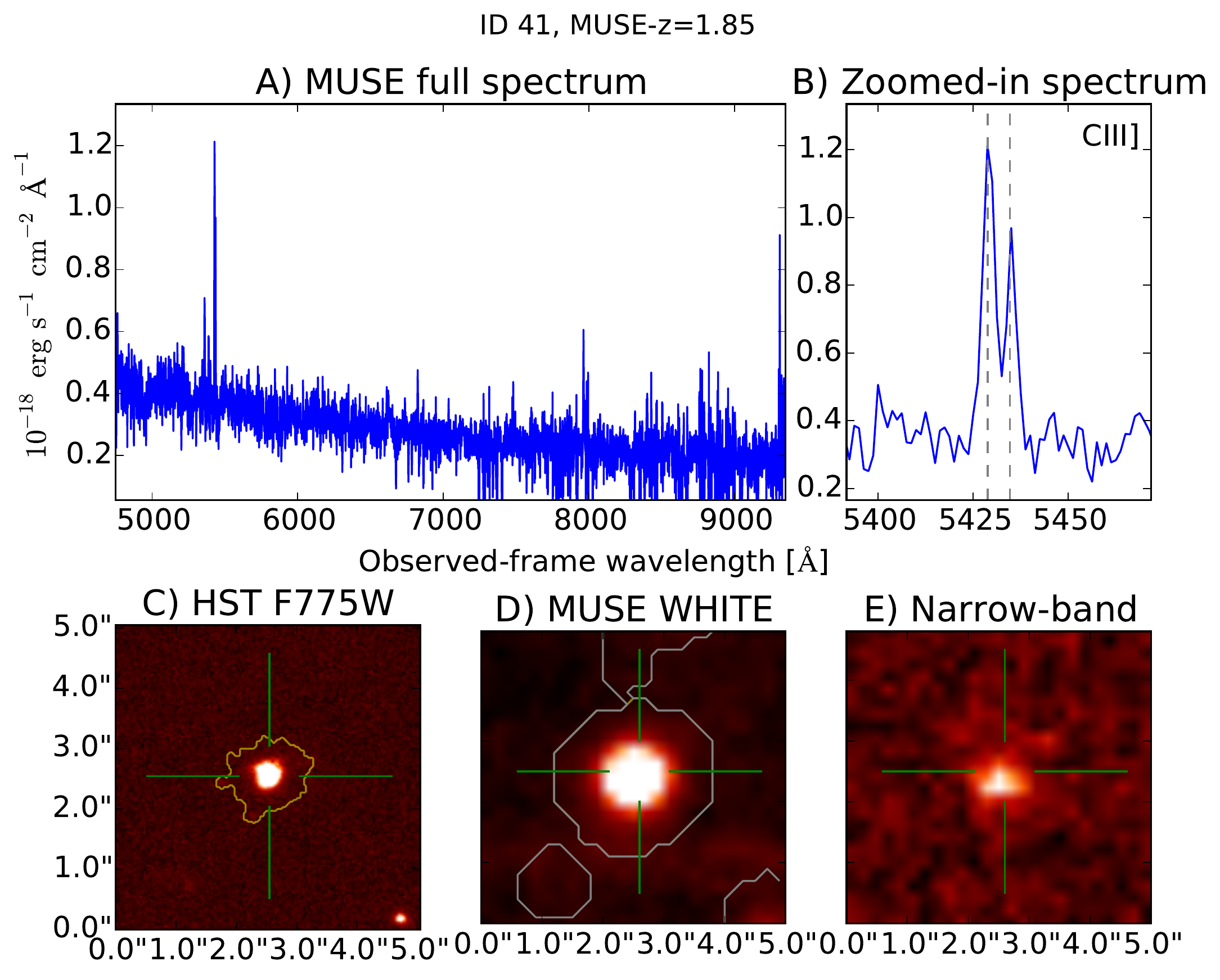}
  \includegraphics[width=\textwidth/2]{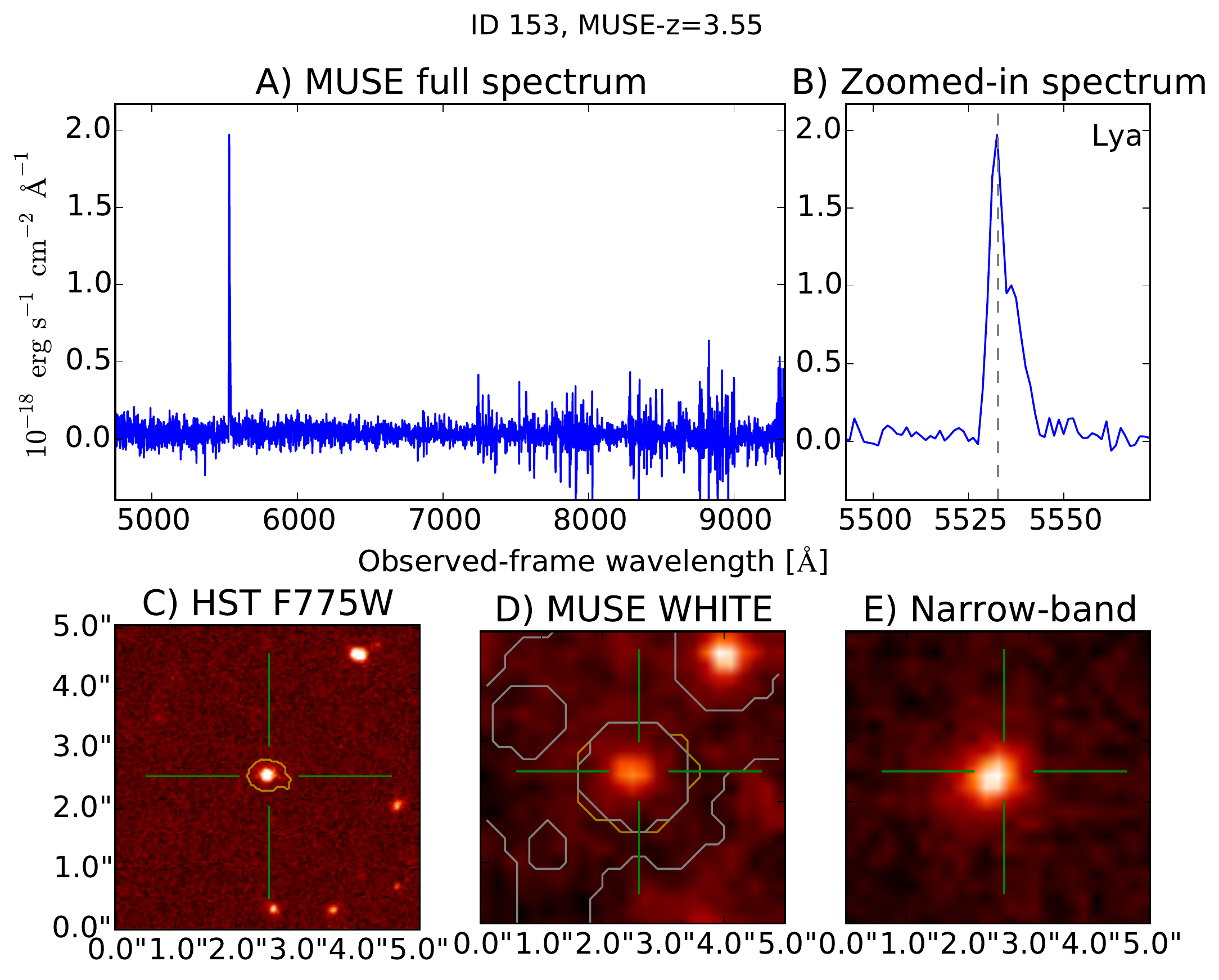}
  \caption{ Full MUSE spectrum and a zoomed-in portion of the
    \oii, \ciii, and \lya detected wavelength region at the top of
    each panel, and the {\it HST} F775W, MUSE white-light, and
    narrowband images of each emission line in the bottom of each
    panel. The yellow contours in the images are the boundary of the
    UVUDF catalog segmentation for the object, but those in the
    MUSE white-light image are convolved with the MUSE beam size.  The
    white boundaries indicate masked objects excluded from local sky
    residual estimates. The green cross indicates the central position
    of the extracted object.  The ID numbers and measured redshift are
    indicated at the top of each panel. All of these redshifts are
    secure ($\rm CONFID = 2 \, or \, 3)$.}
  \label{fig:eml_spec}
\end{figure}

\begin{figure}
  \resizebox{\hsize}{!}
  {\includegraphics{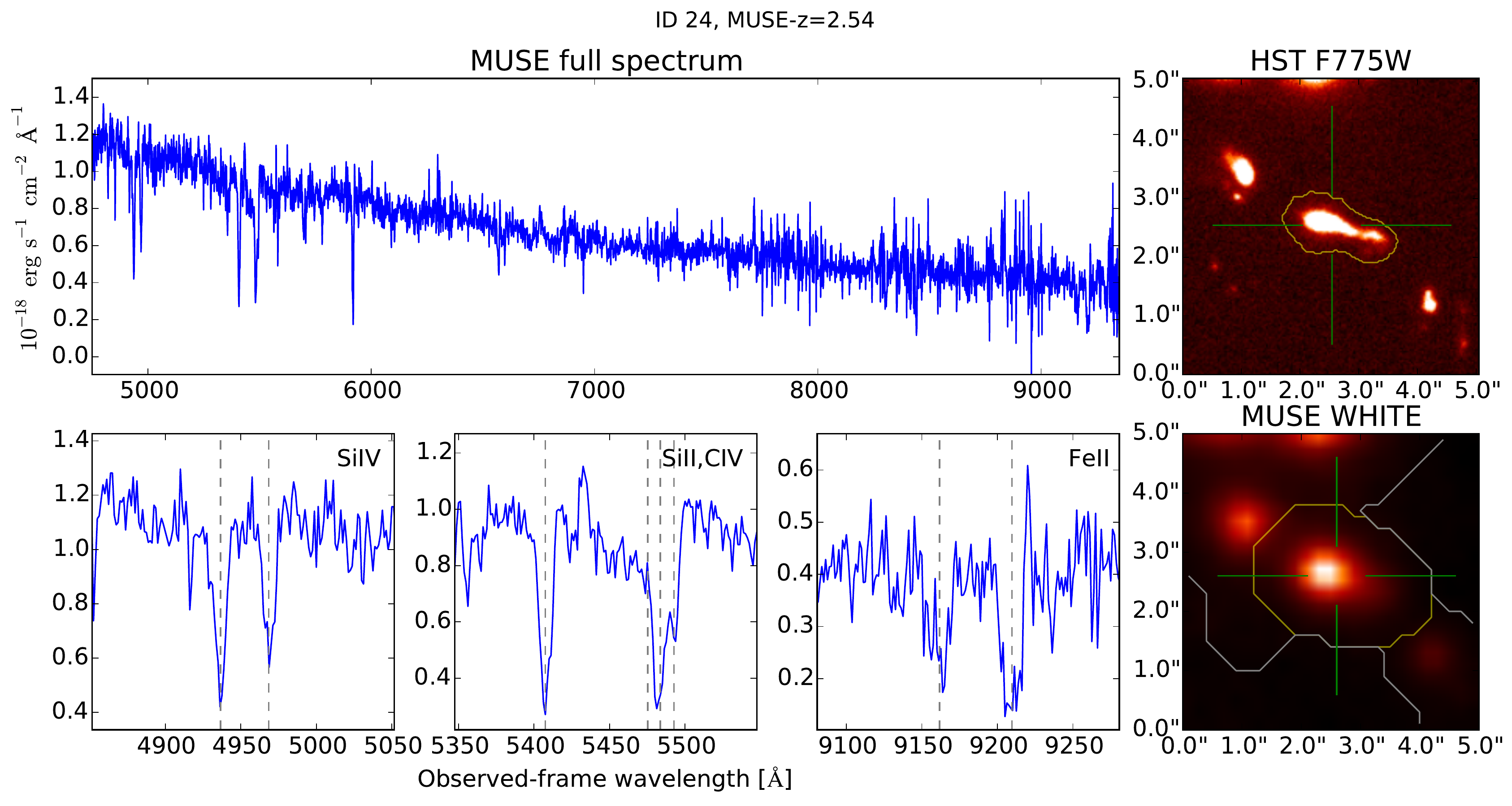}}
  \caption{ Full MUSE spectrum and a zoomed-in portion of the
    SiIV, SiII and CIV, and FeII absorption features. The {\it HST}
    F775W and the MUSE white-light images are shown on the right. The
    yellow contours in the images are the boundary of the UVUDF
    catalog segmentation for the object, but those in the MUSE
    white-light image are convolved with the MUSE beam size.  The white
    boundaries indicate masked objects excluded from local sky
    residual estimates. The green cross indicates the central position
    of the extracted object. The ID number and measured redshift are
    indicated at the top.}
  \label{fig:abs_spec}
\end{figure}

We combined and compared the results from the independent measurements
of the continuum and emission line selected objects to obtain the
final redshifts. All of the determined redshifts are provided in the
catalog (Appendix~\ref{app:cat}). We show some example spectra and
narrowband images of \oii, \ciii, and \lya emitters, and an
absorption line galaxy in Figures~\ref{fig:eml_spec} and
\ref{fig:abs_spec}.

\begin{figure*}
  \begin{center}
    \includegraphics[width=\textwidth]{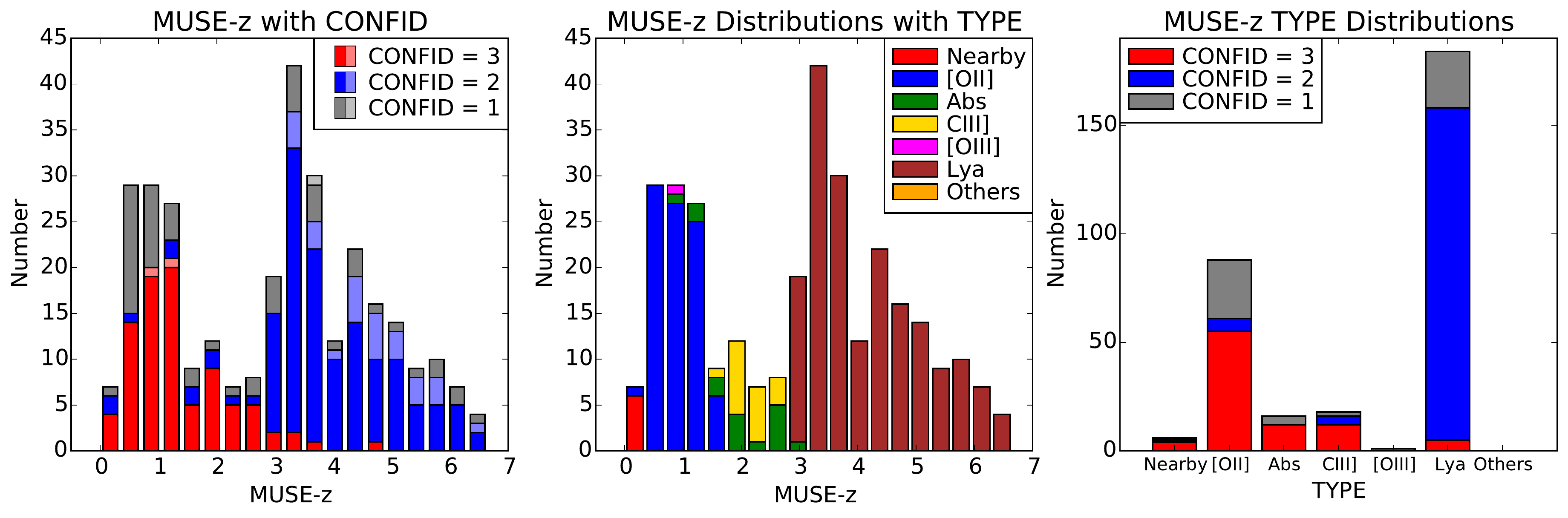}
    \caption{Final MUSE redshift distribution of the {\it unique}
      objects (i.e., overlapping objects are removed) combine both the
      continuum and emission line detected sources in the MUSE Ultra
      Deep Field (\udft). {\bf [Left]} The redshift histogram in bins
      of $\Delta z = 0.35$.  The red, blue, and gray colors represent
      the confidence levels 3, 2, and 1, respectively, for the
      determined redshift. The objects {\it only} found by {\tt
        ORIGIN} are indicated by the faded colors, whereas in
      Fig.~\ref{fig:u10_musez_hstpri_org} {\it all} of the {\tt
        ORIGIN} detected objects are shown. {\bf [Middle]} The same
      redshift distribution as the left planel, but color coded by
      classified type of the objects. {\bf [Right]} The histogram of
      the classified type of the objects color coded by the redshift
      confidence levels. }
    \label{fig:u10_musez}
  \end{center} 
\end{figure*}

We successfully measured $\UDFzALLCone$ and $\UDFzALLCtwo$ redshifts
with $\rm CONFID \geq 1$ and $\geq 2$, respectively, for the {\it
  unique} objects selected with the continuum or line emission
(i.e., overlapping objects are removed).  The final MUSE redshift
distribution in \udft is shown in Figure~\ref{fig:u10_musez} and
summarized in Tables~\ref{tbl:num_z} and \ref{tbl:count}. The shape of
the histogram does not change compared with the individual redshift
assessments (Figure~\ref{fig:u10_musez_hstpri_org}): two peaks at
$z \approx 1$ and $z \approx 3$. The number of $\rm CONFID=2$
redshifts at $z \leq 1.5$ is $69$, at $1.5 < z \leq 3$ (redshift
desert) it is $29$, and at $3 < z \leq 6.7$ it is $155$.

\begin{table*}[htp]
  \caption{Counts of determined redshifts for different source
    extractions in the MUSE Ultra Deep (\udft), Deep
    field (\mosaic), and the unique objects of \udft $+$ the \mosaic }
\begin{center}
\begin{tabular}{cccccccccc}
\hline \hline
                & \multicolumn{3}{c}{\udft ($1\arcmin \times 1\arcmin$)}
                & \multicolumn{3}{c}{\mosaic ($3\arcmin \times 3\arcmin$)}
                & \multicolumn{3}{c}{combined~~\tablefootmark{a}} \\
Confidence level (CONFID)    & $\geq 3$ & $\geq 2$ & $\geq 1$ 
                             & $\geq 3$ & $\geq 2$ & $\geq 1$
                             & $\geq 3$ & $\geq 2$ & $\geq 1$ \\
\hline
{\it HST} continuum selected~~\tablefootmark{b} 
                             & \multirow{2}{*}{$\UDFzHSTCthree$}  & \multirow{2}{*}{$\UDFzHSTCtwo$}  & \multirow{2}{*}{$\UDFzHSTCone$}
                             & $\MOSzHSTCthree$                   & $\MOSzHSTCtwo$                   & $\MOSzHSTCone$
                             & \multirow{2}{*}{$\COMBzHSTCthree$} & \multirow{2}{*}{$\COMBzHSTCtwo$} & \multirow{2}{*}{$\COMBzHSTCone$} \\
With UVUDF counterparts~~\tablefootmark{b}
                             &                       &                     &           
                             & $\MOSzHSTallCthree$   & $\MOSzHSTallCtwo$   & $\MOSzHSTallCone$
                             &                       &                     &                 \\
Emission line selected       & $\UDFzORGCthree$      & $\UDFzORGCtwo$      & $\UDFzORGCone$
                             & $\MOSzORGCthree$      & $\MOSzORGCtwo$      & $\MOSzORGCone$
                             & $\COMBzORGCthree$     & $\COMBzORGCtwo$     & $\COMBzORGCone$ \\
Emission line only~~\tablefootmark{c} & $\UDFzORGonlyCthree$  & $\UDFzORGonlyCtwo$  & $\UDFzORGonlyCone$
                             & $\MOSzORGonlyCthree$  & $\MOSzORGonlyCtwo$  & $\MOSzORGonlyCone$
                             & $\COMBzORGonlyCthree$ & $\COMBzORGonlyCtwo$ & $\COMBzORGonlyCone$ \\
\hline
Total unique objects         & $\UDFzALLCthree$  & $\UDFzALLCtwo$  & $\UDFzALLCone$
                             & $\MOSzALLCthree$  & $\MOSzALLCtwo$  & $\MOSzALLCone$
                             & $\COMBzALLCthree$ & $\COMBzALLCtwo$ & $\COMBzALLCone$ \\
\hline
\end{tabular}
\end{center}
\tablefoot{ 
\tablefoottext{a}{Number of unique objects in \udft and the \mosaic.}
\tablefoottext{b}{For the \mosaic, the {\it HST} continuum selected
  galaxies to inspect are limited to $\rm F775W \leq 27$~mag (\S\ref{subsec:mag_cut}). The rest
  of the galaxies are selected by direct detection of emission lines
  in the data cube. Among the emission line selected galaxies, some galaxies have counterparts
  in the UVUDF catalog with $\rm F775W > 27$~mag. They are listed in the row labeled ``With UVUDF
  counterparts'' in the table.}
\tablefoottext{c}{The objects selected by emission lines that have no
  counterpart in the UVUDF catalog \citep{Rafe15}.} }
\label{tbl:num_z}
\end{table*}%

\begin{table*}[htp]
\caption{Census of the objects in the MUSE Ultra Deep (\udft) and Deep
  field (the \mosaic) sorted by categories}
\begin{center}
\begin{tabular}{lrccrccr}
\hline \hline
                & \multicolumn{3}{c}{\udft ($1\arcmin \times 1\arcmin$)} 
                & \multicolumn{3}{c}{\mosaic ($3\arcmin \times 3\arcmin$)}
                & combined \\
Category/Type   & Counts~~\tablefootmark{a} & Redshift~~\tablefootmark{b} & F775W mag~~\tablefootmark{b} 
                & Counts~~\tablefootmark{a} & Redshift~~\tablefootmark{b} & F775W mag~~\tablefootmark{b}
                & Counts~~\tablefootmark{a} \\
\hline
0. Stars                    &   0   (0) &      ...      &      ...       
                            &   9   (1) &      ...      & $19.0 - 24.8$
                            &   9   (1) \\ 
1. Nearby galaxies          &   5   (1) & $0.21 - 0.31$ & $22.6 - 30.0$  
                            &  47   (2) & $0.10 - 0.42$ & $18.6 - 27.1$
                            &  47   (3) \\ 
2. \oii emitters            &  61  (27) & $0.33 - 1.45$ & $20.4 - 28.8$  
                            & 465  (49) & $0.28 - 1.49$ & $19.4 - 28.3$
                            & 473  (73) \\ 
3. Absorption line galaxies &  12   (4) & $0.95 - 3.00$ & $21.9 - 26.1$  
                            &  57  (22) & $0.60 - 2.95$ & $21.0 - 26.2$
                            &  63  (23) \\ 
4. \ciii emitters           &  16   (2) & $1.55 - 2.54$ & $23.8 - 29.8$  
                            &  41  (18) & $1.55 - 2.86$ & $23.4 - 27.0$
                            &  50  (18) \\ 
5. \oiii emitters           &   1   (0) & $0.71$        &    $27.3$  
                            &   2   (1) & $0.42 - 0.71$ & $27.0 - 27.3$
                            &   2   (1) \\ 
6. \lya  emitters           & 158  (26) & $2.94 - 6.64$ & $25.5 - 31.1+$~~\tablefootmark{c} 
                            & 624  (97) & $2.91 - 6.63$ & $24.4 - 31.2+$~~\tablefootmark{c}
                            & 692 (115) \\ 
7. Others                   &    0  (1) &      ...      &      ...       
                            &    2  (2) & $1.22 - 3.19$ & $21.0 - 24.6$
                            &    2  (2) \\ 
\hline
\end{tabular}
\end{center}
\tablefoot{ 
\tablefoottext{a}{Counts of $\rm CONFID \geq 2$ redshifts. The
  numbers in parentheses indicate redshifts with $\rm CONFID = 1$.}
\tablefoottext{b}{The ranges are for the secure redshifts ($\rm CONFID \geq 2$).}
\tablefoottext{c}{For some \lya emitters, the F775W continuum emission
  is not detected ($> 31.2$~mag).} }
\label{tbl:count}
\end{table*}%

\begin{figure}
  \resizebox{\hsize}{!}
  {\includegraphics{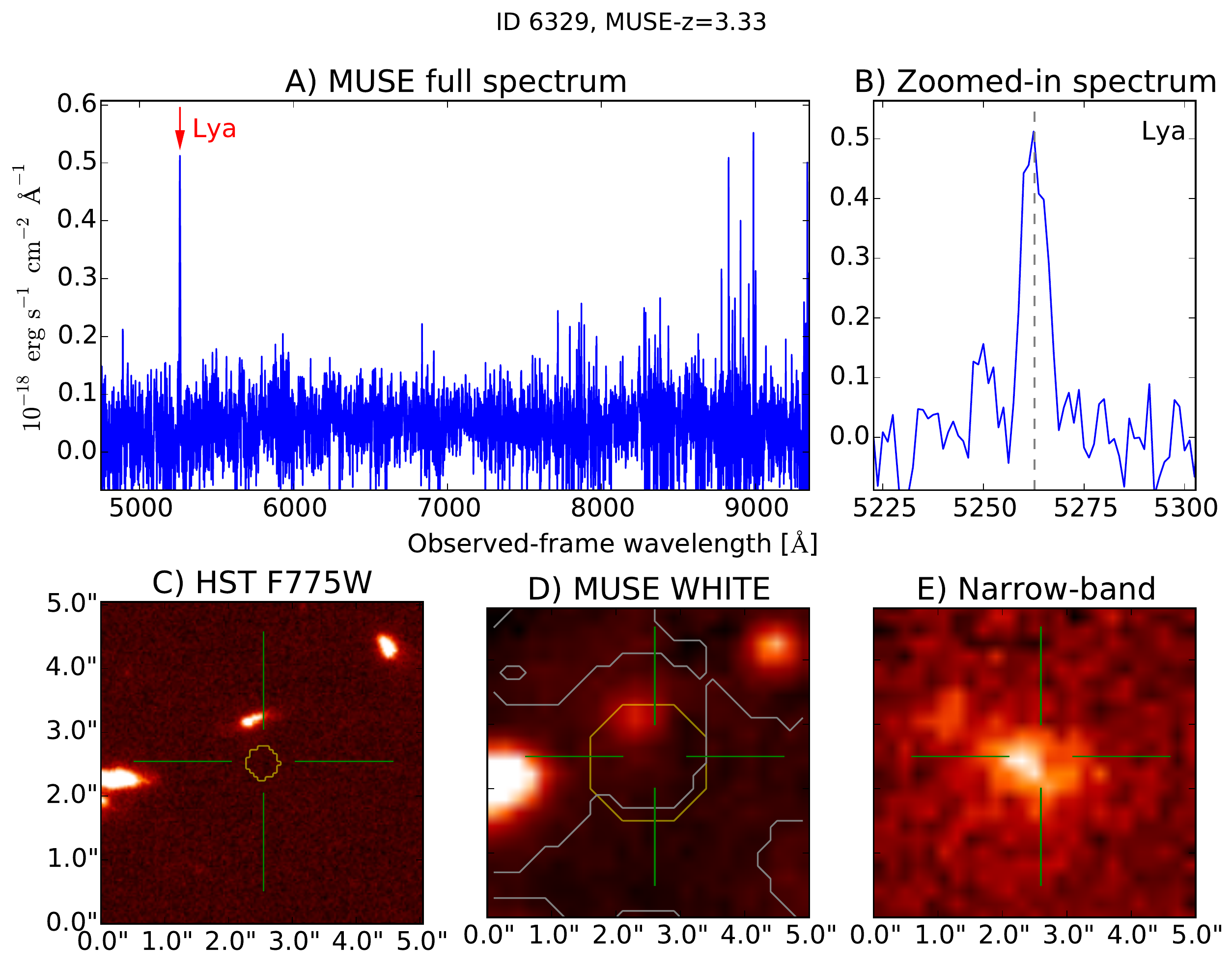}}
  \caption{ Example of an object that is not detected by
    continumm emission but found by {\tt ORIGIN}.  The full MUSE
    spectrum and zoomed-in part of the \lya detected wavelength region are
    presented at the top. The {\it HST} F775W, MUSE white-light,
    and \lya narrowband images are at the bottom.  }
  \label{fig:ORIonly_Lya_spec}
\end{figure}

In Figure~\ref{fig:u10_musez}, it is also immediately noticeable that
beyond $z = 3$, the fraction of confirmed redshifts of the objects
detected {\it only} by {\tt ORIGIN} (the faded color regions)
increases.  By $z \approx 6$, it reaches $\approx 50\%$.  According to
Figure~\ref{fig:u10_Fline_hst_org}, the {\tt ORIGIN} detections do not
seem to favor any specific redshift ranges or line fluxes (except of
course when the line flux is below its detection limit).  Two [OII]
emitters ($\rm CONFID \geq 2$) were only detected by {\tt ORIGIN}:
MUSE ID 6314 and 6315.  The disturbed morphology of ID 6314 in the
{\it HST} imaging likely accounts for the no detection with {\tt
  SExtractor}. The other object, ID 6315, is blended with a nearby
bright source and are difficult to separate using solely imaging data.
For the 28 \lya emitters identified only by {\tt ORIGIN}, most of
these sources have surface brightness continuum emission that is too
low to be detected with {\tt SExtractor} (we measured $\approx 30$~mag
with {\tt NoiseChisel}) or are not visible by eye
(Figure~\ref{fig:ORIonly_Lya_spec}), but there are a small number of
cases that are missed in the UVUDF catalog because of blending or
because they are lying close to nearby bright sources (ID 6313) and
disturbed/complex morphology in the {\it HST} images (ID 6324).  We
emphasize that these 28 objects represent $\sim 20\%$ of \lya emitters
found in this work and they are in a small $1\arcmin \times 1\arcmin$
area of the entire sky where the deepest {\it HST} data exist.

\subsection{Redshift determination in the MUSE Deep Field (the \mosaic)}

\subsubsection{Strategy based on the results in \udft}\label{subsec:mag_cut}

\begin{figure}
  \resizebox{\hsize}{!}
  {\includegraphics{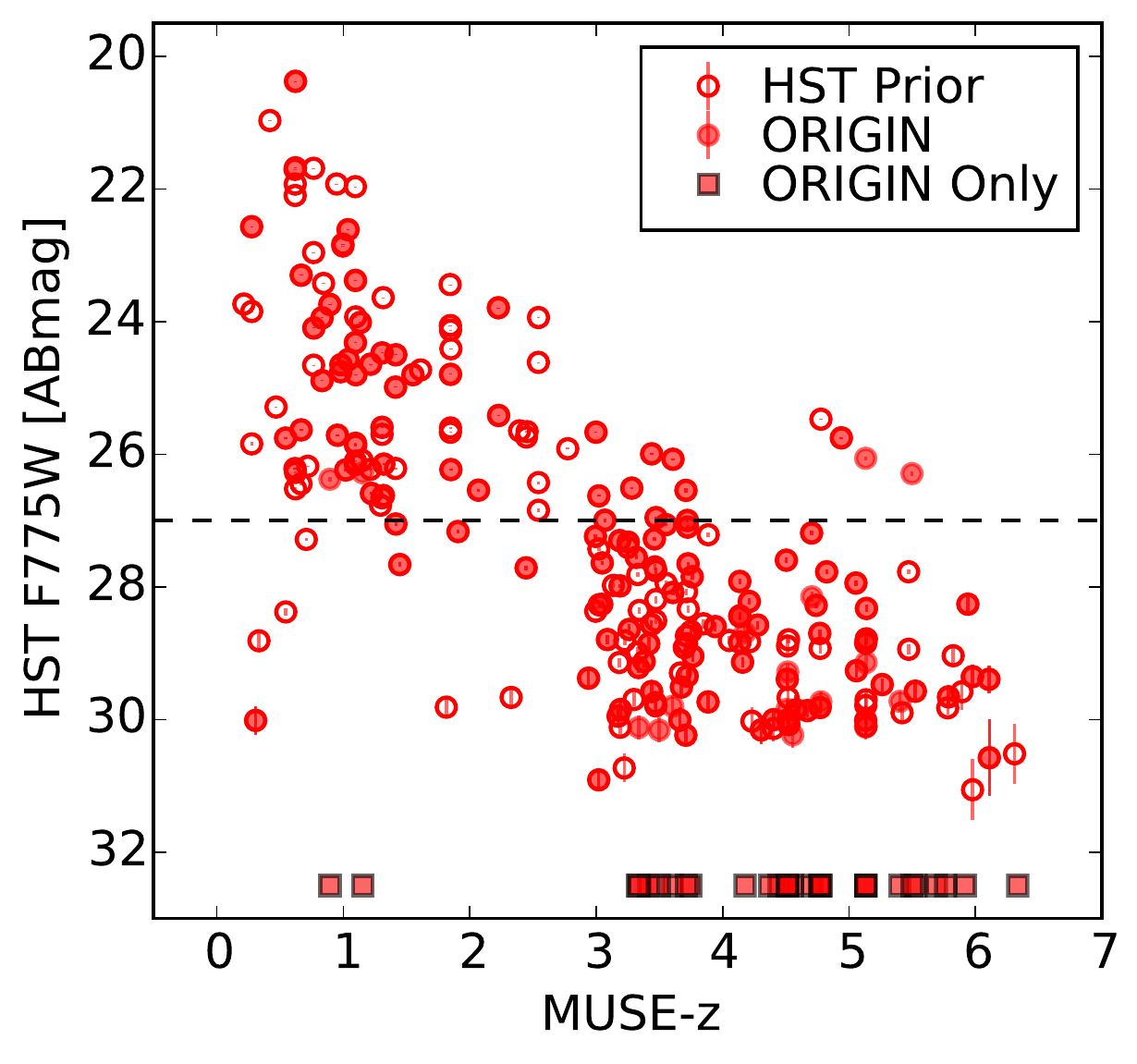}}
  \caption{ Magnitudes of {\it HST} F775W plotted against determined
    redshift ($\rm CONFID \geq 2$) for \udft. The open and filled
    symbols represent the continuum ({\it HST} prior) and emission
    line ({\tt ORIGIN} or {\tt MUSELET}) extracted objects. The filled
    squares are the objects detected only with {\tt ORIGIN} or {\tt
      MUSELET}, and thus their F775W mags are upper limits. The
    horizontal dashed line indicates 27~mag where we make the cut to
    the continuum selected galaxies to perform the redshift
    determination in the \mosaic field (see \S\ref{subsec:mag_cut}).}
  \label{fig:u10_musez_mag}
\end{figure}

\begin{figure}
  \includegraphics[width=\textwidth/2]{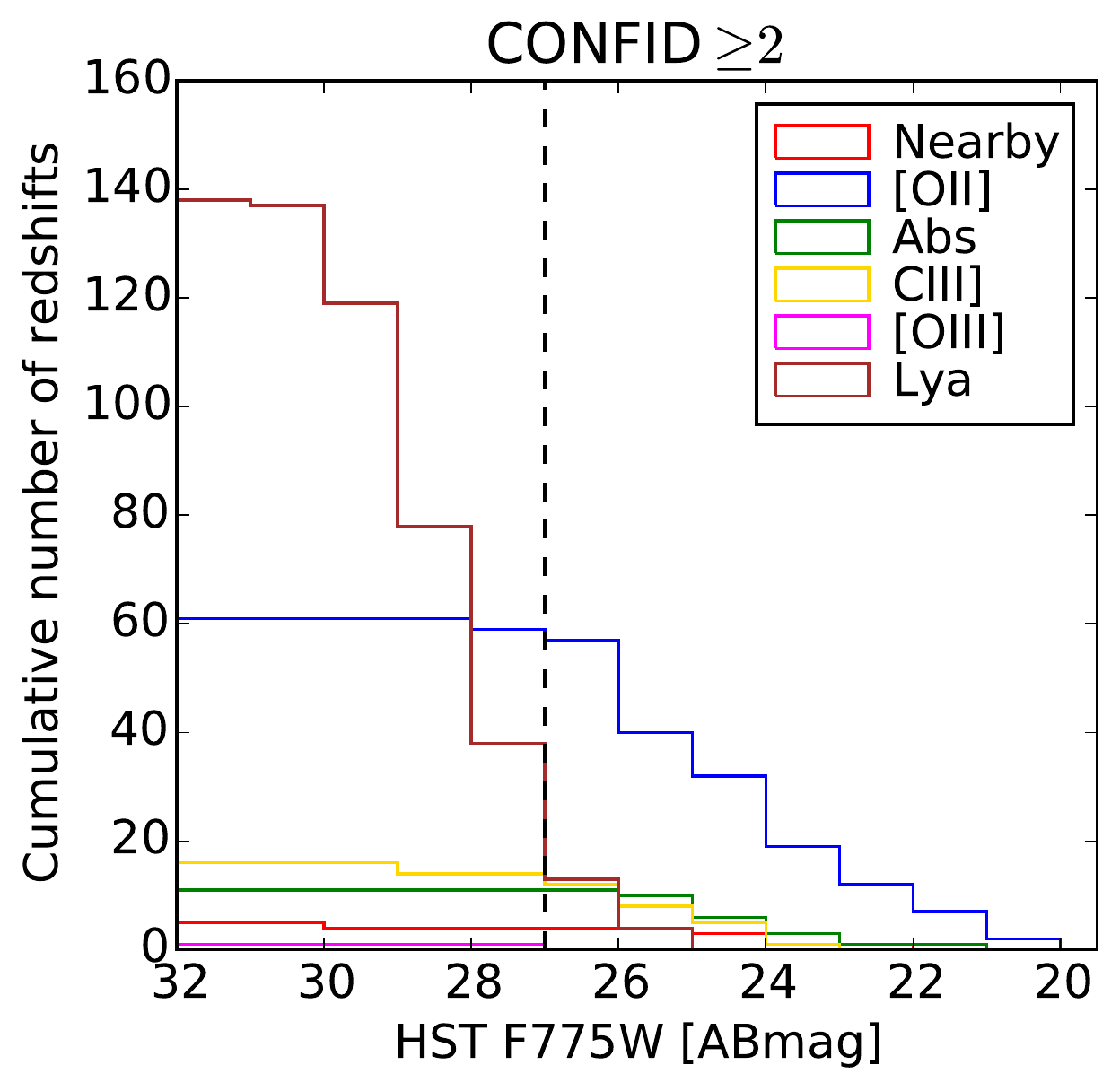}
  \caption{ Cumulative counts of secure redshifts in \udft. The colors
    indicate different types with the same scheme as in
    Figure~\ref{fig:u10_musez_hstpri_org}.}
  \label{fig:u10_accum_type}
\end{figure}

Based on the redshift analysis in our deepest survey field \udft, we
tried to find the best combination of the two extraction methods to
maximize the efficiency of redshift identifications. As shown in
Figure~\ref{fig:u10_musez_mag}, for the objects with redshift
determined, 85 out of 98 objects at $z < 3$ ($\rm CONFID \geq 2$),
mostly \oii emitters, have $\rm F775W \leq 27$~mag (see also
Figure~\ref{fig:u10_accum_type}). While the majority at $z > 3$ are
fainter than 27~mag, $106$ out of $155$ objects ($70\%$) in this
redshift range are detected by line emission directly in the data cube
with {\tt ORIGIN}. In addition, all of the absorption galaxies, which
{\tt ORIGIN} cannot detect, are brighter than $26.1$~mag.

We simulated expected \oii line fluxes to investigate the sudden
reduction of \oii emitters fainter than 27~mag visible in
Figure~\ref{fig:u10_accum_type}. With the spectral energy distribution
(SED) fitting code {\tt FAST}\,\footnote{Fitting and Assessment of
  Synthetic Templates. We assume the dust extinction curve of
  $\tau \approx \lambda^{-1.3}$ from \cite{Char00} together with the
  median value of the measured Balmer decrements of the MUSE sources
  that have two or more Balmer lines detected.}  \citep{Krie09}, we
used all of the galaxies in the \mosaic with photo-$z$ (BPZ) of
$0 - 1.5$ provided in the UVUDF catalog to predict their \oii fluxes.
With the simple assumption that all of the \oii emission originates
from star formation, we can directly translate the star formation
rates obtained from SED fits of the {\it HST} photometry into expected
\oii fluxes.  Then we correct the \oii fluxes for dust extinction
using the median dust correction factor,
which is derived for the galaxies that are detected with MUSE using the
Balmer decrement.  The modeled \oii fluxes decrease with the F775W
magnitude.  From the $3\sigma$ line flux detection limit of the
\mosaic, $\approx 3 \times 10^{-19} \, {\rm erg\,s^{-1}\,cm^{-2}}$,
the fraction of detectable \oii emitters drastically decreases at
$\rm F775W \gtrsim 27$~mag. At 27~mag, we expect $\approx 80\%$ of the \oii
emitters to be detected, but it abruptly decreases to
$\approx 20-50\%$ between $27-28$~mag.  These numbers are
likely to be upper limits because, for simplicity of the model, all
galaxies are assumed to be star forming.

Thus, we make a cut at $\rm F775W \leq 27$~mag to limit the number of
the continuum selected objects based on the {\it HST} priors for which
we need to perform the visual inspection on redshift
determination. This reduces the total number of $\MOSHSTpri$ continuum
selected objects to $\MOSHSTpriMAGcut$ to be inspected.  For the
emission line selected objects, we do not apply any preselection and
examine all of the $\MOSORIGIN$ spectra of the emission line detected
objects.

In the near future, we plan to extend the redshift determination
toward galaxies with $\rm F775W > 27$~mag. At present, we only
release the redshift measured in the $\rm F775W \leq 27$~mag continuum
selected sample and all of the emission line selected sample as the
first version of the MUSE UDF redshift catalog
(Appendix~\ref{app:cat}).

\subsubsection{Redshifts in the \mosaic}

The process of the redshift evaluation in the \mosaic follows the same
procedure as \udft. We inspected the continuum detected and emission
line detected objects individually, and then reconciled these two objects to obtain
the final redshifts.

\begin{figure*}
  \begin{center}
    \includegraphics[width=\textwidth]{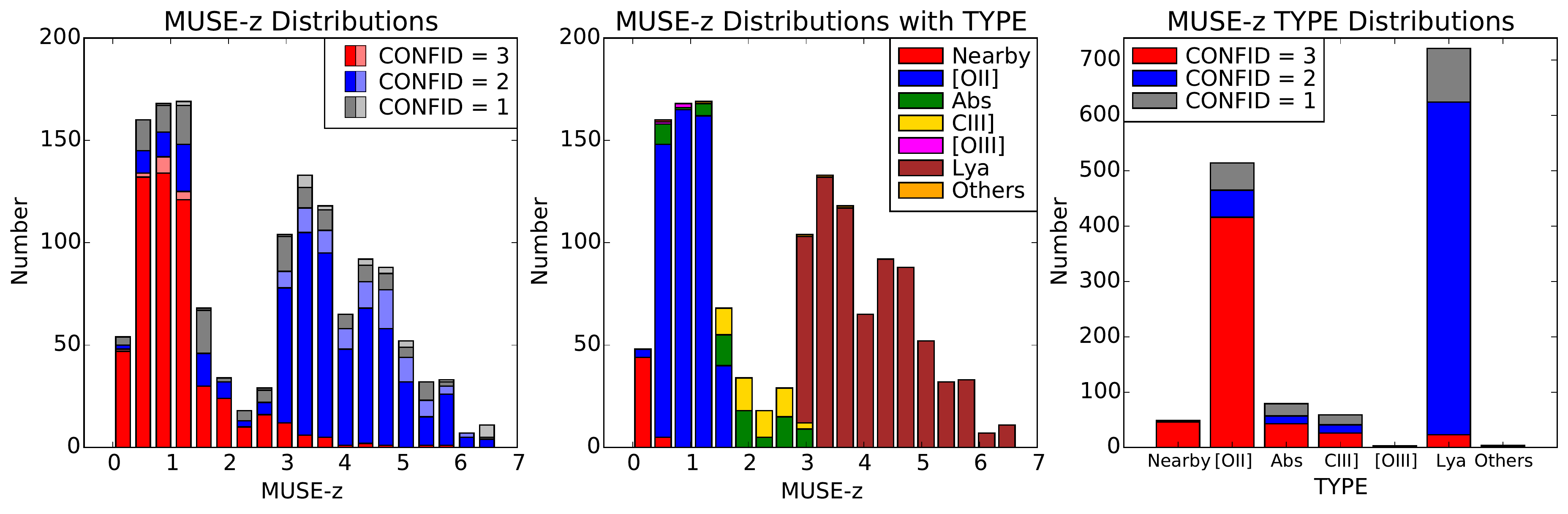}
    \caption{Same histograms as Figure~\ref{fig:u10_musez} but for
      the MUSE Deep Field (the \mosaic). }
    \label{fig:mos_musez}
  \end{center} 
\end{figure*}

The summary and distributions of the determined redshifts are shown in
Tables~\ref{tbl:num_z} and \ref{tbl:count} and
Figure~\ref{fig:mos_musez}. Out of the $\MOSHSTpriMAGcut$ continuum
selected objects ($\rm F775W \leq 27$~mag), redshifts of
$\MOSzHSTCone$ and $\MOSzHSTCtwo$ objects are determined with
confidence levels of $\geq 1$ and $\geq 2$, respectively. When we
include the emission line selected objects, these numbers increase to
$\MOSzALLCone$ and $\MOSzALLCtwo$. Among these, $\MOSzHSTallCone$ and
$\MOSzHSTallCtwo$ objects, respectively, have detectable UVUDF
counterparts.

The fractions of the identified redshifts below and above $z = 3$ are
different compared with those in \udft. Out of $\MOSzALLCtwo$
redshifts with $\rm CONFID \geq 2$ in the \mosaic, $52\%$ ($650$) are
at $z \leq 3$ and $48\%$ ($597$) are at $3 < z \leq 6.7$.  Whereas, in
\udft, there are $39\%$ ($98$) and $61\%$ ($155$), respectively, out
of $\UDFzALLCtwo$ redshifts with $\rm CONFID \geq 2$. A larger
percentage of \lya emitters are detected in the deeper data of \udft
than in the \mosaic, whose integration times are about three times
different.  This trend stays the same, even when we only account for
the continuum detected objects (i.e., excluding the objects detected
{\it only} by {\tt ORIGIN} and {\tt MUSELET}): $56\%$ ($631$) at
$z \leq 3$ and $44\%$ ($502$) at $z > 3$ for the \mosaic, and $43\%$
($96$) and $57\%$ ($127$), respectively, for \udft.  In addition, a
smaller fraction of redshifts ($10\%$) is found in the ``redshift
desert'' at $1.5 < z \leq 3.0$ compared with \udft ($13\%$). These
differences between the \mosaic and \udft, due to the depth of the
data, are also implied by the higher fraction of $\rm CONFID=2$
redshifts at $z < 3$ in the \mosaic.  Interestingly, the deeper data
of the \udft increase the number fraction of $\rm CONFID=1$ redshifts
in the \oii emitters compared to the \mosaic.  Among all of the
determined \oii and \lya emitters, the $\rm CONFID=1$ redshift is 11\%
and 16\% of the $\rm CONFID \geq 2$ redshift, respectively, for the
\mosaic, while it is 44\% and 16\%, respectively, for \udft.  This is
possibly because with higher S/N spectra, it is easier to find more
features.  The depth of the data also affects the detection rate of
the {\tt ORIGIN/MUSELET}-only objects at $z > 3$. We do not see the
same tendency as in \udft, where the fraction of the determined
redshifts of the {\tt ORIGIN/MUSELET}-only objects to the continuum
extracted objects increases with redshift.  One nearby galaxy, 14 \oii
emitters, and 99 \lya emitters are found only by {\tt ORIGIN/MUSELET}
with $\rm CONFID \geq 2$ in the \mosaic.

\begin{figure}
  \resizebox{\hsize}{!}
  {\includegraphics{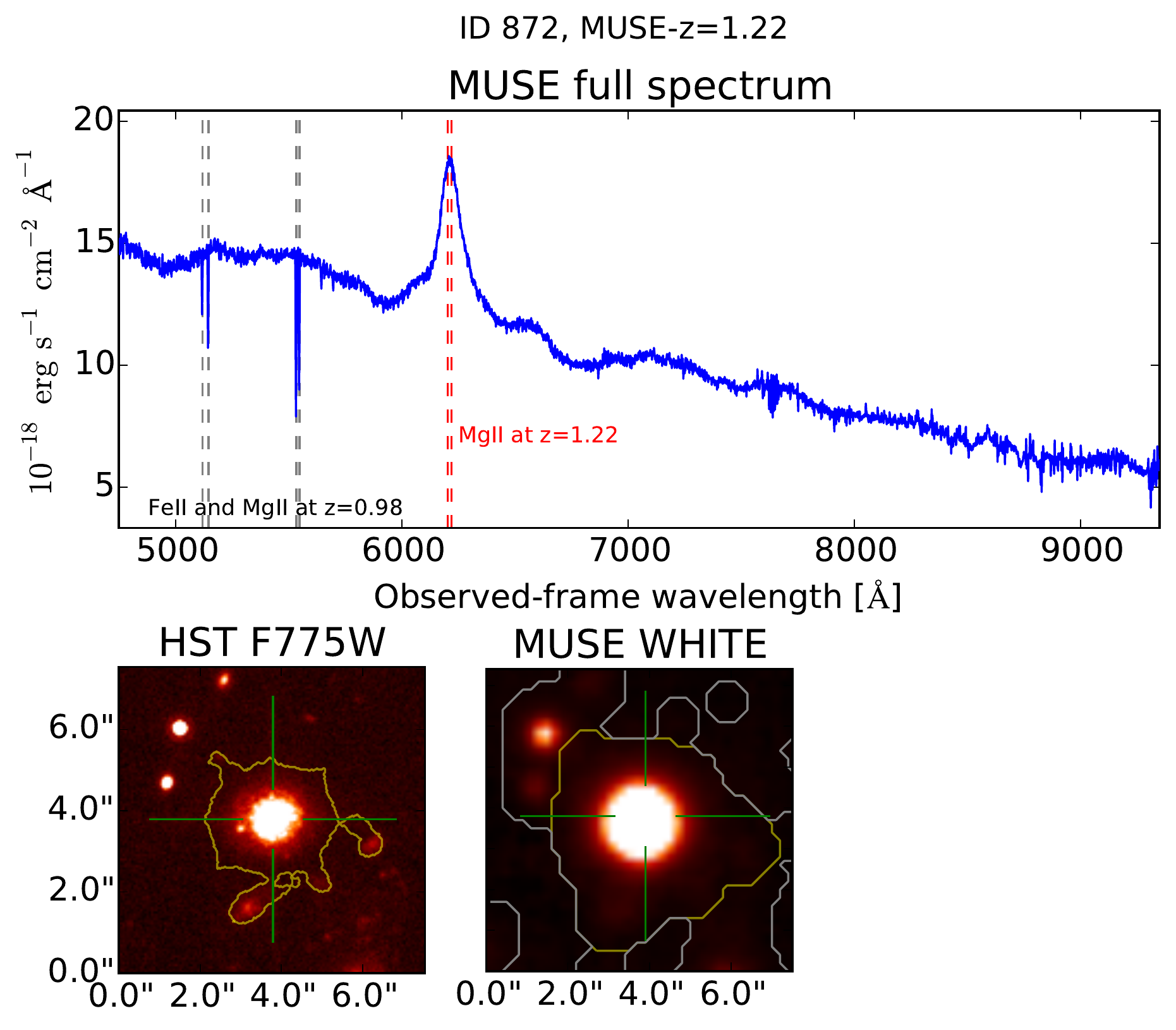}}
  \caption{ Full MUSE spectrum of a quasar at $z=1.22$ detected with a
    broad, prominent \mgii emission feature. An intervening absorber
    (\feii and \mgii) at $z=0.98$ is also detected. The cutout
    ($7.6\arcsec \times 7.6\arcsec$) for this object is larger than
    the standard $5\arcsec \times 5\arcsec$.}
  \label{fig:qso_spec}
\end{figure}

Thanks to the larger survey area of the \mosaic, we find more rare
objects than in \udft. As an example, the one-dimensional spectrum of
a quasar at $z = 1.22$ (MUSE ID 872 or UVUDF ID 23796) with prominent
Mg\,\textsc{ii} emission is shown in
Figure~\ref{fig:qso_spec}\,\footnote{The optical spectrum of this
  object was taken before with {\it HST/ACS} slitless grism
  spectroscopy and Mg\,\textsc{ii} is detected \citep{Stra08}.}. 
This spectrum also exhibits \feii and \mgii absorption features. These
features only appear in the spectrum when it is extracted within the
quasar PSF and are likely to be due to an intervening absorber at
$z=0.98$ \citep[cf.][]{Rigb02}.

\subsection{Redshift comparisons between the overlap region in MUSE
  Ultra Deep (\udft) \& Deep Field (the \mosaic)}\label{subsec:comp_udf10_mosaic}

The \mosaic redshift determination is performed independently from
\udft. In this subsection, we show direct comparisons of the measured
redshifts and the associated parameters in the overlap region of these
two fields.

In the overlapping part of the \mosaic and \udft fields, all of the
redshifts (regardless of CONFID) identified in the \mosaic are also
identified in \udft with the same CONFID or higher, except MUSE ID 275
and four objects detected only by {\tt ORIGIN} or {\tt MUSELET}
(MUSE IDs 6432, 6447, 6865, and 7396). The redshift of ID 275 is
measured to be $z=2.9$ in both \udft and the \mosaic, but CONFID are 1
and 2, respectively. This is because the object lies at the edge of
the \udft field, which gives less confidence. The four objects
detected only by emission lines found in the \mosaic data cube (with
$\rm CONFID \geq 2$) are missed by the {\tt ORIGIN} run in \udft.

Among $\UDFzALLCone$ redshifts identified in \udft, there are 13
redshifts that disagree with the redshifts identified in the \mosaic
by $| \Delta z | > 0.01$. For the object discussed above, MUSE ID 275,
the redshift is measured to be $z=2.899$ and $2.931$ in \udft and the
\mosaic, respectively. Apart from this object, one (MUSE ID 6684) has
the same CONFID of 2 and four (MUSE IDs 44, 49, 90, 127) have the same
CONFID of 1 in both \udft and the \mosaic, but all of the rest (MUSE
IDs 64, 103, 718, 6335, 6339, 6676, 6686) have the high
CONFID of 2 or 3 in the \udft whereas CONFID of 1 in the
\mosaic. Below, we analyze why the obtained redshifts have differences
(CONFID is given in the parentheses):

\begin{description}

\item[\underline{With the same CONFID}]\mbox{}\\

\item[{\bf 6684}   $z_{\rm \udft}=4.740$ (2), $z_{\rm \mosaic}=0.871$ (2)]\mbox{}\\
  A clear emission feature is detected in both \udft and the \mosaic,
  but it is identified as \lya in \udft and as \oii in the
  \mosaic. This is because the line profile in the \mosaic spectrum
  shows double peaks (due to lower S/N) whose separation matches that
  of \oii at $z=0.871$. However, because this object is not detected
  in the UV imaging, it is more likely at higher redshift. Thus, we
  conclude that the feature is \lya and its redshift should be $4.740$
  as determined in \udft.
  \\

\item[{\bf 44}   $z_{\rm \udft}=1.610 $ (1), $z_{\rm \mosaic}=1.436 $ (1)]\mbox{}\\
  Both of the redshifts are from absorption features, but a set of
  multiple \feii absorption features are found in different wavelength
  regions. The CONFID for both of these redshifts are low.
  \\

\item[{\bf 49}   $z_{\rm \udft}=1.864 $ (1), $z_{\rm \mosaic}=1.578 $ (1)]\mbox{}\\
  The \aliii and \feii absorptions are identified in \udft, but a
  different set of absorptions are possibly seen in the \mosaic. Owing
  to low S/N in the continua, they are both not certain.
  \\

\item[{\bf 90}   $z_{\rm \udft}=0.734 $ (1), $z_{\rm \mosaic}=2.389 $ (1)]\mbox{}\\
  The same emission feature detected at $6465$\Ang in \udft and the
  \mosaic is identified as \oii and \ciii, respectively. With the low
  S/N data, it is difficult to confidently distinguish whether it is
  \oii or \ciii.
  \\

\item[{\bf 127}   $z_{\rm \udft}=0.616$ (1), $z_{\rm \mosaic}=4.035$ (1)]\mbox{}\\
  This is an interesting case in which different sets of emission
  features are spotted in the \udft and \mosaic spectra. In \udft,
  emission features at $6025$\Ang, $7855$\Ang, and $8090$\Ang are
  identified as \oii, \hb, and \oiii5007 at $z=0.616$. In contrast,
  none of these features are identified in the \mosaic, but an
  emission line is identified at $6120$\Ang. This feature is
  attributed to \lya. Reviewing the \udft spectrum, this feature is
  clearly detected.  Thus, we think that there are probably two
  objects at $z=0.616$ and $z=4.035$ lying along the sightline.  In
  the combined \udft and \mosaic catalog, we use $z=0.616$ for ID~127
  and add a new object with ID~7582 for the \lya emitter at $z=4.035$.
  \\

\item[\underline{Higher CONFID in \udft}]\mbox{}\\

\item[{\bf 64}   $z_{\rm \udft}=1.847 $ (2), $z_{\rm \mosaic}=1.566 $ (1)]\mbox{}\\
  The \ciii doublet is well detected at $5430$\Ang in \udft. The
  \mosaic redshift was measured by tentative absorption
  features. However, with a closer look, the same \ciii seen in \udft
  is weakly detected in the \mosaic. Thus, the redshift of this object
  is likely to be $1.847$ as measured in \udft.
  \\

\item[{\bf 103}   $z_{\rm \udft}=3.002$ (3), $z_{\rm \mosaic}=2.986$ (1)]\mbox{}\\
  A clear \lya absorption detected in both \udft and the \mosaic
  spectra. In \udft, there are also multiple UV absorption
  lines. Taking into account that the features used to determine the
  redshift are absorption and very broad, $| \Delta z |$ of 0.016 is
  not a significant difference. We use the \udft redshift as the
  final redshift.
  \\

\item[{\bf 718}   $z_{\rm \udft}=4.524$ (2), $z_{\rm \mosaic}=0.801$ (1)]\mbox{}\\
  Similar to ID 6684, the spectrum of the \mosaic is noisier than
  \udft, which causes spurious double peaks of the identified emission
  feature. In the \udft spectrum, the feature shows a clear asymmetric
  profile with a red wing, indicating \lya. Thus, the redshift should
  be $4.524$ as in \udft.
  \\

\item[{\bf 6335}   $z_{\rm \udft}=4.370 $ (2), $z_{\rm \mosaic}=1.098$ (1)]\mbox{}\\
  A clear emission feature is detected in \udft at $6525$\Ang whose
  profile indicates \lya, while the identified feature (\oii) in the
  \mosaic is unclear. We do not see any obvious features around
  $6525$\Ang in the \mosaic spectrum.  This is not surprising because
  the line flux is
  $\approx 1 \times 10^{-18} \, {\rm erg\,s^{-1}\,cm^{-2}}$, just
  around the $3\sigma$ detection limit of the \mosaic.
  \\

\item[{\bf 6339}   $z_{\rm \udft}=5.131 $ (2), $z_{\rm \mosaic}=5.121$ (1)]\mbox{}\\
  The same emission feature is recognized as \lya. It looks as if the
  detected \lya has a blue bump (which is likely to be sky residual)
  in the \mosaic. The blue bump is used to determine redshift, causing
  a redshift difference of 0.01.  However, this blue bump is not
  visible in the deeper data of \udft. Using $z=5.131$ from \udft is
  reasonable.
  \\

\item[{\bf 6676}   $z_{\rm \udft}=3.723$ (2), $z_{\rm \mosaic}=0.541$ (1)]\mbox{}\\
  The emission line clearly looks like \lya in the \udft spectrum and
  the UV continuum emission is not detected. However, in the \mosaic
  spectrum, although the same feature is identified, it is classified
  as \oii. Thus, the \udft redshift should be the correct one.
  \\

\item[{\bf 6686}   $z_{\rm \udft}=0.307$ (2), $z_{\rm \mosaic}=4.383$ (1)]\mbox{}\\
  In the \udft spectrum, two emission features at $4875$\Ang and
  $6545$\Ang correspond to \oii and \oiii5007, which gives $z=0.307$.
  The former feature, in addition to not being convincing in \udft, is
  not seen in the \mosaic.  The latter feature in the \mosaic is
  recognized as \lya because of its distinctive profile. This object
  is not detected up to the F435W band, which indicates that this
  object is likely to be at high redshift. Thus, the correct redshift
  is probably $z=4.383$ measured in the \mosaic.

\end{description}

These results suggest that it is not necessary true that the success
rate of CONFID in the shallower \mosaic region is lower when comparing
the same CONFID. We only found one case (ID 6684) in which the redshift in
the \mosaic field is not correct when both of the redshifts are
$\rm CONFID=2$.  When both redshifts are $\rm CONFID=1$ (IDs 44, 49
90, 127), their redshifts are too uncertain to make a judgement for
the quality.

Because of the shallower depth of the \mosaic, not all of the
redshifts measured in \udft are found in the \mosaic. Excluding the
objects only found by {\tt ORIGIN/MUSELET}, the \mosaic misses 118 out
of $\UDFzHSTCone$ redshifts with $\rm CONFID \geq 1$ and 64 out of
$\UDFzHSTCtwo$ with $\rm CONFID \geq 2$ (night \oii emitters, six
absorption line galaxies, six \ciii emitters, and 43 \lya
emitters). We only visually inspected the spectra of galaxies with
$\rm F775W \leq 27$~mag and for the rest we relied on {\tt ORIGIN}
(and {\tt MUSELET}). Thus, a subset of these missing redshifts may be
recovered if we actually look at the spectra or change the tuning of
{\tt ORIGIN} (and {\tt MUSELET}).  If we include the objects that have
no UVUDF counterpart (i.e., detected only by {\tt ORIGIN/MUSELET}),
then these numbers increase to 133 and 78 for $\rm CONFID \geq 1$ and
$\geq 2$, respectively.  The smaller number of {\tt ORIGIN/MUSELET}
detections in the overlap region in the \mosaic can also be attributed
to the depth of the data.

Together with this paper, we release both the \udft and \mosaic
redshift catalogs (Appendix~\ref{app:cat}). As explained above, the ID
numbers of the objects in the overlapping region are the same in \udft
and the \mosaic. In addition, we compiled a final redshift catalog
combining the \udft and \mosaic results.  For the objects that have
measurements from both of the catalogs, the information is from \udft,
unless there is a disagreement with the \mosaic. For those cases, we
adopted the redshifts discussed above.  For the discussion below, we
used the redshifts and the associated parameters from this combined
catalog.

\subsection{Final redshifts in the entire MUSE UDF survey region}

\begin{figure*}
  \begin{center}
    \includegraphics[width=\textwidth]{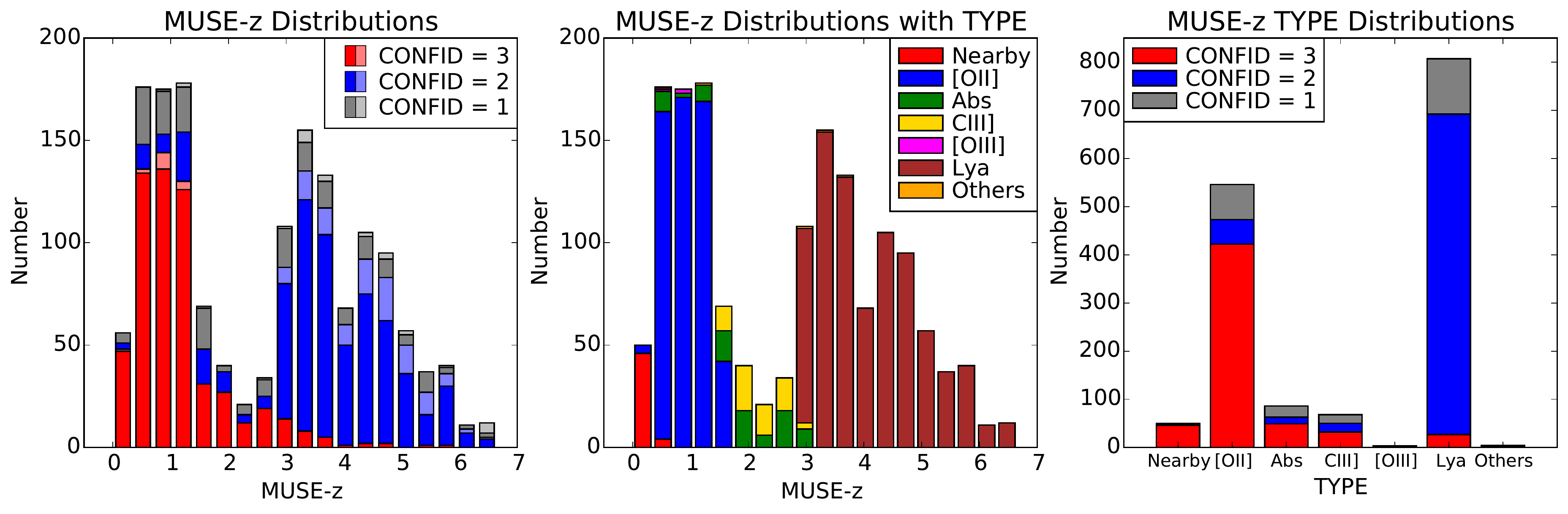}
    \caption{Same histograms as Figures~\ref{fig:u10_musez} and
      \ref{fig:mos_musez}, but for the combined redshifts of the unique
      objects from the MUSE Ultra Deep Field (\udft) and MUSE Deep
      Field (the \mosaic). }
    \label{fig:comb_musez}
  \end{center} 
\end{figure*}

In the entire MUSE UDF survey region (\udft $+$ the \mosaic), there
are $\COMBHSTpri$ UVUDF sources that we used as priors to extract the
continuum selected objects. As shown in Table~\ref{tbl:num_z}, for
$\COMBzHSTCtwo$ (15\%) of them, we successfully obtained secure
redshifts ($\rm CONFID \geq 2$). As discussed above, some of the
objects were merged because of the lower spatial resolution of MUSE
than {\it HST} and we did not investigate a subset
($\rm F775W \leq 27$~mag in the \mosaic where not overlapping with
\udft) of their spectra to determine redshifts.  The direct searches
of emission line objects enabled us to find $\COMBzORGonlyCtwo$ more
redshifts in addition.  Thus, in total, we obtained $\COMBzALLCtwo$
unique redshifts with high confidence ($\rm CONFID \geq 2$). The final
redshift distribution is presented in Figure~\ref{fig:comb_musez}. As
a simple test, we compared the broadband (F775W) luminosity against
the measured MUSE redshifts ($\rm CONFID \geq 2$) to check whether
there are any catastrophic measured redshifts. This test does not show
any extreme outliers.

We used the redshifts determined in both \udft and the \mosaic to
derive the distribution of redshift differences between these two
catalogs.  The standard deviation of the $\Delta z/(1+z)$ distribution
is 0.00017. Assuming that the errors are similar for the \udft and
\mosaic redshifts (i.e., assuming that the depth of the data is not a
dominant factor for redshift errors), then we obtain a global estimate
of the redshift/velocity uncertainty to be $\sigma_z = 0.00012 (1+z)$
or $\sigma_v \approx 40 \, {\rm km\,s^{-1}}$.

\section{Discussion}\label{sec:discussion}

\subsection{Redshift measurement success rates}

Assuming all of the determined redshifts are correct for
$\rm CONFID \geq 2$, we calculated the success rates of the redshift
measurements. Here we define the success rate as the number fraction
of the obtained secure redshifts over the sample size we inspected. We
only used the sample extracted with the {\it HST} priors for this
calculation (i.e., the objects detected only by {\tt ORIGIN/MUSELET}
are excluded).  In the \mosaic, all galaxies with
$\rm F775W > 27$~mag are extracted by {\tt ORIGIN/MUSELET} (or due to
splitting; see \S\ref{subsec:split}). In other words, the success
rates at $\rm F775W > 27$~mag for the \mosaic should be considered as
lower limits.

\begin{figure}
  \begin{center} 
    \resizebox{\hsize}{!}
    {\includegraphics[width=0.48\hsize]{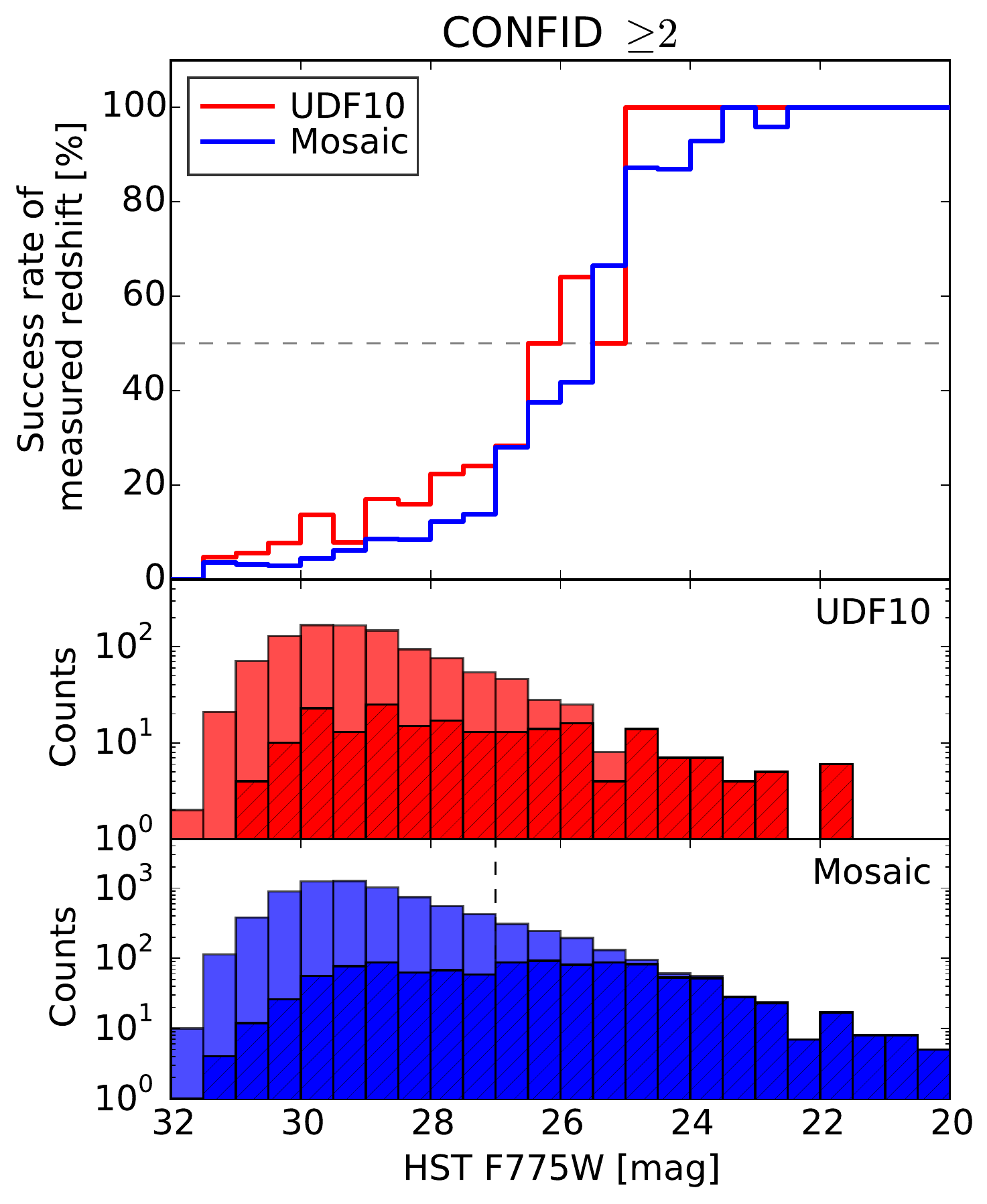}}
    \caption{ Success rates of obtained secure redshifts are shown in
      the top panel for \udft (red) and \mosaic (blue). The horizontal
      dashed line indicates $50\%$ completeness. We only use the
      objects with {\it HST} counterparts in the UVUDF catalog in this
      plot (i.e., the objects detected only by {\tt ORIGIN/MUSELET}
      are excluded). The middle and bottom panels show the counts of
      the total number of the {\it HST} objects (solid bars) and the
      MUSE redshift determined objects (hatched bars). The vertical
      dashed line in the bottom panel indicates the magnitude cut
      where we perform the redshift investigation on the continuum
      detected objects. }
    \label{fig:z_success_rate}
  \end{center} 
\end{figure}

In Figure~\ref{fig:z_success_rate}, the success rates of \udft and
\mosaic are plotted against the {\it HST} F775W magnitude.  In \udft,
we successfully measured secure redshifts ($\rm CONFID \geq 2$) for
all of the galaxies brighter than $25$~mag. At the same magnitude, the
success rate for the \mosaic is $87\%$.  The $100\%$ success rate for
the \mosaic is at $< 22.5$~mag (or $23.5$~mag if we ignore one object
MUSE ID 6934 without MUSE redshift in the $22.5-23.0$~mag bin). The
50\% completeness with respect to the {\it HST} F775W magnitude is
reached at $26.5$~mag and $25.5$~mag, in \udft and the \mosaic,
respectively.  The success rates decrease with the F775W magnitudes,
in particular there is a sudden drop at $\approx 25$~mag for both of
the fields. However, the success rate remains greater than
$\approx 20\%$ at $< 28-29$~mag in \udft and $\lesssim 27$~mag in the
\mosaic.  Except in the $25.0-25.5$~mag bin, at all magnitudes, the
\udft success rate is higher than the \mosaic.

Another MUSE deep survey in the {\it Hubble} Deep Field South
\citep[HDFS;][]{Baco15}, whose coverage is also a single MUSE field of
view as our \udft and has a similar depth ($27$~hours), reaches the
$50\%$ completeness at $26$~mag in the {\it HST} F814W band.  The
achieved slightly higher completeness in \udft is a natural consequence of the
better line flux detection limit, which results from the improved data
reduction and analysis compared to the HDFS data cubes \citep{Baco15}.

The spectroscopic completeness and comparisons with photometric
redshifts are discussed in Paper~III.

\subsection{Objects only detected by {\tt ORIGIN} or {\tt MUSELET}}

Owing to its wide FoV IFU with high sensitivity, MUSE\ enables direct
detection of emission line objects with very faint continuum emission,
which are improbable to target with slit spectroscopy and inefficient
to find with narrowband imaging.  In the entire
$3\arcmin \times 3\arcmin$ survey area, we discovered
$\COMBzORGonlyCtwo$ emission line detected objects with confident
redshifts which are not on the {\it HST} prior list.  In the deepest
region ($1\arcmin \times 1\arcmin$ area of \udft), this number is
$\UDFzORGonlyCtwo$. Among these $\COMBzORGonlyCtwo$ emission line
objects, the majority are \lya emitters (117). The rest are \oii
emitters (14) and a nearby galaxy (1).  Some of these objects in fact
can be visually identified in the images but are
blended\,\footnote{Object MUSE ID 6449 actually has UVUDF ID 4293
  (CANDELS ID 18674) in the UVUDF segmentation map. However, UVUDF ID
  4293 does not exist in the UVUDF catalog. Since the actual
  measurement is not provided in the UVUDF catalog, here we consider
  that it was not successfully extracted in UVUDF.} (see
\S\ref{sec:udf10_finalz}). However, especially for the \lya emitters,
even when they can be visually identified, they have very low surface
brightness or are not detected in the continuum. As mentioned in
\S\ref{subsec:NoiseChisel}, we performed our own flux or upper limit
measurements for these galaxies, but here we look into the most recent
{\it HST} catalogs in more detail to investigate their properties.

Our prior list is based on the UVUDF catalog created by running {\tt
  ColorPro} \citep[a wrapper of {\tt SExtractor};][]{Coe06} with the
detection image obtained by averaging four optical (F435W; F606W,
F775W, F850LP) and four near-infrared (F105W, F125W, F140W, F160W)
images \citep{Rafe15}. Except for the blended galaxies, the continuum
of the MUSE emission line objects was still too faint to be originally
detected in the combination of this deepest detection image.

We also checked if any of the emission line only objects are in the
CANDELS\,\footnote{Cosmic Near-IR Deep Extragalactic Legacy Survey}
GOODS-S multiwavelength catalog \citep{Guo13}, and identified 52
objects (one nearby object, eight \oii emitters, and 43 \lya emitters)
that have potential counterparts within $1''$. This does not
immediately mean that they are the actual corresponding objects
because some {\tt ORIGIN/MUSELET} detected objects are completely
blended with known objects or happen to lie on the same sightline. We
inspected these objects one by one and found one nearby galaxy, seven
\oii emitters, and one \lya emitter that exist in the CANDELS catalog.
For this \lya emitter (MUSE ID 6343 or CANDELS ID 15913), based on the
prospective \lya in its MUSE spectrum, its redshift is determined to
be $5.5$. However, its $U$ magnitude is 28.5~mag in the CANDELS
catalog, which is not probable for a high-$z$ galaxy because of the
Lyman break at 912\AA.  It is more likely that this \lya emitter
happens to lie on the same line of sight as this galaxy.  The
template-fitting technique, {\tt TFIT} \citep{Laid07}, is utilized for
the CANDELS catalog. The {\tt TFIT} tool uses spatial positions and
morphologies in a high-resolution image as the priors to fit objects
in lower resolution images. Until objects cannot be resolved in the
high-resolution image, it can further perform deblending in the lower
resolution images even when object separations are $\lesssim 1.5$
times the PSF FWHM, which is the limit for {\tt SExtractor}.  Thus,
some of these closely separated objects could not be deblended in the
UVUDF catalog but were in CANDELS.  In fact, the one \oii emitter
(MUSE ID 6315) that is not in the CANDELS catalog is not resolved even
with {\it HST,} and the 42 \lya emitters are barely detected or not at
all.

\subsection{Blend and split of merged objects}\label{subsec:split}

\begin{figure}
  \resizebox{\hsize}{!}
  {\includegraphics{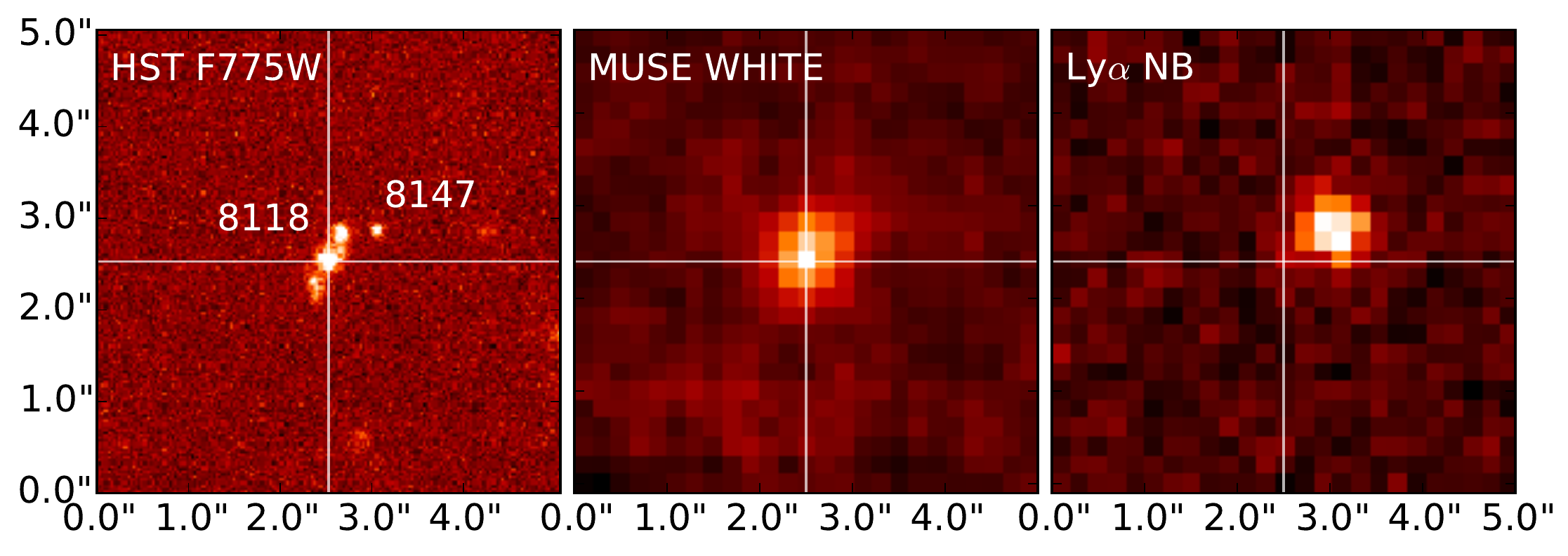}}
  \caption{ Stamp impage of $5\arcsec \times 5\arcsec$ centered at
    $\rm R.A.=53.169458$ and $\rm Dec=-27.778191$. The {\it HST} F775W
    image (white-light, left panel), the MUSE $\lambda$-collapsed
    image (middle), and the continuum subtracted narrowband image
    created at $\lambda=5676$\Ang with a width of $10$\Ang (right).
    While it is not possible to resolve the two nearby {\it HST}
    objects (UVUDF IDs of 8147 and 8118) with the MUSE white-light
    image, the combination of the {\it HST} image with the narrowband
    image of the detected emission line (\lya in this case) makes it
    possible to identify the origin of the emission (UVUDF ID 8147
    corresponds to MUSE ID 6290 with $z=3.67$).}
  \label{fig:split}
\end{figure}

During the redshift evaluation process, we found some cases in which we
were able to split the merged MUSE objects (see \S\ref{sec:analysis_cont}).
Comparisons between a detected emission line narrowband image and the
{\it HST} image provide the spatial information of the corresponding
{\it HST} counterpart of the emission line object. An example is shown
in Figure~\ref{fig:split}. The object shown in the MUSE white
image (the middle panel) is originally extracted as a merged object,
which corresponds to two {\it HST}-detected objects with UVUDF IDs of 8147
and 8118. While these two objects are clearly resolved in the {\it
  HST} F775W image (the left panel), they are not resolved in the MUSE
white image (the middle panel). We managed to associate the determined
redshift (based on \lya) with one of the {\it HST} objects, UVUDF ID
8147, by creating a narrowband image of the detected \lya at $z=3.67$
(the right panel). Although we cannot fully deblend the MUSE emission
because of the limited spatial resolution, we can assign individual MUSE
ID numbers for these objects using their UVUDF coordinates
(``split'').  This \lya emitter is given MUSE ID 6290 and the other is
6749.  This process does not increase or decrease the number
of objects whose redshifts are successfully reported as identified
in the earlier sections, unless both (or multiple) of the split
objects show clear features that can be used to determine redshift.

\begin{figure}
  \resizebox{\hsize}{!}
  {\includegraphics{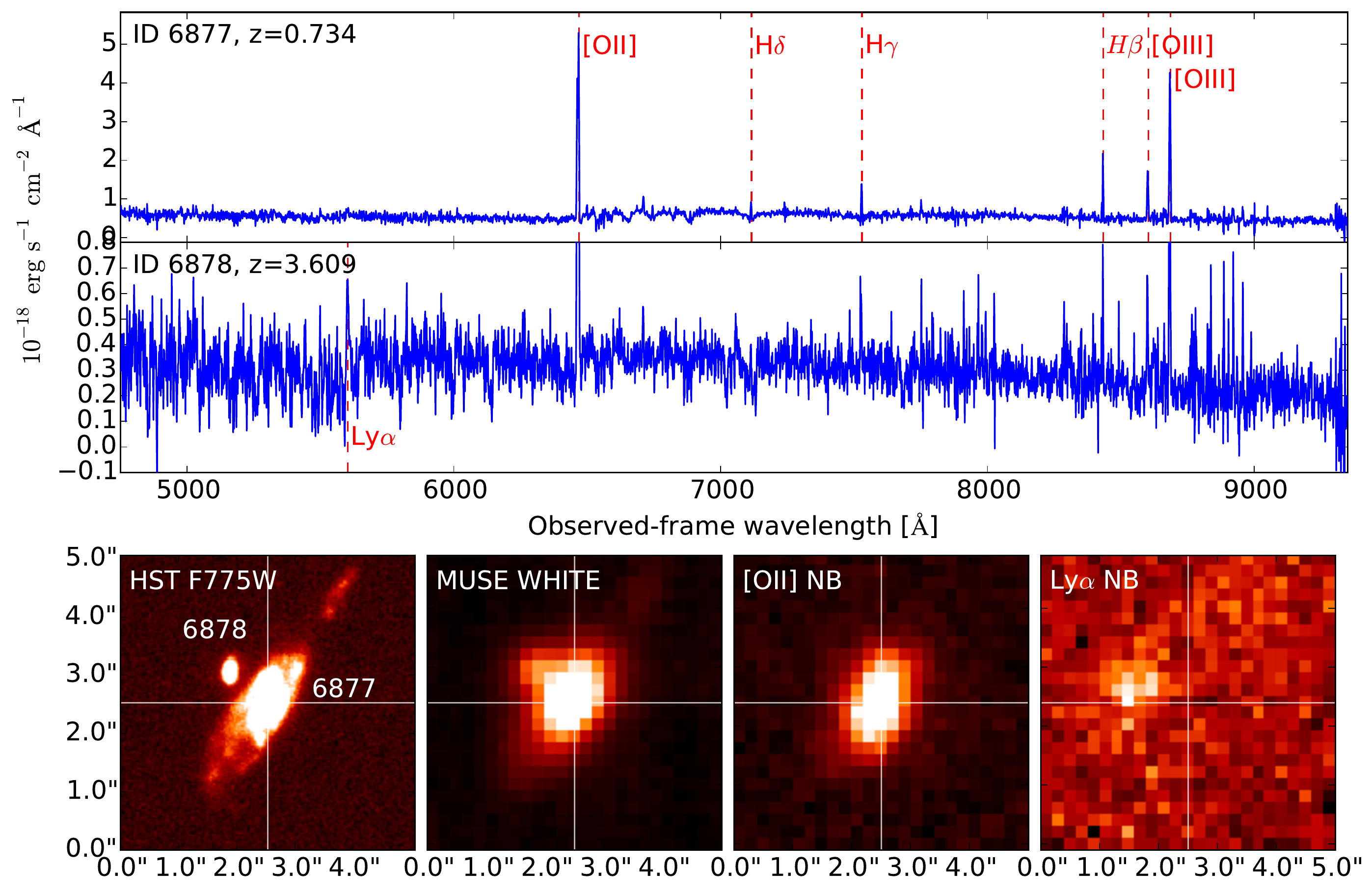}}
  \caption{ Spectra (extracted by PSF weighting) of MUSE IDs 6877
    ($z=0.734$) and 6878 ($z=3.609$) at the top. The images shown at
    the bottom are {\it HST} F775W, MUSE white-light, and narrowband
    images of \oii at $z=0.734$ and \lya at $z=3.609$, from left to
    right. Because these two objects are barely resolved with MUSE,
    they were originally treated as a single ``merged''
    object. However, their spectra and a comparison of the narrowband
    images with the {\it HST} image help to ``split'' and associate
    the MUSE redshifts with the corresponding {\it HST} counterparts.}
  \label{fig:split_bothz}
\end{figure}

Although small in number, there are also cases in which multiple
redshifts are found in a single merged object.
The MUSE IDs of 6877 and 6878 are originally merged and treated as a
single MUSE extracted object. However, clear detections of the \oii
doublet, \oiii$\lambda\lambda4959,5007$, and several Balmer lines at
$z=0.734$ for ID 6877 and \lya at $z=3.609$ for ID 6878 allow us to
assign confident redshifts for both of the objects in addition to
splitting them (Figure~\ref{fig:split_bothz}). 

\begin{figure}
  \resizebox{\hsize}{!}
  {\includegraphics{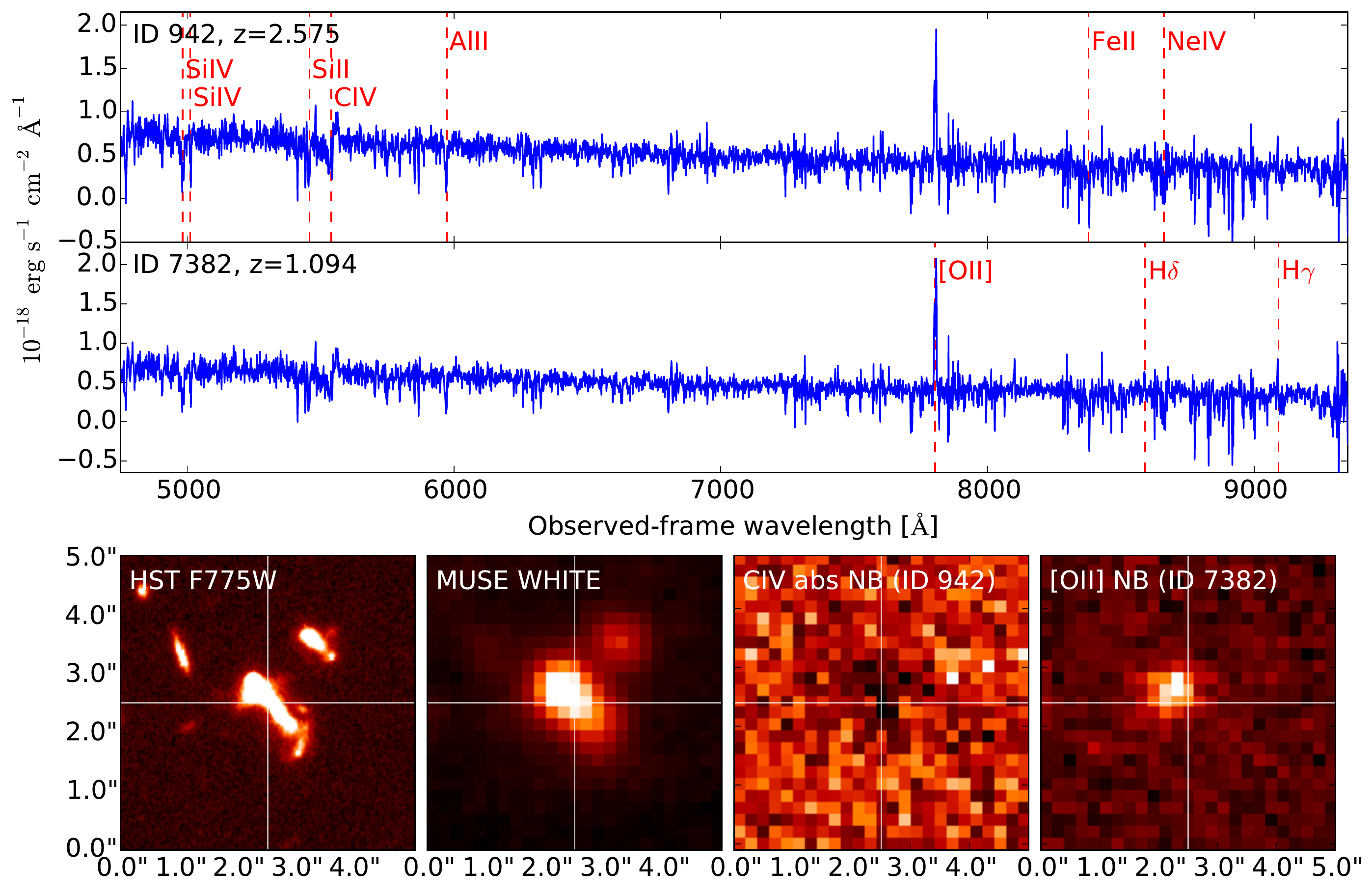}}
  \caption{ Similar to Figure~\ref{fig:split_bothz}, but for MUSE IDs
    942 ($z=2.575$) and 7382 ($z=1.094$). The extracted spectra
    clearly show that there are at least two objects lying on the same
    sight line. A small offset is also seen between the narrowband
    images of the C\,\textsc{iv} absorption feature at $z=2.575$ and
    \oii at $z=1.094$. However, there is no obvious {\it HST}
    object to provide the counterpart for the \oii emitter. }
  \label{fig:split_bothz_noID}
\end{figure}

There are also some rare cases in which even though two or more of the
merged objects have more than one redshift identified, they cannot be
associated with any of the known objects owing to complex morphology,
too close separation even for the {\it HST}, or no continuum
detection. As shown in Figure~\ref{fig:split_bothz_noID}, in the
one-dimensional spectrum of an absorption galaxy at $z=2.575$, MUSE ID
942, multiple emission lines are visible at $7805$\Ang, $8100$\Ang,
$8590$\Ang, and $9090$\Ang. These features do not correspond to any of
the features that can be seen at $z=2.575$ of ID 942, but are
consistent with \oii, \neiii$3869$, \hd, and \hg at $z=1.094$ (MUSE ID
7382), respectively. We use the center of the \oii narrowband image as
the coordinates of this object. This particular case was found by
hand, but {\tt ORIGIN} has played an important role in finding these
kinds of blended objects that are hard to find by humans. For example,
multiple pronounced emission features seen in MUSE ID 945 immediately
reveal its redshift to be $z=0.605$. It is very easy to neglect a
weaker emission feature detected at $7230$\Ang in the same spectrum,
which is not associated with ID 945. {\tt ORIGIN} successfully found
this feature and we identify it as \lya at $z=4.947$ (MUSE ID 6470, no
UVUDF counterpart).

In the end, we split about 150 of the merged objects.  Among
$\COMBzALLCone$ objects with MUSE-$z$ ($\rm CONFID \geq 1$), there are
79 that remain as merged objects.  We stress that we do not use or
provide the photometric measurements for the merged objects.  When
merged objects have been successfully split during the redshift
determination, the split objects have been associated with the
corresponding photometries from the UVUDF catalog.  If we find
multiple redshifts in an object that is not merged (the case of
Figure~\ref{fig:split_bothz_noID}), only the object with the centroid
in the narrowband images of the detected features closest to the UVUDF
coordinates or the object with the reasonable {\it HST} color is
assigned to have the UVUDF photometries.

\subsection{Comparisons with previous spectroscopic redshifts}

\begin{figure}
  \begin{center} 
    \includegraphics[width=\textwidth/2]{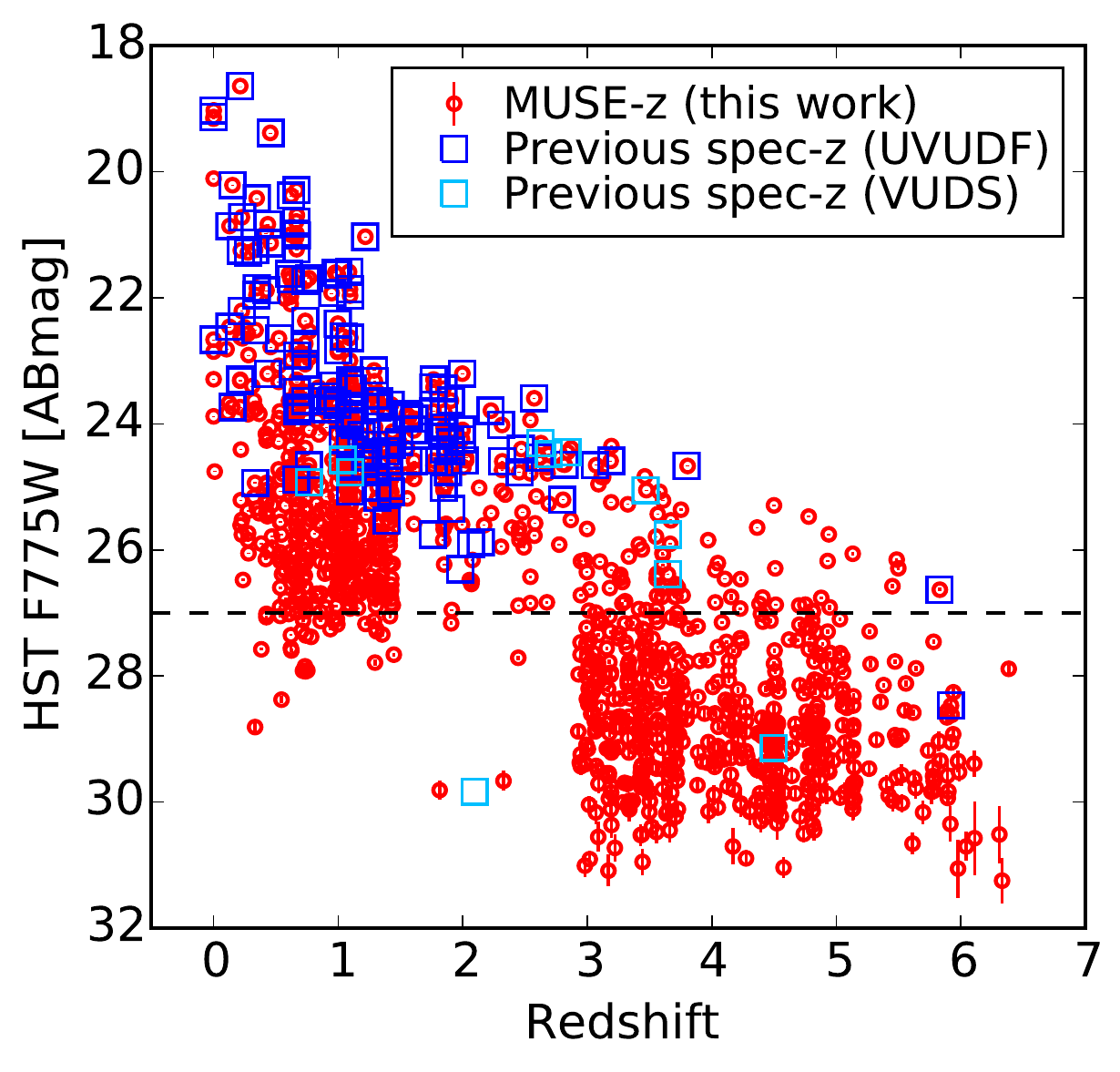}
    \caption{ Magnitude vs. redshift space of secure MUSE-$z$
      ($\rm CONFID \geq 2$; red circles) and previous reliable
      spec-$z$ for the unique objects in the entire MUSE deep survey
      region (\udft $+$ \mosaic).  The previously known spec-$z$ taken
      from the UVUDF and VUDS catalogs are indicated with blue and
      cyan squares, respectively.  The horizontal dashed line
      indicates 27~mag where we make the cut to the continuum selected
      galaxies to perform the redshift determination in the \mosaic
      field (see \S\ref{subsec:mag_cut}).}
    \label{fig:prev_specz}
  \end{center} 
\end{figure}

There has been a large effort to measure spectroscopic redshifts
(spec-$z$) in the UDF and the surrounding regions, such as GOODS,
VVDS, and VUDS.  According to the list of spec-$z$ compiled in the
UVUDF catalog by \cite{Rafe15}\,\footnote{The spectroscopic redshifts
  provided in the UVUDF catalog are gathered from the following
  surveys: VVDS \citep{LeFe04}, Szokoly \citep{Szok04}, K20
  \citep{Mign05}, GRAPES \citep{Dadd05}, Vanzella GOODS
  \citep{Vanz05,Vanz06,Vanz08,Vanz09}, Popesso GOODS \citep{Pope09},
  Balestra GOODS \citep{Bale10}, GMASS \citep{Kurk13}, and 3D-HST
  \citep{Morr15}.}, there are $\COMBprevzUVUDF$ previously known {\it
  secure} spec-$z$ in our survey region ($3' \times 3'$). The total
number of published redshifts in the UVUDF catalog is 169 over the
entire HUDF area.  In addition, within a $0.5''$ search radius, we
find that there are $\COMBprevzVUDS$ MUSE objects\,\footnote{MUSE ID
  6945 has a previously known spec-$z$ in both of the UVUDF and VUDS
  catalogs (1.098 and 1.096, respectively). These redshifts agree, and
  thus we do not count this as an additional spec-$z$ and only use the
  spec-$z$ from UVUDF.}  that have previously known spec-$z$ measured
by VUDS with high reliability \citep[redshift flags of 4, 3, or
2;][]{Tasc17}. Thus, in total, we compile $\COMBprevz$ previously
known reliable spec-$z$ in our survey field. It is noteworthy that
MUSE delivers a drastic improvement not only in the number of
spec-$z$, but also in the range of redshift and the limiting magnitude
of the objects with confident spec-$z$ (Figure~\ref{fig:prev_specz}).

The majority (all for \udft) of the published spec-$z$ are in the
redshift range of $0 < z < 3$ and F775W magnitude of $< 25$. The three
objects at $z > 3.5$ in the UVUDF catalog are from the spectroscopic
observations of Lyman break galaxies \citep{Vanz09}\,\footnote{Their
  object IDs are J033236.83-274558.0, J033240.01-274815.0, and
  J033239.06-274538.7}.

\begin{figure}
  \begin{center} 
    \includegraphics[width=\hsize]{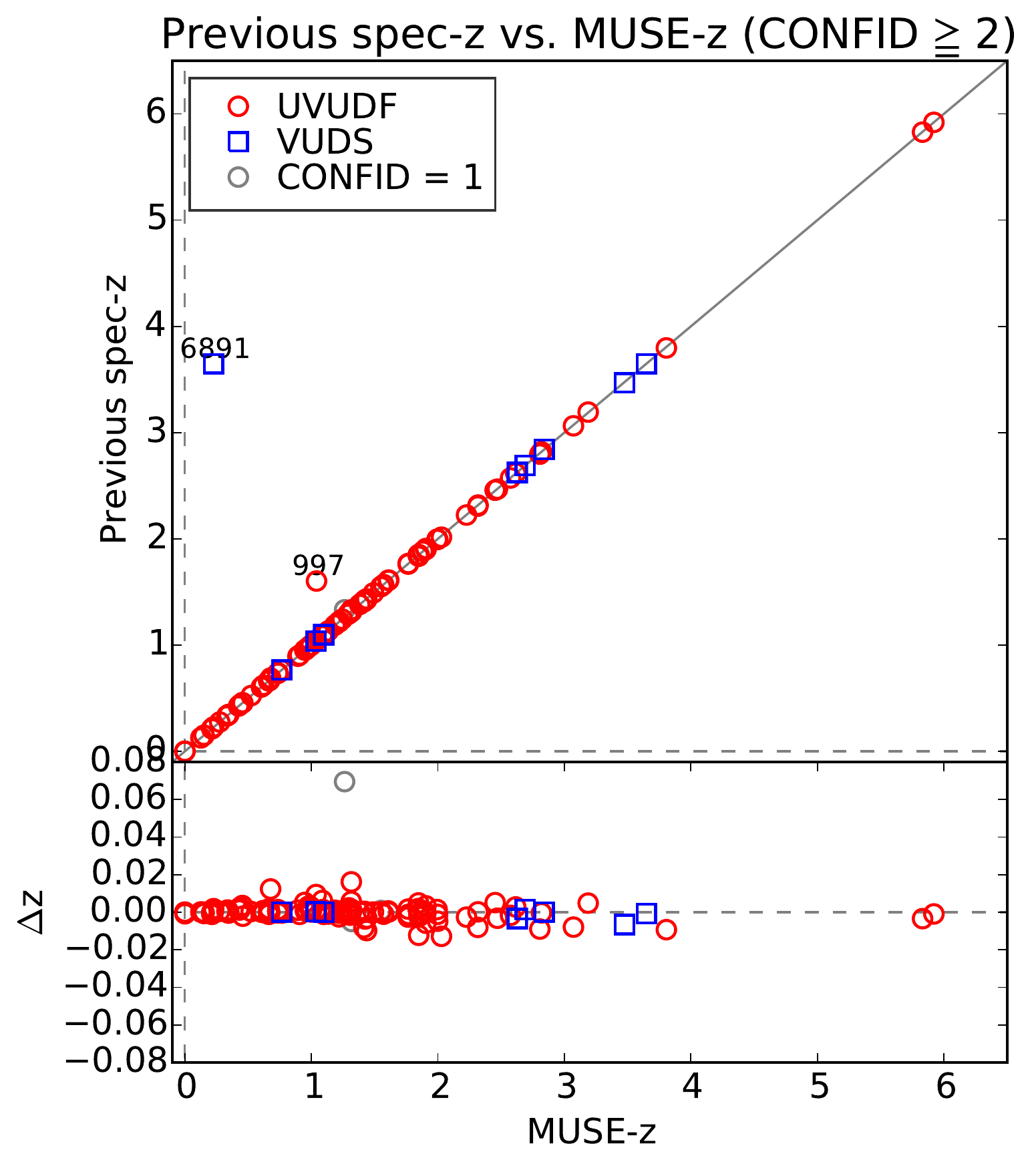}
    \caption{ Comparisons of MUSE-$z$ to the previous spectroscopic
      redshifts from the other ground- and space-based telescopes
      compiled in the UVUDF catalog. The high and low confidence
      MUSE-$z$ are shown as red and gray symbols, respectively.
      MUSE ID 997 indicated in the figure is the only object with a
      catastrophic difference in the spec-$z$ comparisons. See the
      text for more details.}
    \label{fig:u10_comp_specz}
  \end{center} 
\end{figure}

A simple check of MUSE redshifts is to compare them with the published
spectroscopic redshifts.  In Fig.~\ref{fig:u10_comp_specz}, our
MUSE spectroscopic redshifts are compared to these previously known
spectroscopic redshifts.  When we only use the robust redshifts
(CONFID $\geq 2$), we find six objects with $|\Delta z| > 0.01$ (MUSE
IDs 29, 949, 957, 997, and 1048 from UVUDF and MUSE ID 6891 from
VUDS), but all except IDs 997 and 6891 have $0.01 < |\Delta z| < 0.02,$
which is expected to happen owing to observations with different
spectroscopic resolutions.  We investigate the two objects with
catastrophic differences below (CONFID of MUSE-$z$ is given in the
parentheses):

\begin{description}

\item[{\bf 997} $z_{\rm MUSE}=1.041$ (3), $z_{\rm known}=1.603$ \citep{Morr15}]\mbox{}\\
  The previous spec-$z$ is measured from WFC3 grism (G141) data by
  identifying \hb and \oiii (with a good quality, 3 out of 4), while
  MUSE-$z$ is determined by a well-detected \oii doublet,
  \neiii$3869$, \hd, and \hg (with CONFID of 3, the
  highest). Reinspecting the WFC3 grism spectrum verifies that
  $z=1.603$ is not as convincing as originally determined. In
  addition, there is an extra complication in extracting the grism
  slitless spectrum due to multiple nearby objects. Thus, we conclude
  that our MUSE redshift is likely to be correct for MUSE ID 997
  (UVUDF ID 8592).
  \\

\item[{\bf 6891} $z_{\rm MUSE}=0.227$ (3), $z_{\rm known}=3.647$ \citep{Tasc17}]\mbox{}\\
  Although the redshift of ID 6891 (UVUDF ID 4864) is identified by
  good detections of \hb , \oiii, and \ha, its spectrum is blended
  with ID 6892 (UVUDF ID 4863), which shows a \lya emission line at
  $z_{\rm MUSE} = 3.648$ ($\rm CONFID = 2$).  The separation of these
  two objects is $0.3''$. The previously measured spec-$z$ corresponds
  to MUSE ID 6892, but the pointing coordinates are closer to MUSE ID
  6891. Also, these two objects were not successfully
  deblended in the CANDELS catalog (CANDELS ID 13745 for both of the
  objects). Thus, we think that our MUSE redshift measurement for MUSE
  ID 6891 (and 6892) has no problem.

\end{description}

When including $\rm CONFID=1$ redshifts, one MUSE object (MUSE ID
1143) has previously known spectroscopic redshifts but disagrees with
our MUSE redshifts \citep[$z_{\rm MUSE}=1.262$ and
$z_{\rm known}=1.332$ from][]{Morr15}.  The previously known redshift
1.332 is measured in the near-infrared spectrum obtained with HST/WFC3
grism (G141) spectroscopy. \cite{Morr15} identified \hb and \oiii. In
our MUSE spectrum, we identified \oii$\lambda3729$ at $8435$\Ang, which
gives $z=1.262$. Its CONFID is 1 because \oii$\lambda3726$ is not
detected probably because it overlies a sky line. The difference in
redshift is $\delta z = 0.07,$ which cannot be explained by the low
spectral resolution ($R \sim 130$) of the grism spectroscopy.  We found
that the 3D-HST survey \citep{Bram12} also published a redshift for
this object of $z=1.299$ \citep[by identifying \ha;][]{Momc16}, which is
consistent with our measurement. The slight discrepancy
($|\Delta z| = 0.037$) can be attributed to the fact that the object
is spatially extended in the dispersion direction for the grism. Thus,
we assessed that the MUSE redshift for this object is more feasible.

There are 10 objects whose spec-$z$ are reported in the UVUDF (seven
objects) and VUDS (three objects) catalogs but are missed by our
redshift determination. These objects are listed below with MUSE ID,
UVUDF ID in the parentheses, previously measured spec-$z$, and the
reference:

\begin{description}

\item[{\bf 55 (24380)}: $z = 1.39 \pm 0.01$ \citep{Dadd05}]\mbox{}\\
  The spectrum was obtained with {\it HST}/ACS grism low-resolution
  spectroscopy (G800L, $R \approx 100$ at $8000$\Ang for a point
  source), which detected a broad \mgii absorption line.  This feature
  is not seen in our MUSE spectrum even with smoothing probably due to
  much higher noise level in the continuum. This object is located in
  the overlapping region of the \mosaic and \udft, but we cannot find
  a redshift from the spectrum extracted from the deeper \udft data.
  \\

\item[{\bf 1199 (3482)}: $z = 1.91 \pm 0.01$ \citep{Dadd05}]\mbox{}\\
  \cite{Dadd05} found a broad blended \mgii$\lambda\lambda2796,2803$
  doublet absorption feature (and likely the \feii multiplet
  absorption) in their spectrum taken with HST/ACS grism
  spectroscopy. This feature can be identified in our MUSE spectrum
  too, but only after knowing the redshift or heavily smoothing the
  spectrum.
  \\

\item[{\bf 1308 (10544)}: $z = 1.313$ \citep{Morr15}]\mbox{}\\
  The known redshift was determined using the spectrum taken with {\it
    HST}/WFC3 grism spectroscopy (G141, $R \approx 130$ at
  $1.1 \leq \lambda \, {\rm [\mu m]} \leq 1.7$). The emission features
  \hb, \oiii, and \ha were identified to obtain $z=1.313$.  With this
  redshift, we confirm the \oii emission at $8630$\Ang in our MUSE
  spectrum. It is located in the wavelength region where sky emission
  dominates, which made it easily missed during our redshift
  determination process. This redshift is added into the combined
  catalog by hand ($z_{\rm MUSE}=1.315$ with $\rm CONFID = 3$).
  \\

\item[{\bf 1312 (21773)}: $z = 1.76 \pm 0.02$ \citep{Dadd05}]\mbox{}\\
  The {\it HST}/ACS grism spectrum detected a broad \mgii absorption
  line. With the previously determined redshift, we can confirm
  the same feature in the MUSE data. However, it lies in a wavelength
  region crowded with sky lines, which makes it even harder to identify
  the feature without the a priori information.
  \\

\item[{\bf 1351 (4562)}: $z = 2.145$ \citep{Bale10}]\mbox{}\\
  The spectrum was taken with VLT/VIMOS.  The previously known
  redshift was determined based on the O\,\textsc{i} ($1302.20$\Ang)
  and C\,\textsc{ii} ($1335.10$\Ang) absorption features detected at
  $4095$\Ang and $4200$\Ang, respectively.  These features are both
  outside the MUSE wavelength coverage, preventing us from recovering
  its redshift with MUSE. At this redshift, \ciii and some absorption
  features are covered by the MUSE wavelength range, but none of these
  features are visible.
  \\

\item[{\bf 1364 (8292)}: $z = 2.067$ \citep{Morr15}]\mbox{}\\
  The redshift was measured in the spectrum taken with the {\it
    HST}/WFC3 near-infrared grism. These authors detected \hb and
  \oiii. We can indeed recognize the \mgii$\lambda\lambda2796,2803$
  absorption lines (and possibly \feii$\lambda2344$ and
  \feii$\lambda\lambda2374,2382$) in the MUSE data with this
  redshift. It is difficult to notice these features in our data
  because at this redshift, the \mgii features lie in the low S/N
  region due to many sky lines.
  \\

\item[{\bf 1520 (21730)}: $z = 1.98 \pm 0.02$ \citep{Dadd05}]\mbox{}\\
  A broad \mgii absorption feature (and possibly the \feii multiplet
  absorption) was detected in their {\it HST}/ACS grism spectrum.
  This feature might be visible in the MUSE data, but is not obvious
  because of crowded sky lines. It is difficult to use our MUSE data
  solely to determine this redshift.
  \\

\item[{\bf 4070 (635)}: $z = 4.498$ ($z_{\rm flag}=3$) \citep{Tasc17}]\mbox{}\\
  The VUDS redshift (VUDS ID 532000222) may have been determined by a
  tentative \lya emission at 6685\Ang.  In our MUSE spectrum we also
  discern an emission line at 6685\Ang, but this is contamination from
  MUSE ID 1185 (\lya at $z_{\rm MUSE}=4.50$), which is $2.38''$
  away. It is possible that the VUDS slit also covered the extended
  \lya emission from ID 1185, which led to $z = 4.498$.
  \\

\item[{\bf 5216 (8384)}: $z = 2.095$ ($z_{\rm flag}=2$) \citep{Tasc17}]\mbox{}\\
  It looks as though the VUDS redshift (VUDS ID 539990803) was
  derived from a good cross-correlation signal. However, our
  cross-correlation with {\tt MARZ} does not provide a good signal and
  we are not able to find the expected absorption features at this
  redshift in our MUSE spectrum.
  \\

\item[{\bf 5494 (7285,7374)}: $z = 0.768$ ($z_{\rm flag}=3$) \citep{Tasc17}]\mbox{}\\
  At the VUDS redshift (VUDS ID 532000256), we might expect to detect
  \feii, \mgii, \hb, and \oiii in the MUSE spectrum, but none of these
  features are seen. The VUDS redshift is likely obtained through a
  good cross-correlation signal, but we are not able to do so with the
  {\tt MARZ} cross-correlation.
  \\

\end{description}

Of these 10 redshifts that were missed by our redshift determination process,
four were obtained via ground-based spectroscopy.  However,
for one of these redshifts (MUSE ID 1351), the detected features are not
accessible with MUSE, and for the other three, we cannot confirm the
previously measured redshifts in the MUSE spectra (IDs 4070, 5216,
and 5494). The rest were taken with sky emission free space-based
spectroscopy. The four optical spectra ({\it HST}/ACS) provide much
higher sensitivity, especially in the continua, which facilitates
detections of absorption features (\mgii in this case). An extremely
broad feature is also easily washed out in our higher resolution
spectra without smoothing.  The remaining two redshifts were measured
with the near-infrared spectra from space ({\it HST}/WFC3). We can
confirm the expected features in the MUSE spectra with the known
redshift, although they are difficult to identify because their
wavelengths are in the region of many strong sky emission lines.

\subsection{Comparisons with photometric redshifts}

\begin{figure}
  \resizebox{\hsize}{!}
  {\includegraphics{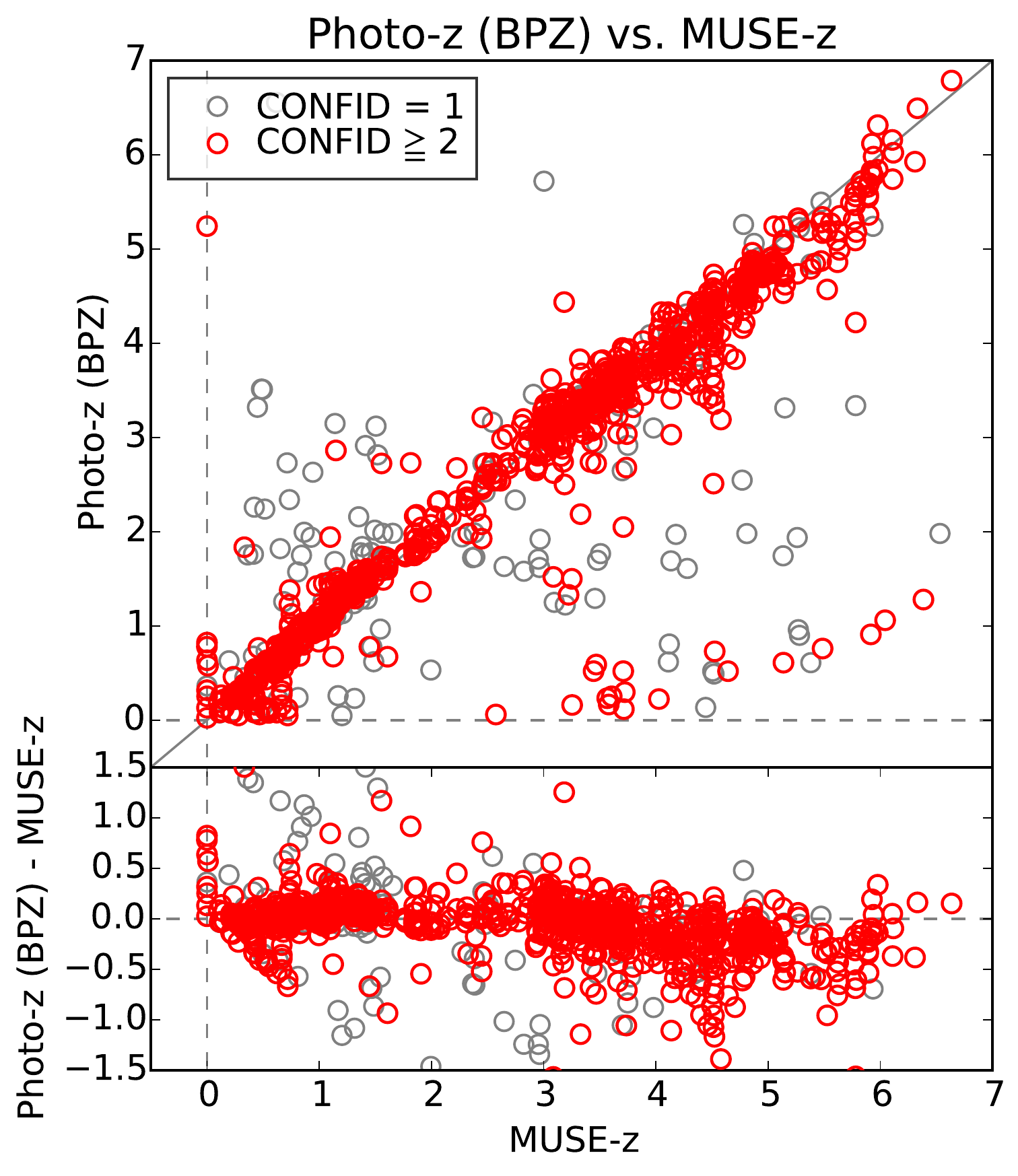}
    \includegraphics{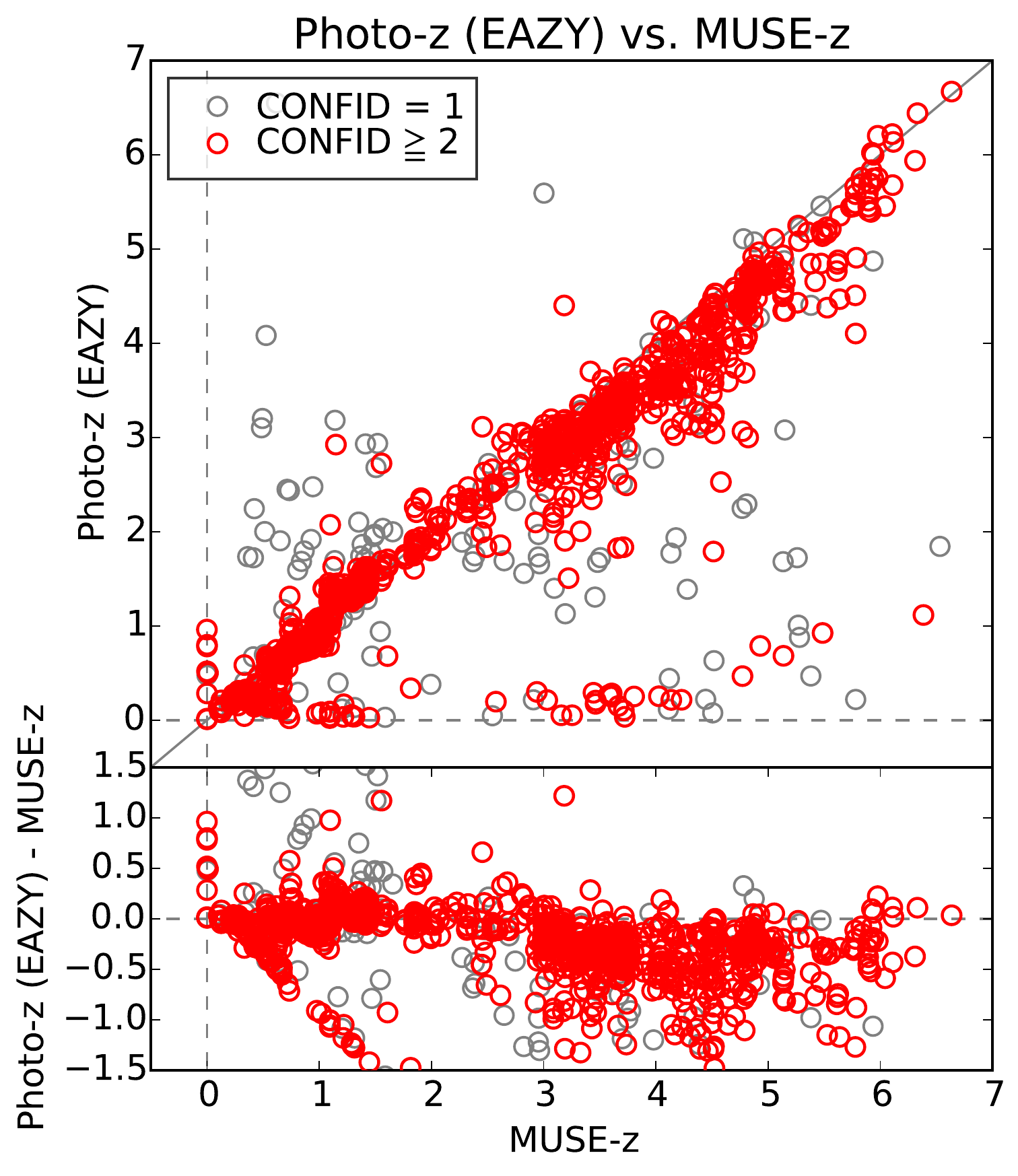}}
  \caption{ Comparisons of redshifts measured from the MUSE data in
    \udft $+$ \mosaic compared with photometric redshifts, computed
    with BPZ (left) and EAZY (right). The redshifts with
    $\rm CONFID \geq 2$ and $\rm CONFID = 1$ are plotted with the red
    and gray circles, respectively. }
  \label{fig:u10_comp_photz}
\end{figure}

We compared our MUSE determined spec-$z$ against the published
photometric redshifts (photo-$z$) from \cite{Rafe15} in
Figure~\ref{fig:u10_comp_photz}. The merged objects
(extracted with {\it HST} priors that cannot be resolved by MUSE) are
not included in the plots. \cite{Rafe15} provide two sets of
photo-$z$, based on the Bayesian photometric redshift (BPZ) estimation
\citep{Beni00} and the EAZY software \citep{Bram08}.

\begin{figure}
  \resizebox{\hsize}{!}
  {\includegraphics{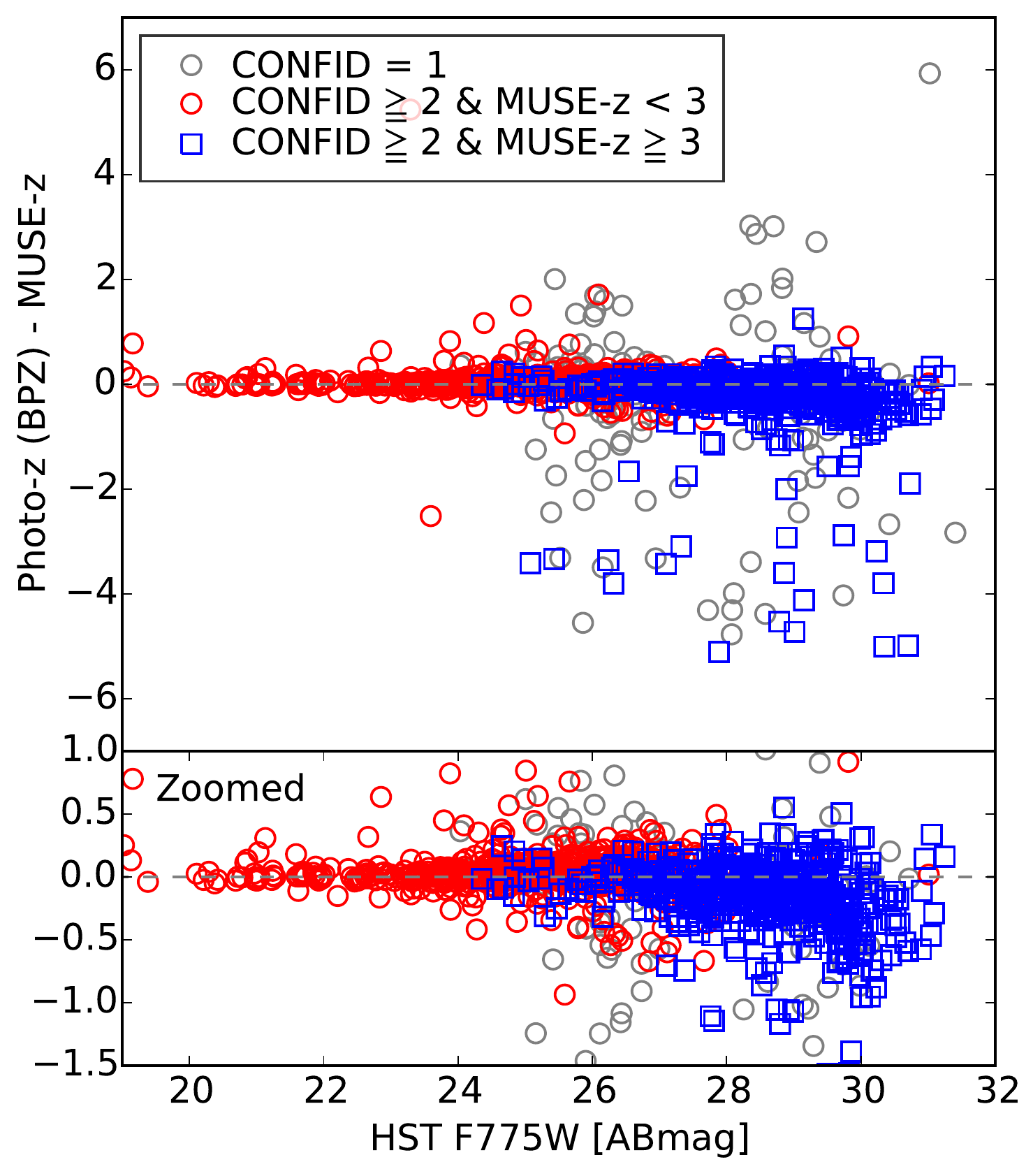}
    \includegraphics{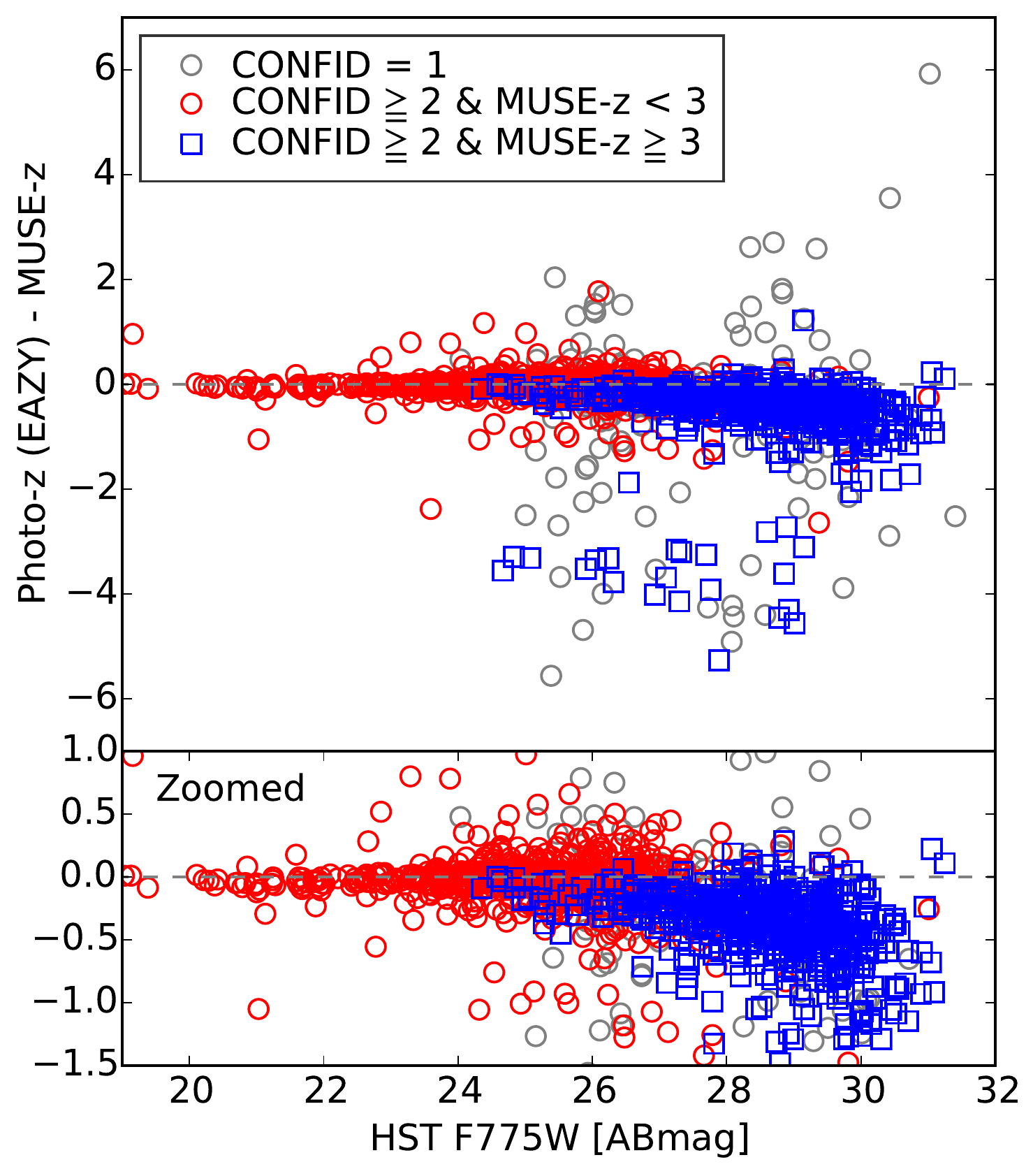}}
  \caption{ Difference in the photometric redshift and MUSE
    redshift (\udft $+$ \mosaic) against the F775W magnitude. The
    redshifts with $\rm CONFID = 1$ are plotted with gray
    circles. The secure redshifts ($\rm CONFID \geq 2$) are separated
    into two groups: $z < 3$ (red circles) and $z \geq 3$ (blue
    squares).}
  \label{fig:u10_comp_photz_diff}
\end{figure}

Although MUSE spec-$z$ and photo-$z$ agree well with both BPZ and EAZY
up to $z \approx 3$, a systematic difference appears at $z \gtrsim 3$.
Both of the photo-$z$ measurements tend to underestimate the redshift.
All of the MUSE redshifts at $z > 3$ are identified by \lya, which is
known to cause the apparent offset toward higher redshift \citep[of a
few hundred $\rm km\,s^{-1}$;][]{Shap03,Stei10,Hash13} attributed to
galactic-scale outflows or absorption by the intergalactic
medium. However, this \lya apparent offset is not the major cause of
the photo-$z$ offset because the median of the measured offsets
($\Delta z$) between MUSE-$z$ ($\rm CONFID \geq 2$) and photo-$z$ is
significantly larger than the expected \lya velocity offset
($\approx -0.13$ and $\approx -0.32$ for BPZ and EAZY, respectively).
In Figure~\ref{fig:u10_comp_photz_diff}, the redshift difference is
plotted against the {\it HST} F775W magnitude. It is clear that the
photo-$z$ measurements are biased at the faint end.

Further analyses and discussions including the redshift completeness
can be found in Paper~III. In this paper, we checked the
  catastrophic redshift outliers and systematic biases.  We
  found that changes in model of treatments of inter-galactic medium
  (IGM) absorption for \lya-forest and Lyman continuum absorption can
  reduce the photo-$z$ and MUSE-$z$ discrepancy. We also adopted
photometric redshifts from the {\tt BEAGLE} software \citep{Chev16}.

\subsection{Color selections of high-z galaxies}

The presence of the 912\Ang Lyman break enables selection of
high-redshift galaxies using their colors. Here we compared MUSE-$z$
to some commonly used color-color selection diagrams to discuss the
fraction of successful color selections. The combination of the MUSE
wavelength coverage and the {\it HST} filter sets observed in our deep
field facilitates some tests on the color selection technique of
$2 \lesssim z \lesssim 6$ candidate galaxies. Known stars are excluded
from all of these plots.

\begin{figure*}
  \begin{center}
    \resizebox{\hsize}{!}
    {\includegraphics[width=0.98\hsize]{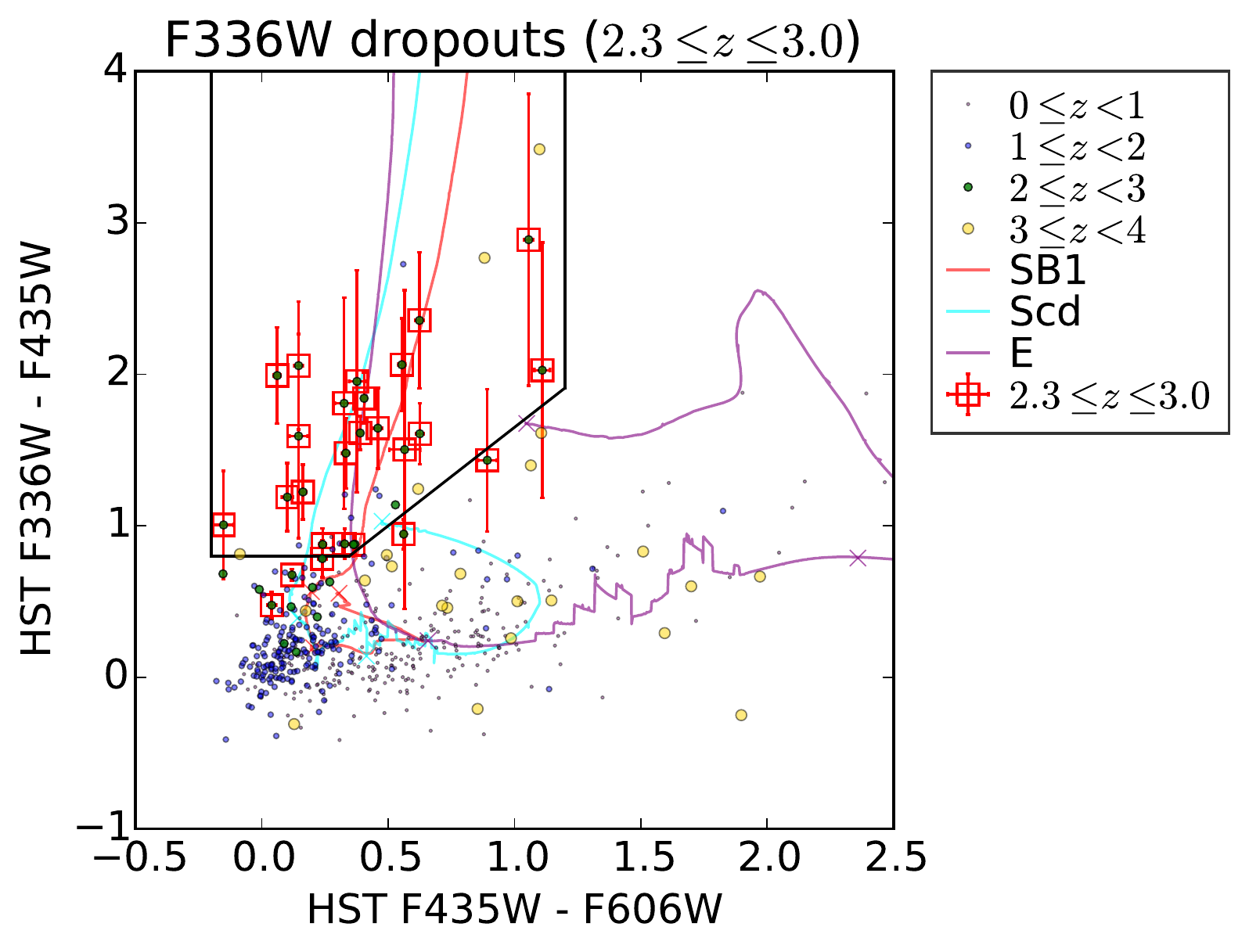}
      \includegraphics[width=0.75\hsize]{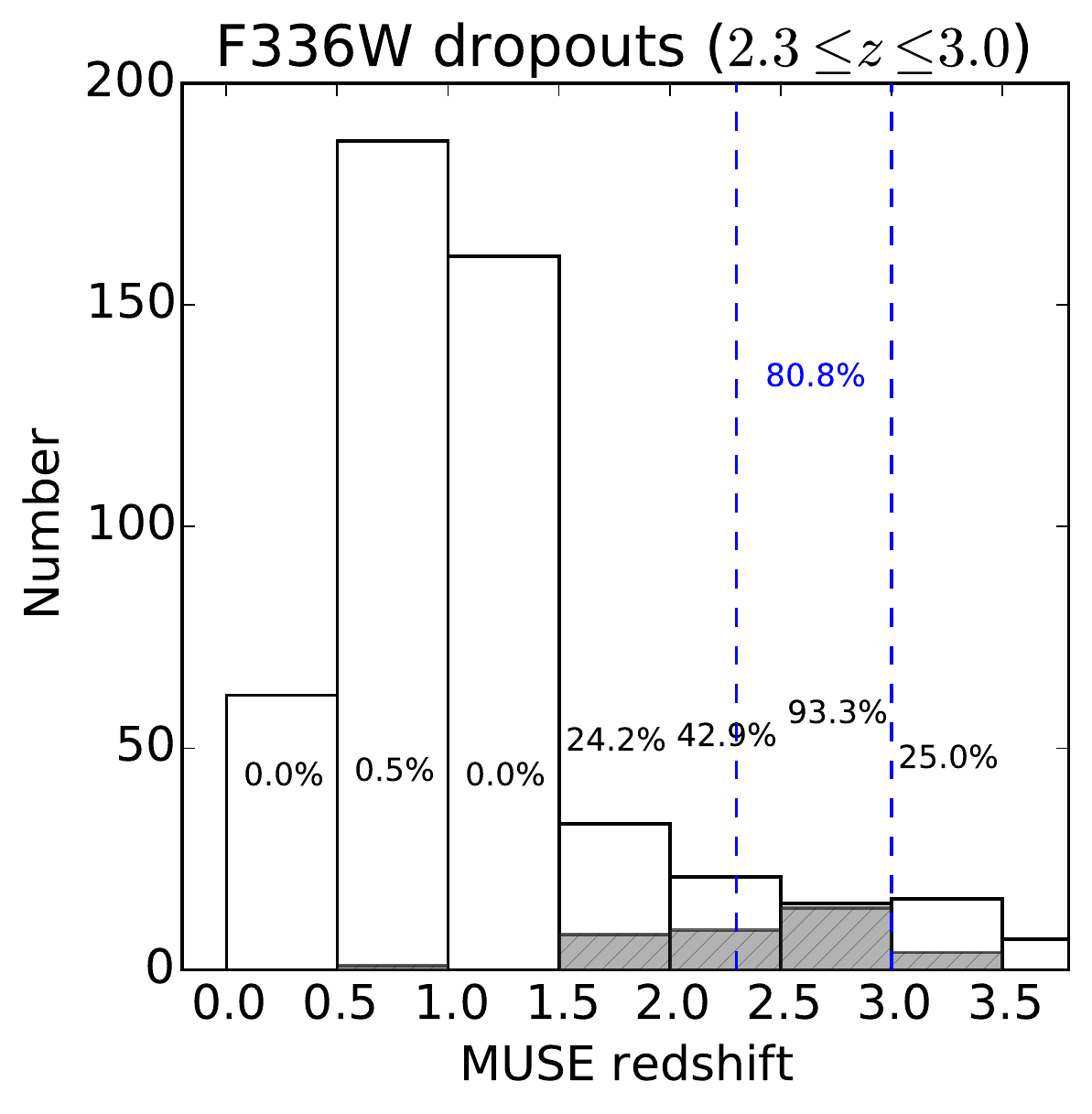}}
    \caption{ {\bf [Left]} Color-color diagram of $F336W - F435W$
      vs.  $F435W - F606W$ to select $z \sim 2.7$ galaxies (F336W
      dropouts). Only the MUSE spectroscopic redshifts with
        $\rm CONFID \geq 2$ are plotted. The galaxies in the targeted
      redshift range ($2.3 \leq z \leq 3.0$) are shown with red
      open squares. The violet, blue, green, gold, orange, magenta,
      and red filled circles, with sizes from small to large,
      indicate galaxies in $0 \leq z < 4$ with steps
      of $\Delta z=1$, respectively.  The model tracks of the
        starburst template \citep[SB1;][]{Kinn96} and the CWW
        templates \citep[Scd and E,][]{Cole80} are overlaid. The
        crosses on the model tracks indicate redshifts from $z=0$ in
        increments of $\Delta z=1$.  {\bf [Right]} The counts of
      galaxies in each redshift bin (only the galaxies detected in all
      of the F336W, F435W, and F606W bands are used here). The empty
      bars are all of the galaxies with MUSE redshifts and the filled
      bars are a subset of the MUSE redshifts lying within the F336W dropout
      selection boundary (the left panel).  The percentages shown
      above each bar are the fractions of the galaxies within the
      color-color boundary for each bin.  The vertical dashed blue
      lines indicate the targeted redshift range,
      $2.3 \leq z \leq 3.0$. The percentage indicated in between these
      lines in blue is the fraction of the galaxies within the
      color-color boundary for the targeted redshift range. }
    \label{fig:F336W_dropout}
  \end{center} 
\end{figure*}

In Figure~\ref{fig:F336W_dropout} (left panel), we overplot all of the
MUSE-$z$ on the F336W dropout selection diagram ($\rm F336W - F435W$
versus  $\rm F435W - F606W$). The bandwidth of F336W favors Lyman break
galaxies (LBGs) at $2.3 \lesssim z \lesssim 3.0$.  Although
\cite{Hath10} and \cite{Tepl13} also show the F225W and F275W dropout
selections, we exclude these from this paper because they
preferentially select $z \sim 1.7$ and $z \sim 2.1$, respectively,
which is a range in which MUSE does not recover a large number of
redshifts.  We follow the selection criteria of \cite{Hath10}, i.e.,
\\

\noindent
F336W dropouts:
\[
  \begin{cases}
    & {\rm F336W - F435W > 0.8} \\
    & {\rm F435W - F606W < 1.2} \\
    & {\rm F435W - F606W > -0.2} \\
    & {\rm F336W - F435W > 0.35 + [1.3 \times (F435W - F606W)]}
  \end{cases}
\]

The original selection conditions also apply a magnitude cut of
${\rm F435W \leq 26.5}$~mag and S/N limits of ${\rm S/N(F435W) > 3}$,
${\rm S/N(F336W) < 3}$, ${\rm S/N(F275W) < 1}$, and
${\rm S/N(F225W) < 1}$ to securely find high-$z$ candidates. We do not
adopt these limits because our purpose here is to demonstrate where
our galaxies with measured redshift lie in the diagrams. However, we
do require solid detections, and thus no limit is shown in the
diagrams.

The fraction of galaxies that meet these selection criteria in each
redshift bin is shown in the right panel of
Figure~\ref{fig:F336W_dropout}. This plot indeed identifies galaxies
at $2.3 \lesssim z \lesssim 3.0$ among the galaxies with secure
redshift ($81\%$). If we limit to $2.5 < z < 3.0$, $93\%$ of the
galaxies in this redshift range are classified as F336W dropouts.
Those missed (the filled green circles in the left panel of
Figure~\ref{fig:F336W_dropout}) lie very close to the selection
boundary where lower-$z$ galaxies start to get intermingled, in
particular around $\rm 0.5 \lesssim F336W - F435W \lesssim 0.8$ and
$\rm -0.2 \lesssim F435W - F606W \lesssim 0.4$.  There is only one
interloper from galaxies at $0 \leq z < 1$ at
$\rm F336W - F435W = 0.93$ and $\rm F435W - F606W = 0.27$, almost on
the borderline.  This object is MUSE ID 2537 ($z=0.717$) whose
redshift is identified by \hd and \oiii$\lambda\lambda4959,5007$.

\begin{figure*}
  \begin{center}
    \includegraphics[width=0.98\hsize]{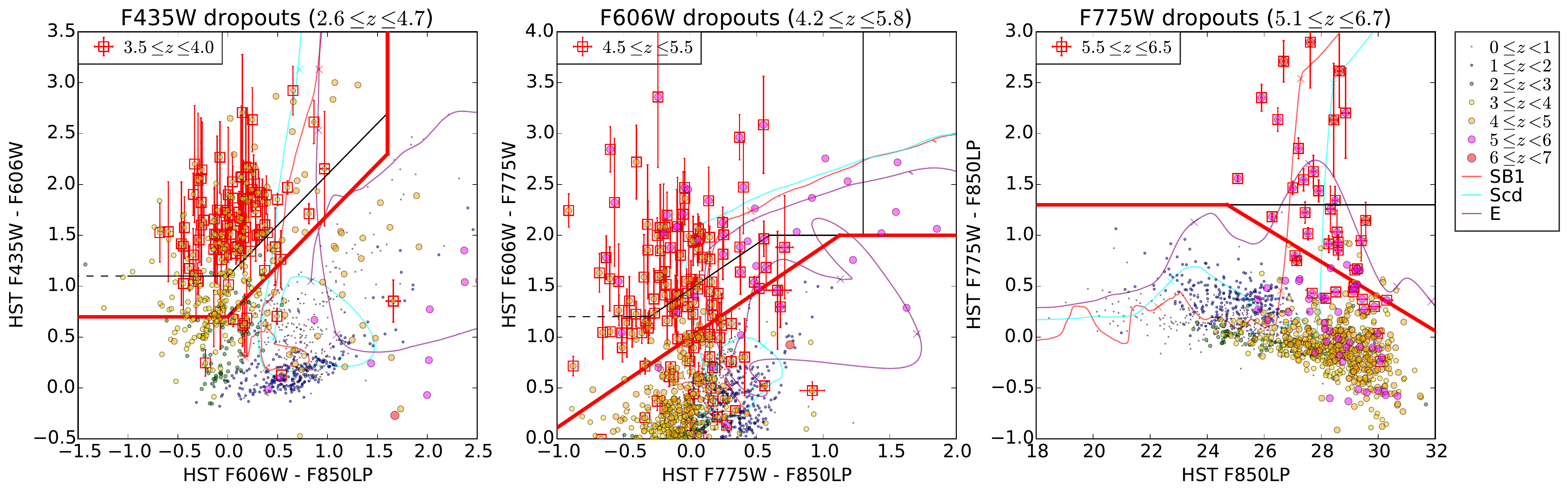}
    \caption{ Similar to the left panel of
      Figure~\ref{fig:F336W_dropout} but selecting $z \sim 3.5$
      (F435W dropouts), $z \sim 5.0$ (F606W dropouts), and
      $z \sim 6.0$ (F775W dropouts), shown in the left, middle, and
      right panels, respectively. For the F435W and
          F606W dropouts, we do not plot the objects at $z > 5.0$
          and $z > 6.0$, respectively. The galaxies
      indicated with the red squares are not in the full targeted
      redshift range, but are restricted to narrower ranges (see also
      Figure~\ref{fig:hiz_dropout_hist}).  The solid and dashed black
      lines are the boundaries of the selection criteria from
      \cite{Star09,Star10}. The dashed black lines are shown because
      they are at the outside of the plotting regions in
      \cite{Star09,Star10}.  The solid red lines are our empirically
      redefined new selection boundaries to gain more galaxies in the
      targeted redshift ranges. The model tracks for the F850LP
        magnitude in the F775W dropout diagram is computed for an
        absolute magnitude of $-22$~mag. The 5$\sigma$ limit of the
        F850LP magnitude is 28.9 mag \citep{Rafe15}.}
  \label{fig:hiz_dropout}
  \end{center} 
\end{figure*}

For even higher redshifts, we utilize the F435W, F606W, and F775W
dropout techniques to inspect $z \sim 3.5$, $z \sim 5.0$, and
$z \sim 6.0$ galaxy selections, respectively \citep{Star09,Star10}, as
shown in Figure~\ref{fig:hiz_dropout}. The selection boundaries are
listed below.  Again, here we do not include any S/N cuts on the
dropout conditions.
\\

\noindent
F435W dropouts:
\[
  \begin{cases}
    & {\rm F435W - F606W > 1.1} \\
    & {\rm F606W - F850LP < 1.6} \\
    & {\rm F435W - F606W > 1.1 + F606W - F850LP}
  \end{cases}
\]

\noindent
F606W dropouts:
\[
  \begin{cases}
    & {\rm F606W - F775W > 1.2} \\
    & {\rm F775W - F850LP < 1.3} \\
    & {\rm F606W - F775W > 1.47 + (F775W - F850LP) ~ or ~ 2.0}
  \end{cases}
\]

\noindent
F775W dropouts:
\[
  \begin{cases}
    & {\rm F775W - F850LP > 1.3}
  \end{cases}
\]

\begin{figure*}
  \begin{center}
    \includegraphics[width=0.98\hsize]{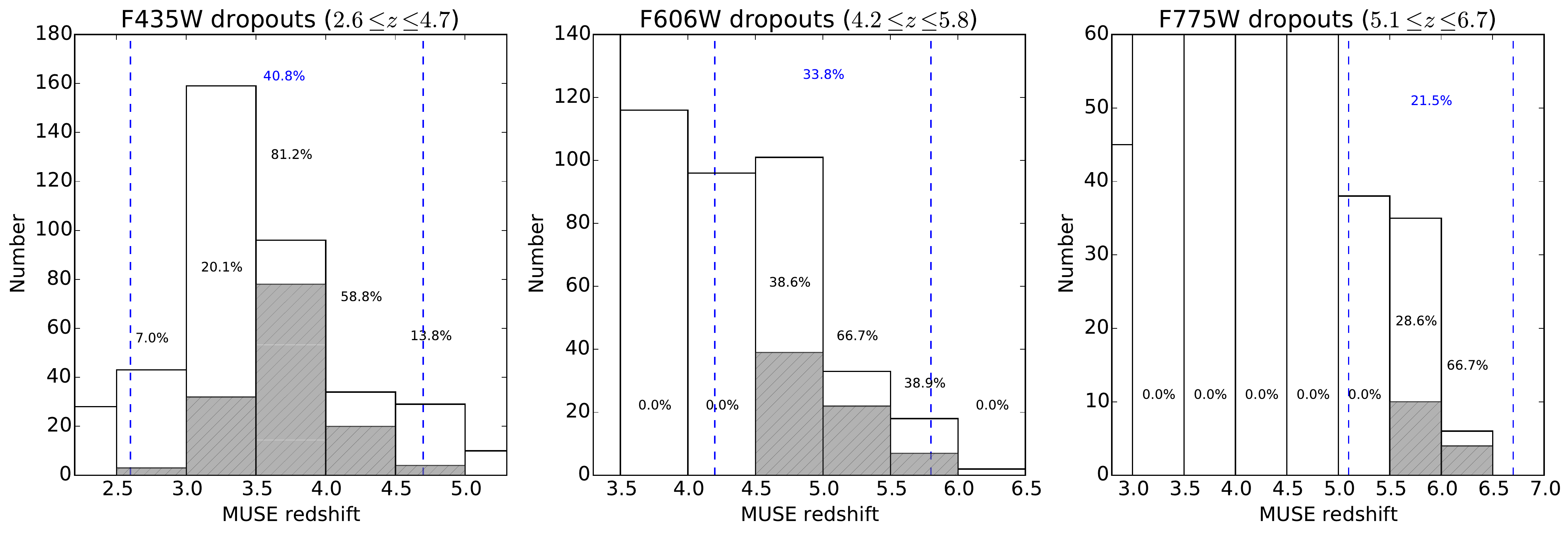}
    \caption{ Similar to the right panel of
      Figure~\ref{fig:F336W_dropout} but for investigating the
      fractions of galaxies that meet the dropout conditions at
      $z \sim 3.5$, $z \sim 5.0$, and $z \sim 6.0$, shown in the
      left, middle, and right panels, respectively.  }
  \label{fig:hiz_dropout_hist}
  \end{center} 
\end{figure*}

In all of these dropout selections, the targeted dropout galaxies are
well captured. No low-$z$ interloper is found in any of these
selections.  The fraction of galaxies that lie within the boundaries
is much smaller compared the F336W selection
(Figure~\ref{fig:hiz_dropout_hist}). This fraction also decreases from
the lower-$z$ to higher-$z$ dropout selections ($41\%$, $34\%$, and
$22\%$ for the F435W, F606, and F775W dropouts, respectively). If we
limit the redshift bin to $3.5 < z < 4.0$, which is the most sensitive
redshift range for the F435W filter to detect the dropouts, then
$81\%$ of galaxies in this redshift bin meet the dropout
criteria. Similarly, for selecting F606W and F775W dropouts, the
fractions are highest ($67\%$ for both) in the redshift bins of
$5.0 < z < 5.5$ and $6.0 < z < 6.5$. Although error bars are not shown
in the diagrams (in order to make them readable), the errors of the
objects outside of the selection boundaries are almost always smaller
than those in the boundaries. Their positions are reliable in the
color-color planes.

From these results, we can empirically adjust the selection boundary
to increase the fraction of the candidate galaxies at the targeted
redshift ranges. Based on visual modifications, we relax the selection
conditions based on our MUSE redshifts as follows:
\\

\noindent
Redefined F435W dropouts from MUSE-$z$:
\[
  \begin{cases}
    & {\rm F435W - F606W > 0.7} \\
    & {\rm F606W - F850LP < 1.6} \\
    & {\rm F435W - F606W > 0.7 + F606W - F850LP} \\
  \end{cases}
\]

\noindent
Redefined F606W dropouts from MUSE-$z$:
\[
  \begin{cases}
    & {\rm F606W - F775W > 1.0 + 0.89 \times (F775W - F850LP) ~ or ~ 2.0} \\
  \end{cases}
\]

\noindent
Redefined F775W dropouts from MUSE-$z$:
\[
  \begin{cases}
    & {\rm F775W - F850LP > -0.17 \times F850LP + 5.5 ~ or ~ 1.3} \\
  \end{cases}
\]

\begin{figure*}
  \begin{center}
    \includegraphics[width=0.98\hsize]{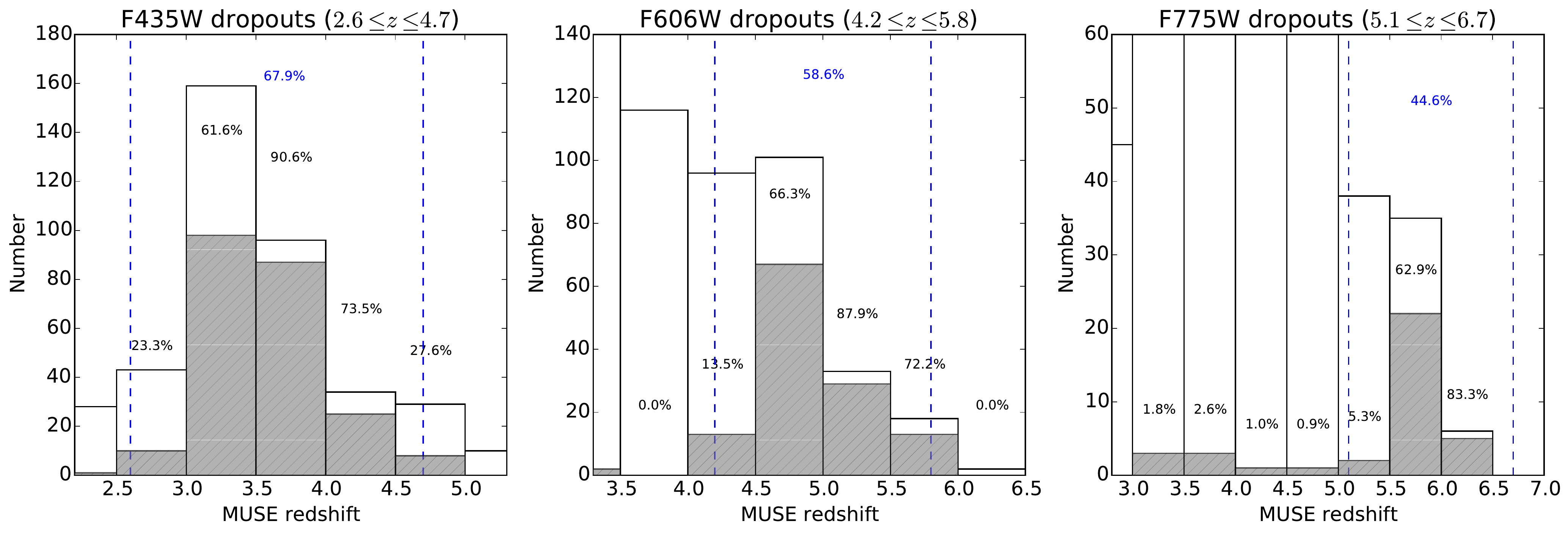}
    \caption{ Same plots as Figure~\ref{fig:hiz_dropout_hist}, but
      for our empirically redefined new selection boundaries shown in
      Figure~\ref{fig:hiz_dropout} with the red lines.}
  \label{fig:hiz_dropout_hist_new}
  \end{center} 
\end{figure*}

These updated boundaries are shown with the solid red lines in
Figure~\ref{fig:hiz_dropout}. The histograms of the selected galaxies
with these new criteria are shown in
Figure~\ref{fig:hiz_dropout_hist_new}. We are trying this experiment,
but not all of our sample galaxies in the diagrams (in particular
those lying between the original and new boundaries) necessarily show
a clear Lyman break because we only use \lya emission to determine
their redshifts.

We allow a slightly bluer $\rm F435W - F606W$ color for selecting
$z \sim 3$ galaxies (F435W dropouts) because there is a crowd of
$3 \leq z < 4$ galaxies lying right below the original boundary. The
new boundary significantly improves the $z \sim 3$ galaxy selection
without picking up contaminants. In particular, now it finds
$\approx 40\%$ more galaxies at $3.0 < z < 3.5$.

For the F606W dropouts, the targeted $z \sim 5$ galaxies also extend
toward bluer $\rm F606W - F775W$ colors. Thus, we lower the color
limit of $\rm F606W - F775W$. In addition, we remove the border at
$\rm F775W - F850LP = 1.3$ because some of galaxies at $5.5 < z < 6.0$
have redder colors. These new conditions successfully increase the
fraction of $z \sim 5$ galaxies lying within the boundary from $34\%$
to $59\%$ and now capture $88\%$ of the galaxies in the
$5.0 < z < 5.5$ bin and $66\%$ in the $4.5 < z < 5.0$ bin.  This also
adds one low-$z$ interloper and two galaxies at
$3.0 < z < 4.0$\,\footnote{The low-$z$ interloper is MUSE ID 873
  ($z=0.66$). The other galaxies are MUSE IDs 493 ($z=3.18$) and 7377
  ($z=3.42$).}.

It is more difficult to enhance the $z \sim 6$ galaxy selection. We
double the number of galaxies at $5 \lesssim z \lesssim 6.5$ to meet
the new conditions, but this also increases the number of galaxies at
lower redshifts. With the newly suggested criteria, we cover galaxies
at $6.0 < z < 6.5$ well ($83\%$) and improve a lot for galaxies
at $5.5 < z < 6.0$ (from $29\%$ to $63\%$). There are more
galaxies below $z=5$ in the new selection box, but only one is below
$z=1$ (MUSE ID 2478 $z=0.73$). The remaining eight lower-$z$ galaxies
have redshifts between $3$ and $5$.  In total, we successfully select
29 galaxies at $5.0 < z < 6.5$. It is challenging to purely select
galaxies at $5.0 < z < 5.5$ with this color-magnitude diagram because
galaxies at $3 < z < 5$ have similar colors and magnitudes.  We only
collect two galaxies ($5.3\%$) in this bin using the new diagram
(0 in the original one).

\begin{figure}
  \begin{center}
    \includegraphics[width=0.98\hsize]{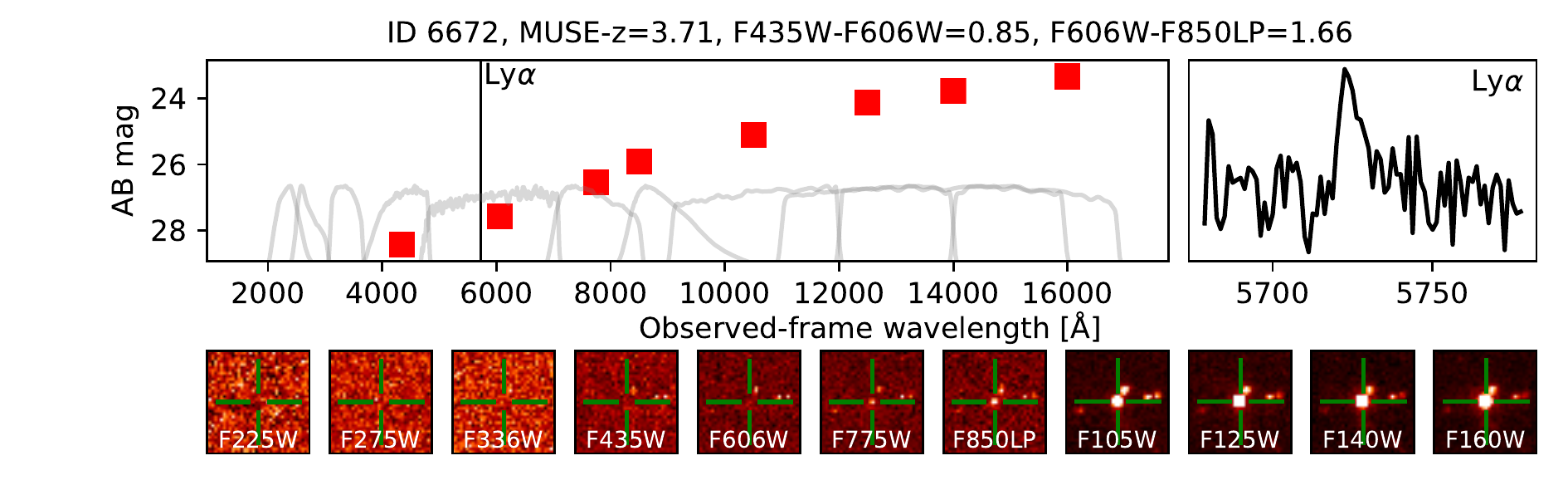}
    \includegraphics[width=0.98\hsize]{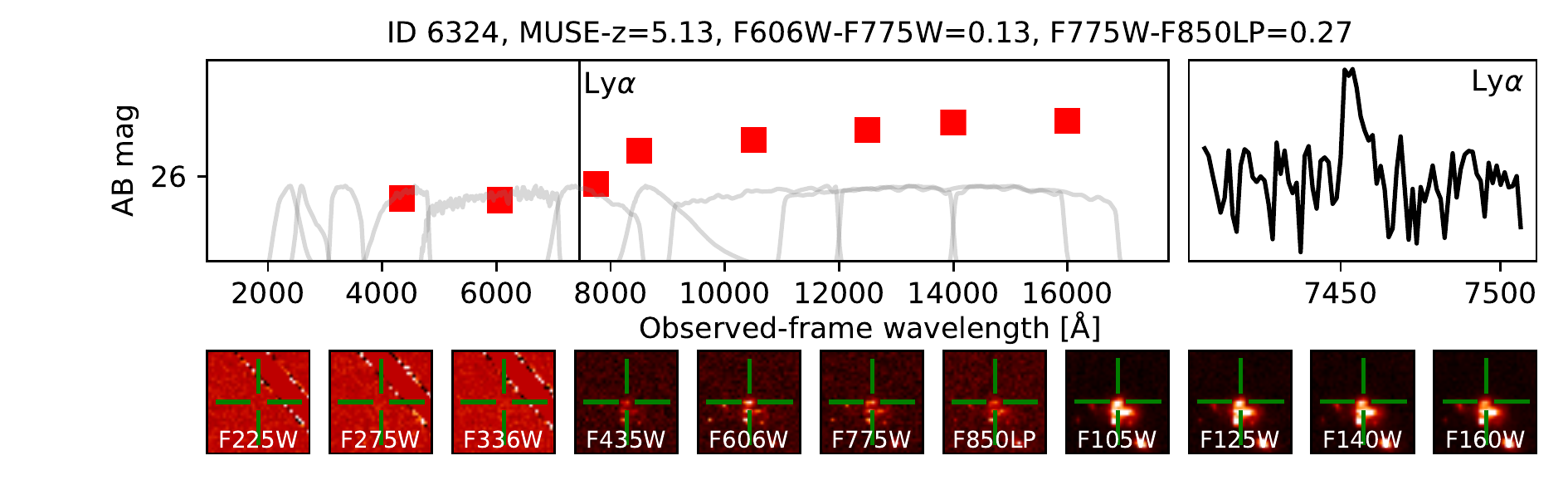}
    \includegraphics[width=0.98\hsize]{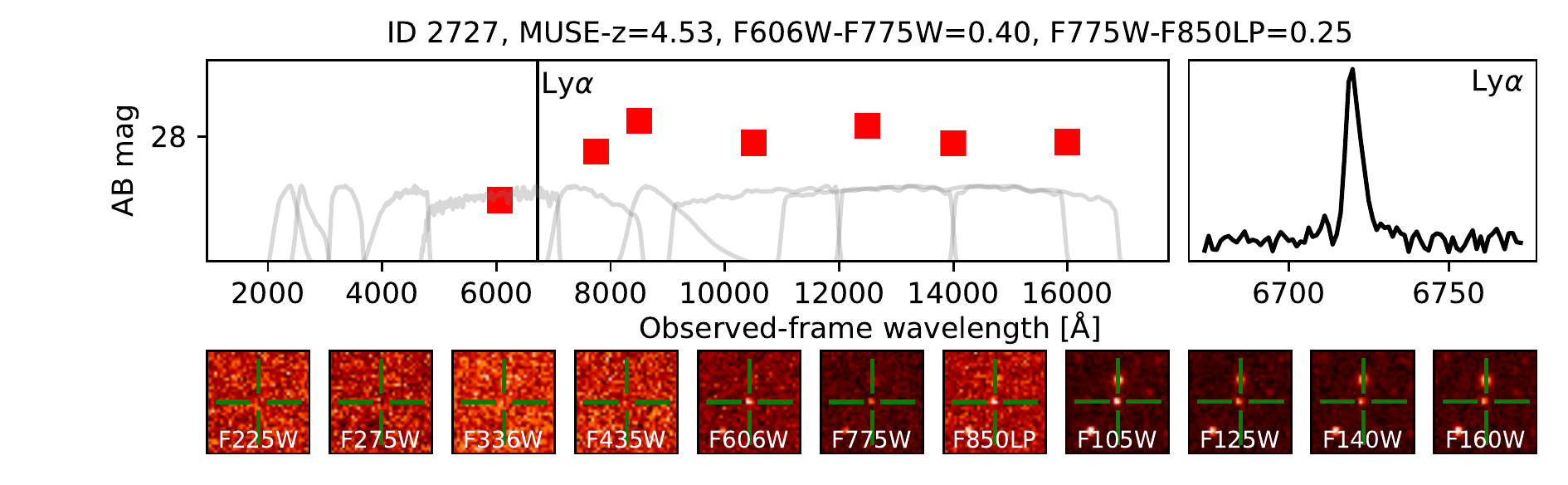}
    \includegraphics[width=0.98\hsize]{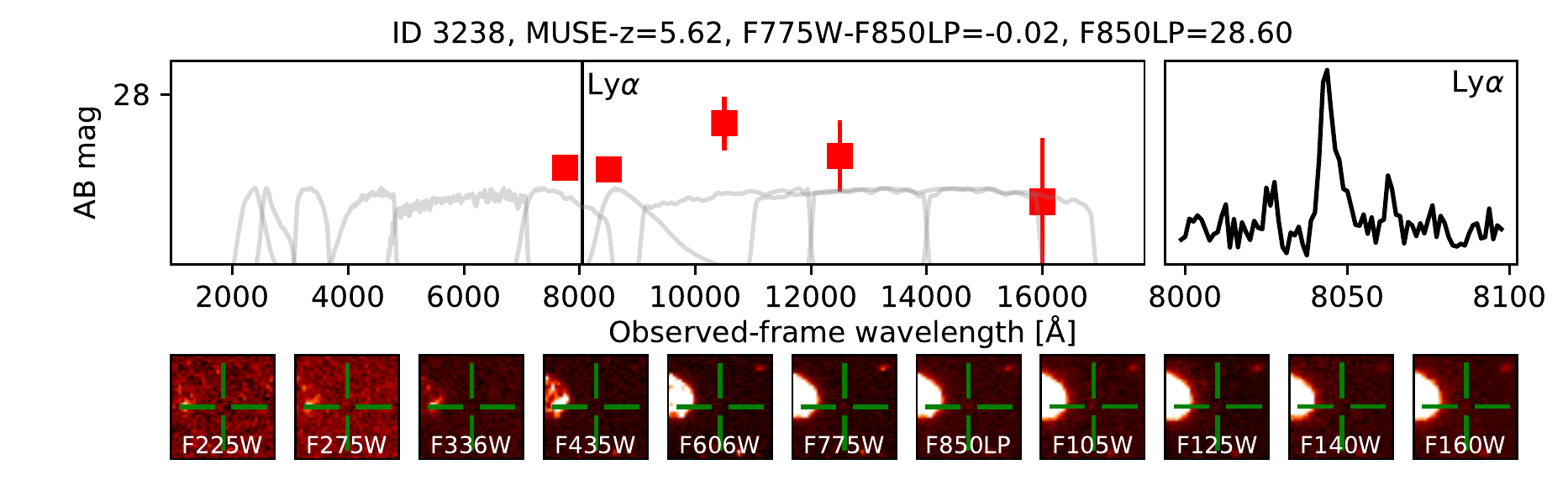}
    \caption{ Examples of the outliers in the F435W
          (top panel), F606W (two middle panels), and F775W (bottom
          panel) dropout selections. In each panel, a spectral energy
          distribution using the {\it HST} photometries is shown at
          the top left, a zoomed-in part of the \lya detected
          wavelength region is presented at the top right, and
          thumbnail {\it HST} images of the object are shown at
          the bottom. The object ID, MUSE spec-$z$, and the colors of
          each galaxy can be found at the top of each panel. }
  \label{fig:dropout_outliers}
  \end{center} 
\end{figure}

There are still significant numbers of high-$z$ galaxies lying outside
of the selection boxes. We have checked these galaxies individually to
determine whether their bluer colors are due to blending or
photometric issues. In the F435W dropout diagram, we find 10 objects
with $3.5 \leq z \leq 4.0$ for $\rm F435W-F606W < 0.9$.  Among these
objects, one may be blended with a nearby object, one is a dusty
galaxy (ID 6672, see Figure~\ref{fig:dropout_outliers}) and one may
have an ambiguous \lya feature. In the remaining seven objects, two
have good photometries lying near the revised selection box, and five
have $S/N {\rm (F435W)} < 5$ but the error budgets of their
photometries make them agree with the revised selection box. For the
F606W dropout diagram, we checked 12 galaxies with
$4.5 \leq z \leq 5.5$ for $\rm 0 \lesssim F606W-F775W \lesssim 0.4$.
There is one (ID 6324, Figure~\ref{fig:dropout_outliers}) whose
photometries may be contaminated by a neighboring object, but the rest
of the objects have good photometries (seven with
$S/N {\rm (F606W)} \gtrsim 5$ and four with $S/N {\rm (F606W)} > 10$;
e.g., ID 2727, Figure~\ref{fig:dropout_outliers}). However, we find
that all of these objects have the \lya emission line lying in the
F606W coverage, which is likely to produce bluer $\rm F606W-F775W$
colors. This effect is visible in the diagrams in
Figures~\ref{fig:hiz_dropout_hist_new} and
\ref{fig:hiz_dropout_hist}. Allowing the boundary of $\rm F606W-F775W$
to be bluer increases more lower-z selections in the targeted redshift
range of $4.2 < z < 5.8$. In fact, this trend is also seen in the 13
outlier galaxies with $5.5 \leq z \leq 6.5$ for
$\rm 0 \lesssim F775W-F850LP \lesssim 0.5$ in the F775W dropout
diagram (e.g., ID 3238, Figure~\ref{fig:dropout_outliers}). All of
these galaxies have $S/N {\rm (F775W)} \gtrsim 5$ and the detected
\lya line is lying in the F775W coverage; three of these galaxies are
starting to enter the F850LP coverage, but at the wavelength where the
F850LP filter throughput is still $< 50\%$. Similar to the outliers in
the F606W dropout diagram, the enhanced blue $\rm F775W-F850LP$ colors
of these objects can be ascribed to \lya emission contributing to the
F775W flux.  The newly defined selection boundaries help to increase
the completeness of $3 \lesssim z \lesssim 7$ galaxy selections. It is
not surprising that we find outliers as color selection criteria are
not designed for comprehensive detection.  Although rest-frame UV
spectra of high-$z$ galaxies are affected by IGM transmission, the
high equivalent width of \lya emission also causes bluer colors, which
makes the apparent \lya break less significant (particularly in the
F606W and F775W dropout selections).  In addition, color selections
are known to exclude some high-$z$ passive and dust-obscured galaxies
\citep[e.g.,][]{Dadd04, Redd05, vanD06}.  The outliers in the dropout
diagrams are also found in the VVDS survey \citep{LeFe05c, Palt07}.

Here we limit to the comparisons of measured redshifts against the
dropout selections. The galaxy populations plotted here are likely to
be different from the traditional dropout selected galaxies because
the majority of the redshifts are determined with emission lines.
However, candidate galaxies lying between the original and new
selection conditions may be able to serve as second highest priority
targets for multi-object spectroscopy to increase the efficiency in
finding high-$z$ galaxies.  Also, we do not adopt any S/N or magnitude
cuts here.  For future work, we will discuss in more detail why some
galaxies fail to meet the dropout conditions (with respect to their
spectroscopic and photometric redshifts), the relationship between
\lya emitters and dropout galaxies and their physical properties, and
the evolution of the fraction of \lya emitters.

\section{Summary and conclusions}\label{sec:summary}

We have conducted a deep spectroscopic survey in the HUDF with
MUSE. The $3\arcmin \times 3\arcmin$ deep survey ($\approx 10$~hour
depth, the \mosaic) region almost covers the entire HUDF with the
$1\arcmin \times 1\arcmin$ ultra deep survey ($\approx 30$~hour depth,
\udft) region enclosed.  We used two different spectral extraction
methods for the redshift identification: the {\it HST} prior continuum
selected objects \citep[based on the UVUDF catalog;][]{Rafe15} and the
emission line objects selected directly in the cubes without prior
information.  We visually inspected these spectra to determine
redshifts via redshift analysis software.  We also measured the line
fluxes for the objects with redshifts determined. Along with this
paper, we release the redshift and line flux catalogs.  Here we
summarize our findings in the redshift assessments:

\begin{enumerate}

\item In \udft, we obtained $\UDFzALLCtwo$ secure redshifts
  ($\rm CONFID \geq 2$) in the redshift range $0.21 \leq z \leq 6.64$.
  Among these, $\UDFzORGonlyCtwo$ are not in the {\it HST} UVUDF
  catalog that we used for the prior based source extraction. The
  majority of these redshifts (28) are \lya emitters. When we only count {\it
    HST} prior extracted objects, we managed to retrieve $\UDFzHSTCtwo$
  redshifts out of $\UDFHSTpri$ continuum selected objects
  ($26\%$). We reach a $50\%$ completeness at $26.5$~mag (F775W), and
  the completeness stays around $20\%$ up to $28-29$~mag.

\item In the \mosaic, in addition to investigating all of the emission
  line galaxies, we performed visual inspections for those with
  $\rm F775W \leq 27$~mag for the continuum selected galaxies. This
  arbitrary cut is made based on our findings for \udft that
  $\approx 90\%$ of galaxies at $z < 3$ are brighter than this
  magnitude. At $z > 3$, although most of the galaxies are fainter
  than $27$~mag, we are able to detect $70\%$ of these galaxies with
  emission features directly in the data cube without any continuum
  information.  Together with the direct search for emission line
  objects in the cube, this helps to increase the efficiency of
  redshift determination.

\item Out of the $\MOSHSTpriMAGcut$ continuum selected objects with
  $\rm F775W \leq 27$~mag in the \mosaic field, we determined secure
  redshift of $\MOSzHSTCtwo$ objects. When we included the emission
  line detected objects (regardless of magnitude), we obtained
  $\MOSzALLCtwo$ unique redshifts with confidence. The $50\%$
  completeness with respect to F775W mag is at $25.5$~mag.

\item In the overlapping region of \udft and \mosaic, all of the
  redshifts measured in  \mosaic were also measured in the \udft
  with at least the same confidence level or higher, except one object
  lying at the edge of the \udft field and four objects detected
  only through emission lines in the \mosaic cube.  On the other hand,
  78 secure redshifts were missing in \mosaic when compared with
  \udft. There are a few discrepancies but the measurements in the
  \udft, which has higher S/N spectra, are usually more convincing.

\item Among $\COMBHSTpri$ unique continuum selected galaxies in the
  entire MUSE UDF survey region (\udft $+$ the \mosaic), we recovered
  redshifts for $\COMBzHSTCtwo$ (15\%) objects. We also found
  $\COMBzORGonlyCtwo$ objects only via emission line searches
  directly in the data cubes (no counterparts in the {\it HST} UVUDF
  catalog). Thus, in total, we obtained $\COMBzALLCtwo$ unique
  redshifts with high confidence.

\item Compared with the {\it HST} beamsize, the MUSE data suffer from
  source confusion. However, we were able to partly solve this issue
  when an emission/absorption feature is identified. The location of
  the detected feature in its narrowband image can be associated with
  the location of the corresponding source in the {\it HST}
  images. There are also a few cases where two (or more) galaxies are
  completely blended even with the {\it HST} resolution, but we
  identified both (or all) of these cases with two (or more) different
  sets of emission lines at different redshifts.  Finally, among
  $\COMBzALLCone$ objects with MUSE-$z$ ($\rm CONFID \geq 1$), there
  are 79 that remain as merged objects.

\item Of the previously known $\COMBprevz$ spectroscopic redshifts in
  our survey field, we recovered all except 10 objects. Of these, four
  were measured in spectra taken with ground-based
  spectroscopy. However, for one of these redshifts, the detected
  features are at wavelengths shorter than the blue end of MUSE, and
  for the other three, we cannot confirm the previously measured
  redshifts based on the expected detectable features. The other six
  spectra were taken with {\it HST} grism spectroscopy. The features
  used for determining their redshifts either were a very broad
  absorption feature or lay on the red side of MUSE spectra where sky
  emission dominates.

\item The comparison of MUSE-$z$ and photometric redshifts revealed
  that they agree well up to $z \approx 3$. At higher redshift, an
  offset appears. Although most of the $z > 3$ redshifts are
  determined with \lya, the median offset between MUSE-$z$ and
  photo-$z$ is much larger than the expected \lya velocity offset to
  the systematic redshift due to gas motion. We observed a trend that
  the photo-$z$ offset increases with fainter continuum emission in
  {\it HST}/F775W.

\item We investigated all of the galaxies with secure MUSE-$z$ in some
  common color selection (dropout) diagrams of high-$z$ galaxies. With
  the F336W dropout selection criteria, $81\%$ of targeted
  $z \sim 2.7$ galaxies are captured. This fraction decreases for
  higher-$z$ selections: $41\%$ for F435W dropouts ($z \sim 3$),
  $34\%$ for F606 dropouts ($z \sim 5$), and $22\%$ for F775W dropouts
  ($z \sim 6$). We empirically redefined the selection boundaries to
  increase the fractions to $68\%$, $59\%$, and $45\%$, but our galaxy
  populations are likely to be different from the traditional dropout
  selected galaxies (continuum selected) because the majority of the
  redshifts were determined using emission lines.

\end{enumerate}

With deep MUSE spectroscopic observations in HUDF, we dramatically
improve the redshift completeness.  The improvements are not only the
increase in number from $\COMBprevz$ to $\COMBzALLCtwo$, but also the
coverage of the redshift range well beyond $z > 3$ and depths up to
the 30th magnitude (F775W).  Together with existing large telescopes
and planned future observatories such as {\it JWST} and 30 m class
telescopes, it opens new horizons for exploring the early universe.

In the near future, we plan further advances in the analysis, such as
investigating spectra extracted at the positions of continuum selected
objects fainter than $27$~mag in the \mosaic, fine tuning {\tt ORIGIN}
(and {\tt MUSELET}) to further improve the direct detection of
emission lines without enhancing spurious sources, upgrading {\tt
  MARZ} to increase the accuracy of automated redshift determinations
with newly built spectral templates, and developing deblending
techniques based on prior information.

\begin{acknowledgements} 

  The authors are grateful for useful suggestions by the referee that
  improved the manuscript.  We thank Laure Piqueras for her support
  with analysis and observation software.  This work is supported by
  the ERC advanced grant 339659-MUSICOS (R. Bacon).  JB acknowledges
  support by Funda\c{c}\~ao para a Ci\^encia e a Tecnologia (FCT)
  through national funds (UID/FIS/04434/2013) and by FEDER through
  COMPETE2020 (POCI-01-0145-FEDER-007672) and was in part supported by
  FCT through Investigador FCT contract IF/01654/2014/CP1215/CT0003.
  JR and BC acknowledge support from the ERC starting grant
  336736-CALENDS. TC acknowledges support of the ANR FOGHAR
  (ANR-13-BS05-0010-02), the OCEVU Labex (ANR-11-LABX-0060), and the
  A*MIDEX project (ANR-11-IDEX-0001-02) funded by the
  ``Investissements d'avenir'' French government program managed by
  the ANR. JS acknowledges support of the ERC Grant agreement
  278594-GasAroundGalaxies. LW acknowledges funding by the Competitive
  Fund of the Leibniz Association through grant SAW-2015-AIP-2. RAM
  acknowledges support by the Swiss National Science Foundation.

\end{acknowledgements}

\bibliography{bib}

\begin{thebibliography}{73}
\expandafter\ifx\csname natexlab\endcsname\relax\def\natexlab#1{#1}\fi

\bibitem[{{Akhlaghi} \& {Ichikawa}(2015)}]{Akhl15}
{Akhlaghi}, M. \& {Ichikawa}, T. 2015, \apjs, 220, 1

\bibitem[{{Bacon} {et~al.}(2010){Bacon}, {Accardo}, {Adjali}, {Anwand},
  {Bauer}, {Biswas}, {Blaizot}, {Boudon}, {Brau-Nogue}, {Brinchmann},
  {Caillier}, {Capoani}, {Carollo}, {Contini}, {Couderc}, {Daguis{\'e}},
  {Deiries}, {Delabre}, {Dreizler}, {Dubois}, {Dupieux}, {Dupuy}, {Emsellem},
  {Fechner}, {Fleischmann}, {Fran{\c c}ois}, {Gallou}, {Gharsa}, {Glindemann},
  {Gojak}, {Guiderdoni}, {Hansali}, {Hahn}, {Jarno}, {Kelz}, {Koehler},
  {Kosmalski}, {Laurent}, {Le Floch}, {Lilly}, {Lizon}, {Loupias}, {Manescau},
  {Monstein}, {Nicklas}, {Olaya}, {Pares}, {Pasquini}, {P{\'e}contal-Rousset},
  {Pell{\'o}}, {Petit}, {Popow}, {Reiss}, {Remillieux}, {Renault}, {Roth},
  {Rupprecht}, {Serre}, {Schaye}, {Soucail}, {Steinmetz}, {Streicher}, {Stuik},
  {Valentin}, {Vernet}, {Weilbacher}, {Wisotzki}, \& {Yerle}}]{Baco10}
{Bacon}, R., {Accardo}, M., {Adjali}, L., {et~al.} 2010, in \procspie, Vol.
  7735, Ground-based and Airborne Instrumentation for Astronomy III, 773508

\bibitem[{{Bacon} {et~al.}(2015){Bacon}, {Brinchmann}, {Richard}, {Contini},
  {Drake}, {Franx}, {Tacchella}, {Vernet}, {Wisotzki}, {Blaizot}, {Bouch{\'e}},
  {Bouwens}, {Cantalupo}, {Carollo}, {Carton}, {Caruana}, {Cl{\'e}ment},
  {Dreizler}, {Epinat}, {Guiderdoni}, {Herenz}, {Husser}, {Kamann}, {Kerutt},
  {Kollatschny}, {Krajnovic}, {Lilly}, {Martinsson}, {Michel-Dansac},
  {Patricio}, {Schaye}, {Shirazi}, {Soto}, {Soucail}, {Steinmetz}, {Urrutia},
  {Weilbacher}, \& {de Zeeuw}}]{Baco15}
{Bacon}, R., {Brinchmann}, J., {Richard}, J., {et~al.} 2015, \aap, 575, A75

\bibitem[{{Bacon} {et~al.}(2017){Bacon}, {Conseil}, {Mary},
  {et~al.}}]{subm_Baco17}
{Bacon}, R., {Conseil}, D., {Mary}, D., {et~al.} 2017, \aap, in press (MUSE UDF
  SI paper I)

\bibitem[{{Baldry} {et~al.}(2014){Baldry}, {Alpaslan}, {Bauer},
  {Bland-Hawthorn}, {Brough}, {Cluver}, {Croom}, {Davies}, {Driver},
  {Gunawardhana}, {Holwerda}, {Hopkins}, {Kelvin}, {Liske},
  {L{\'o}pez-S{\'a}nchez}, {Loveday}, {Norberg}, {Peacock}, {Robotham}, \&
  {Taylor}}]{Bald14}
{Baldry}, I.~K., {Alpaslan}, M., {Bauer}, A.~E., {et~al.} 2014, \mnras, 441,
  2440

\bibitem[{{Balestra} {et~al.}(2010){Balestra}, {Mainieri}, {Popesso},
  {Dickinson}, {Nonino}, {Rosati}, {Teimoorinia}, {Vanzella}, {Cristiani},
  {Cesarsky}, {Fosbury}, {Kuntschner}, \& {Rettura}}]{Bale10}
{Balestra}, I., {Mainieri}, V., {Popesso}, P., {et~al.} 2010, \aap, 512, A12

\bibitem[{{Beck} {et~al.}(2017){Beck}, {Lin}, {Ishida}, {Gieseke}, {de Souza},
  {Costa-Duarte}, {Hattab}, {Krone-Martins}, \& {for the COIN
  Collaboration}}]{Beck17}
{Beck}, R., {Lin}, C.-A., {Ishida}, E.~E.~O., {et~al.} 2017, ArXiv e-prints
  [\eprint[arXiv]{1701.08748}]

\bibitem[{{Beckwith} {et~al.}(2006){Beckwith}, {Stiavelli}, {Koekemoer},
  {Caldwell}, {Ferguson}, {Hook}, {Lucas}, {Bergeron}, {Corbin}, {Jogee},
  {Panagia}, {Robberto}, {Royle}, {Somerville}, \& {Sosey}}]{Beck06}
{Beckwith}, S.~V.~W., {Stiavelli}, M., {Koekemoer}, A.~M., {et~al.} 2006, \aj,
  132, 1729

\bibitem[{{Ben{\'{\i}}tez}(2000)}]{Beni00}
{Ben{\'{\i}}tez}, N. 2000, \apj, 536, 571

\bibitem[{{Bertin} \& {Arnouts}(1996)}]{Bert96}
{Bertin}, E. \& {Arnouts}, S. 1996, \aaps, 117, 393

\bibitem[{{Bonnett} {et~al.}(2016){Bonnett}, {Troxel}, {Hartley}, {Amara},
  {Leistedt}, {Becker}, {Bernstein}, {Bridle}, {Bruderer}, {Busha}, {Carrasco
  Kind}, {Childress}, {Castander}, {Chang}, {Crocce}, {Davis}, {Eifler},
  {Frieman}, {Gangkofner}, {Gaztanaga}, {Glazebrook}, {Gruen}, {Kacprzak},
  {King}, {Kwan}, {Lahav}, {Lewis}, {Lidman}, {Lin}, {MacCrann}, {Miquel},
  {O'Neill}, {Palmese}, {Peiris}, {Refregier}, {Rozo}, {Rykoff}, {Sadeh},
  {S{\'a}nchez}, {Sheldon}, {Uddin}, {Wechsler}, {Zuntz}, {Abbott}, {Abdalla},
  {Allam}, {Armstrong}, {Banerji}, {Bauer}, {Benoit-L{\'e}vy}, {Bertin},
  {Brooks}, {Buckley-Geer}, {Burke}, {Capozzi}, {Carnero Rosell}, {Carretero},
  {Cunha}, {D'Andrea}, {da Costa}, {DePoy}, {Desai}, {Diehl}, {Dietrich},
  {Doel}, {Fausti Neto}, {Fernandez}, {Flaugher}, {Fosalba}, {Gerdes},
  {Gruendl}, {Honscheid}, {Jain}, {James}, {Jarvis}, {Kim}, {Kuehn},
  {Kuropatkin}, {Li}, {Lima}, {Maia}, {March}, {Marshall}, {Martini},
  {Melchior}, {Miller}, {Neilsen}, {Nichol}, {Nord}, {Ogando}, {Plazas},
  {Reil}, {Romer}, {Roodman}, {Sako}, {Sanchez}, {Santiago}, {Smith},
  {Soares-Santos}, {Sobreira}, {Suchyta}, {Swanson}, {Tarle}, {Thaler},
  {Thomas}, {Vikram}, {Walker}, \& {Dark Energy Survey Collaboration}}]{Bonn16}
{Bonnett}, C., {Troxel}, M.~A., {Hartley}, W., {et~al.} 2016, \prd, 94, 042005

\bibitem[{{Brammer} {et~al.}(2008){Brammer}, {van Dokkum}, \& {Coppi}}]{Bram08}
{Brammer}, G.~B., {van Dokkum}, P.~G., \& {Coppi}, P. 2008, \apj, 686, 1503

\bibitem[{{Brammer} {et~al.}(2012){Brammer}, {van Dokkum}, {Franx},
  {Fumagalli}, {Patel}, {Rix}, {Skelton}, {Kriek}, {Nelson}, {Schmidt},
  {Bezanson}, {da Cunha}, {Erb}, {Fan}, {F{\"o}rster Schreiber}, {Illingworth},
  {Labb{\'e}}, {Leja}, {Lundgren}, {Magee}, {Marchesini}, {McCarthy},
  {Momcheva}, {Muzzin}, {Quadri}, {Steidel}, {Tal}, {Wake}, {Whitaker}, \&
  {Williams}}]{Bram12}
{Brammer}, G.~B., {van Dokkum}, P.~G., {Franx}, M., {et~al.} 2012, \apjs, 200,
  13

\bibitem[{{Brinchmann} {et~al.}(2017){Brinchmann}, {Inami}, {Bacon},
  {et~al.}}]{subm_Brin17}
{Brinchmann}, J., {Inami}, H., {Bacon}, R., {et~al.} 2017, \aap, submitted
  (MUSE UDF SI paper III)

\bibitem[{{Brinchmann} {et~al.}(2008){Brinchmann}, {Kunth}, \&
  {Durret}}]{Brin08b}
{Brinchmann}, J., {Kunth}, D., \& {Durret}, F. 2008, \aap, 485, 657

\bibitem[{{Bruzual} \& {Charlot}(2003)}]{BC03}
{Bruzual}, G. \& {Charlot}, S. 2003, \mnras, 344, 1000

\bibitem[{{Charlot} \& {Fall}(2000)}]{Char00}
{Charlot}, S. \& {Fall}, S.~M. 2000, \apj, 539, 718

\bibitem[{{Chevallard} \& {Charlot}(2016)}]{Chev16}
{Chevallard}, J. \& {Charlot}, S. 2016, \mnras, 462, 1415

\bibitem[{{Coe} {et~al.}(2006){Coe}, {Ben{\'{\i}}tez}, {S{\'a}nchez}, {Jee},
  {Bouwens}, \& {Ford}}]{Coe06}
{Coe}, D., {Ben{\'{\i}}tez}, N., {S{\'a}nchez}, S.~F., {et~al.} 2006, \aj, 132,
  926

\bibitem[{{Coleman} {et~al.}(1980){Coleman}, {Wu}, \& {Weedman}}]{Cole80}
{Coleman}, G.~D., {Wu}, C.-C., \& {Weedman}, D.~W. 1980, \apjs, 43, 393

\bibitem[{{Daddi} {et~al.}(2004){Daddi}, {Cimatti}, {Renzini}, {Fontana},
  {Mignoli}, {Pozzetti}, {Tozzi}, \& {Zamorani}}]{Dadd04}
{Daddi}, E., {Cimatti}, A., {Renzini}, A., {et~al.} 2004, \apj, 617, 746

\bibitem[{{Daddi} {et~al.}(2005){Daddi}, {Renzini}, {Pirzkal}, {Cimatti},
  {Malhotra}, {Stiavelli}, {Xu}, {Pasquali}, {Rhoads}, {Brusa}, {di Serego
  Alighieri}, {Ferguson}, {Koekemoer}, {Moustakas}, {Panagia}, \&
  {Windhorst}}]{Dadd05}
{Daddi}, E., {Renzini}, A., {Pirzkal}, N., {et~al.} 2005, \apj, 626, 680

\bibitem[{{Drake} {et~al.}(2017){Drake}, {Garel}, {Hashimoto},
  {et~al.}}]{subm_Drak17}
{Drake}, A., {Garel}, T., {Hashimoto}, T., {et~al.} 2017, \aap, submitted (MUSE
  UDF SI paper VI)

\bibitem[{{Finley} {et~al.}(2017){Finley}, {Bouch\'e}, {Contini},
  {et~al.}}]{subm_Finl17}
{Finley}, H., {Bouch\'e}, N., {Contini}, T., {et~al.} 2017, \aap, submitted
  (MUSE UDF SI paper VII)

\bibitem[{{Gallego} {et~al.}(2017){Gallego}, {Cantalupo}, S.,
  {et~al.}}]{subm_Gall17}
{Gallego}, S., {Cantalupo}, S., S., L., {et~al.} 2017, \mnras, submitted

\bibitem[{{Gu\'erou} {et~al.}(2017){Gu\'erou}, A., {Epinat},
  {et~al.}}]{subm_Guer17}
{Gu\'erou}, A., K.~D., {Epinat}, B., {et~al.} 2017, \aap, in press (paper MUSE
  UDF SI paper V)

\bibitem[{{Guo} {et~al.}(2013){Guo}, {Ferguson}, {Giavalisco}, {Barro},
  {Willner}, {Ashby}, {Dahlen}, {Donley}, {Faber}, {Fontana}, {Galametz},
  {Grazian}, {Huang}, {Kocevski}, {Koekemoer}, {Koo}, {McGrath}, {Peth},
  {Salvato}, {Wuyts}, {Castellano}, {Cooray}, {Dickinson}, {Dunlop}, {Fazio},
  {Gardner}, {Gawiser}, {Grogin}, {Hathi}, {Hsu}, {Lee}, {Lucas}, {Mobasher},
  {Nandra}, {Newman}, \& {van der Wel}}]{Guo13}
{Guo}, Y., {Ferguson}, H.~C., {Giavalisco}, M., {et~al.} 2013, \apjs, 207, 24

\bibitem[{{Hashimoto} {et~al.}(2017){Hashimoto}, {Garel}, {Guiderdoni},
  {et~al.}}]{subm_Hash17}
{Hashimoto}, T., {Garel}, T., {Guiderdoni}, B., {et~al.} 2017, \aap, submitted
  (MUSE UDF SI paper X)

\bibitem[{{Hashimoto} {et~al.}(2013){Hashimoto}, {Ouchi}, {Shimasaku}, {Ono},
  {Nakajima}, {Rauch}, {Lee}, \& {Okamura}}]{Hash13}
{Hashimoto}, T., {Ouchi}, M., {Shimasaku}, K., {et~al.} 2013, \apj, 765, 70

\bibitem[{{Hathi} {et~al.}(2010){Hathi}, {Ryan}, {Cohen}, {Yan}, {Windhorst},
  {McCarthy}, {O'Connell}, {Koekemoer}, {Rutkowski}, {Balick}, {Bond},
  {Calzetti}, {Disney}, {Dopita}, {Frogel}, {Hall}, {Holtzman}, {Kimble},
  {Paresce}, {Saha}, {Silk}, {Trauger}, {Walker}, {Whitmore}, \&
  {Young}}]{Hath10}
{Hathi}, N.~P., {Ryan}, Jr., R.~E., {Cohen}, S.~H., {et~al.} 2010, \apj, 720,
  1708

\bibitem[{{Hinton} {et~al.}(2016){Hinton}, {Davis}, {Lidman}, {Glazebrook}, \&
  {Lewis}}]{Hint16}
{Hinton}, S.~R., {Davis}, T.~M., {Lidman}, C., {Glazebrook}, K., \& {Lewis},
  G.~F. 2016, Astronomy and Computing, 15, 61

\bibitem[{{Kinney} {et~al.}(1996){Kinney}, {Calzetti}, {Bohlin}, {McQuade},
  {Storchi-Bergmann}, \& {Schmitt}}]{Kinn96}
{Kinney}, A.~L., {Calzetti}, D., {Bohlin}, R.~C., {et~al.} 1996, \apj, 467, 38

\bibitem[{{Kriek} {et~al.}(2009){Kriek}, {van Dokkum}, {Labb{\'e}}, {Franx},
  {Illingworth}, {Marchesini}, \& {Quadri}}]{Krie09}
{Kriek}, M., {van Dokkum}, P.~G., {Labb{\'e}}, I., {et~al.} 2009, \apj, 700,
  221

\bibitem[{{Kurk} {et~al.}(2013){Kurk}, {Cimatti}, {Daddi}, {Mignoli},
  {Pozzetti}, {Dickinson}, {Bolzonella}, {Zamorani}, {Cassata}, {Rodighiero},
  {Franceschini}, {Renzini}, {Rosati}, {Halliday}, \& {Berta}}]{Kurk13}
{Kurk}, J., {Cimatti}, A., {Daddi}, E., {et~al.} 2013, \aap, 549, A63

\bibitem[{{Laidler} {et~al.}(2007){Laidler}, {Papovich}, {Grogin}, {Idzi},
  {Dickinson}, {Ferguson}, {Hilbert}, {Clubb}, \& {Ravindranath}}]{Laid07}
{Laidler}, V.~G., {Papovich}, C., {Grogin}, N.~A., {et~al.} 2007, \pasp, 119,
  1325

\bibitem[{{Le F{\`e}vre} {et~al.}(2005){Le F{\`e}vre}, {Paltani}, {Arnouts},
  {Charlot}, {Foucaud}, {Ilbert}, {McCracken}, {Zamorani}, {Bottini},
  {Garilli}, {Le Brun}, {Maccagni}, {Picat}, {Scaramella}, {Scodeggio},
  {Tresse}, {Vettolani}, {Zanichelli}, {Adami}, {Bardelli}, {Bolzonella},
  {Cappi}, {Ciliegi}, {Contini}, {Franzetti}, {Gavignaud}, {Guzzo}, {Iovino},
  {Marano}, {Marinoni}, {Mazure}, {Meneux}, {Merighi}, {Pell{\`o}}, {Pollo},
  {Pozzetti}, {Radovich}, {Zucca}, {Arnaboldi}, {Bondi}, {Bongiorno},
  {Busarello}, {Gregorini}, {Lamareille}, {Mathez}, {Mellier}, {Merluzzi},
  {Ripepi}, \& {Rizzo}}]{LeFe05c}
{Le F{\`e}vre}, O., {Paltani}, S., {Arnouts}, S., {et~al.} 2005, \nat, 437, 519

\bibitem[{{Le F{\`e}vre} {et~al.}(2015){Le F{\`e}vre}, {Tasca}, {Cassata},
  {Garilli}, {Le Brun}, {Maccagni}, {Pentericci}, {Thomas}, {Vanzella},
  {Zamorani}, {Zucca}, {Amorin}, {Bardelli}, {Capak}, {Cassar{\`a}},
  {Castellano}, {Cimatti}, {Cuby}, {Cucciati}, {de la Torre}, {Durkalec},
  {Fontana}, {Giavalisco}, {Grazian}, {Hathi}, {Ilbert}, {Lemaux}, {Moreau},
  {Paltani}, {Ribeiro}, {Salvato}, {Schaerer}, {Scodeggio}, {Sommariva},
  {Talia}, {Taniguchi}, {Tresse}, {Vergani}, {Wang}, {Charlot}, {Contini},
  {Fotopoulou}, {L{\'o}pez-Sanjuan}, {Mellier}, \& {Scoville}}]{LeFe15}
{Le F{\`e}vre}, O., {Tasca}, L.~A.~M., {Cassata}, P., {et~al.} 2015, \aap, 576,
  A79

\bibitem[{{Le F{\`e}vre} {et~al.}(2004){Le F{\`e}vre}, {Vettolani}, {Paltani},
  {Tresse}, {Zamorani}, {Le Brun}, {Moreau}, {Bottini}, {Maccagni}, {Picat},
  {Scaramella}, {Scodeggio}, {Zanichelli}, {Adami}, {Arnouts}, {Bardelli},
  {Bolzonella}, {Cappi}, {Charlot}, {Contini}, {Foucaud}, {Franzetti},
  {Garilli}, {Gavignaud}, {Guzzo}, {Ilbert}, {Iovino}, {McCracken}, {Mancini},
  {Marano}, {Marinoni}, {Mathez}, {Mazure}, {Meneux}, {Merighi}, {Pell{\`o}},
  {Pollo}, {Pozzetti}, {Radovich}, {Zucca}, {Arnaboldi}, {Bondi}, {Bongiorno},
  {Busarello}, {Ciliegi}, {Gregorini}, {Mellier}, {Merluzzi}, {Ripepi}, \&
  {Rizzo}}]{LeFe04}
{Le F{\`e}vre}, O., {Vettolani}, G., {Paltani}, S., {et~al.} 2004, \aap, 428,
  1043

\bibitem[{{Leclercq} {et~al.}(2017){Leclercq}, {Bacon}, {Wisotzki},
  {et~al.}}]{subm_Lecl17}
{Leclercq}, F., {Bacon}, R., {Wisotzki}, L., {et~al.} 2017, \aap, submitted
  (MUSE UDF SI paper VIII)

\bibitem[{{Lilly} {et~al.}(2007){Lilly}, {Le F{\`e}vre}, {Renzini}, {Zamorani},
  {Scodeggio}, {Contini}, {Carollo}, {Hasinger}, {Kneib}, {Iovino}, {Le Brun},
  {Maier}, {Mainieri}, {Mignoli}, {Silverman}, {Tasca}, {Bolzonella},
  {Bongiorno}, {Bottini}, {Capak}, {Caputi}, {Cimatti}, {Cucciati}, {Daddi},
  {Feldmann}, {Franzetti}, {Garilli}, {Guzzo}, {Ilbert}, {Kampczyk}, {Kovac},
  {Lamareille}, {Leauthaud}, {Borgne}, {McCracken}, {Marinoni}, {Pello},
  {Ricciardelli}, {Scarlata}, {Vergani}, {Sanders}, {Schinnerer}, {Scoville},
  {Taniguchi}, {Arnouts}, {Aussel}, {Bardelli}, {Brusa}, {Cappi}, {Ciliegi},
  {Finoguenov}, {Foucaud}, {Franceschini}, {Halliday}, {Impey}, {Knobel},
  {Koekemoer}, {Kurk}, {Maccagni}, {Maddox}, {Marano}, {Marconi}, {Meneux},
  {Mobasher}, {Moreau}, {Peacock}, {Porciani}, {Pozzetti}, {Scaramella},
  {Schiminovich}, {Shopbell}, {Smail}, {Thompson}, {Tresse}, {Vettolani},
  {Zanichelli}, \& {Zucca}}]{Lill07}
{Lilly}, S.~J., {Le F{\`e}vre}, O., {Renzini}, A., {et~al.} 2007, \apjs, 172,
  70

\bibitem[{{Mary et al.}(in prep.)}]{prep_Mary17}
{Mary et al.} in prep.

\bibitem[{{Maseda} {et~al.}(2017){Maseda}, {Brinchmann}, {Franx},
  {et~al.}}]{subm_Mase17}
{Maseda}, M., {Brinchmann}, J., {Franx}, M., {et~al.} 2017, \aap, in press
  (MUSE UDF SI paper IV)

\bibitem[{{Mignoli} {et~al.}(2005){Mignoli}, {Cimatti}, {Zamorani}, {Pozzetti},
  {Daddi}, {Renzini}, {Broadhurst}, {Cristiani}, {D'Odorico}, {Fontana},
  {Giallongo}, {Gilmozzi}, {Menci}, \& {Saracco}}]{Mign05}
{Mignoli}, M., {Cimatti}, A., {Zamorani}, G., {et~al.} 2005, \aap, 437, 883

\bibitem[{{Momcheva} {et~al.}(2016){Momcheva}, {Brammer}, {van Dokkum},
  {Skelton}, {Whitaker}, {Nelson}, {Fumagalli}, {Maseda}, {Leja}, {Franx},
  {Rix}, {Bezanson}, {Da Cunha}, {Dickey}, {F{\"o}rster Schreiber},
  {Illingworth}, {Kriek}, {Labb{\'e}}, {Ulf Lange}, {Lundgren}, {Magee},
  {Marchesini}, {Oesch}, {Pacifici}, {Patel}, {Price}, {Tal}, {Wake}, {van der
  Wel}, \& {Wuyts}}]{Momc16}
{Momcheva}, I.~G., {Brammer}, G.~B., {van Dokkum}, P.~G., {et~al.} 2016, \apjs,
  225, 27

\bibitem[{{Morris} {et~al.}(2015){Morris}, {Kocevski}, {Trump}, {Weiner},
  {Hathi}, {Barro}, {Dahlen}, {Faber}, {Finkelstein}, {Fontana}, {Ferguson},
  {Grogin}, {Gr{\"u}tzbauch}, {Guo}, {Hsu}, {Koekemoer}, {Koo}, {Mobasher},
  {Pforr}, {Salvato}, {Wiklind}, \& {Wuyts}}]{Morr15}
{Morris}, A.~M., {Kocevski}, D.~D., {Trump}, J.~R., {et~al.} 2015, \aj, 149,
  178

\bibitem[{{Newman} {et~al.}(2013){Newman}, {Cooper}, {Davis}, {Faber}, {Coil},
  {Guhathakurta}, {Koo}, {Phillips}, {Conroy}, {Dutton}, {Finkbeiner}, {Gerke},
  {Rosario}, {Weiner}, {Willmer}, {Yan}, {Harker}, {Kassin}, {Konidaris},
  {Lai}, {Madgwick}, {Noeske}, {Wirth}, {Connolly}, {Kaiser}, {Kirby},
  {Lemaux}, {Lin}, {Lotz}, {Luppino}, {Marinoni}, {Matthews}, {Metevier}, \&
  {Schiavon}}]{Newm13}
{Newman}, J.~A., {Cooper}, M.~C., {Davis}, M., {et~al.} 2013, \apjs, 208, 5

\bibitem[{{Paltani} {et~al.}(2007){Paltani}, {Le F{\`e}vre}, {Ilbert},
  {Arnouts}, {Bardelli}, {Tresse}, {Zamorani}, {Zucca}, {Bottini}, {Garilli},
  {Le Brun}, {Maccagni}, {Picat}, {Scaramella}, {Scodeggio}, {Vettolani},
  {Zanichelli}, {Adami}, {Bolzonella}, {Cappi}, {Charlot}, {Ciliegi},
  {Contini}, {Foucaud}, {Franzetti}, {Gavignaud}, {Guzzo}, {Iovino},
  {McCracken}, {Marano}, {Marinoni}, {Mazure}, {Meneux}, {Merighi},
  {Pell{\`o}}, {Pollo}, {Pozzetti}, {Radovich}, {Bondi}, {Bongiorno},
  {Brinchmann}, {Cucciati}, {de la Torre}, {Lamareille}, {Mellier}, {Merluzzi},
  {Temporin}, {Vergani}, \& {Walcher}}]{Palt07}
{Paltani}, S., {Le F{\`e}vre}, O., {Ilbert}, O., {et~al.} 2007, \aap, 463, 873

\bibitem[{{Patr{\'{\i}}cio} {et~al.}(2016){Patr{\'{\i}}cio}, {Richard},
  {Verhamme}, {Wisotzki}, {Brinchmann}, {Turner}, {Christensen}, {Weilbacher},
  {Blaizot}, {Bacon}, {Contini}, {Lagattuta}, {Cantalupo}, {Cl{\'e}ment}, \&
  {Soucail}}]{Patr16}
{Patr{\'{\i}}cio}, V., {Richard}, J., {Verhamme}, A., {et~al.} 2016, \mnras,
  456, 4191

\bibitem[{{Piqueras et al.}(2017)}]{subm_Piqu17}
{Piqueras et al.} 2017, ASP Conf. Ser., ADASS XXVI

\bibitem[{{Popesso} {et~al.}(2009){Popesso}, {Dickinson}, {Nonino}, {Vanzella},
  {Daddi}, {Fosbury}, {Kuntschner}, {Mainieri}, {Cristiani}, {Cesarsky},
  {Giavalisco}, {Renzini}, \& {GOODS Team}}]{Pope09}
{Popesso}, P., {Dickinson}, M., {Nonino}, M., {et~al.} 2009, \aap, 494, 443

\bibitem[{{Rafelski} {et~al.}(2015){Rafelski}, {Teplitz}, {Gardner}, {Coe},
  {Bond}, {Koekemoer}, {Grogin}, {Kurczynski}, {McGrath}, {Bourque}, {Atek},
  {Brown}, {Colbert}, {Codoreanu}, {Ferguson}, {Finkelstein}, {Gawiser},
  {Giavalisco}, {Gronwall}, {Hanish}, {Lee}, {Mehta}, {de Mello},
  {Ravindranath}, {Ryan}, {Scarlata}, {Siana}, {Soto}, \& {Voyer}}]{Rafe15}
{Rafelski}, M., {Teplitz}, H.~I., {Gardner}, J.~P., {et~al.} 2015, \aj, 150, 31

\bibitem[{{Reddy} {et~al.}(2005){Reddy}, {Erb}, {Steidel}, {Shapley},
  {Adelberger}, \& {Pettini}}]{Redd05}
{Reddy}, N.~A., {Erb}, D.~K., {Steidel}, C.~C., {et~al.} 2005, \apj, 633, 748

\bibitem[{{Rigby} {et~al.}(2002){Rigby}, {Charlton}, \& {Churchill}}]{Rigb02}
{Rigby}, J.~R., {Charlton}, J.~C., \& {Churchill}, C.~W. 2002, \apj, 565, 743

\bibitem[{{S{\'a}nchez-Bl{\'a}zquez} {et~al.}(2006){S{\'a}nchez-Bl{\'a}zquez},
  {Peletier}, {Jim{\'e}nez-Vicente}, {Cardiel}, {Cenarro},
  {Falc{\'o}n-Barroso}, {Gorgas}, {Selam}, \& {Vazdekis}}]{Sanc06}
{S{\'a}nchez-Bl{\'a}zquez}, P., {Peletier}, R.~F., {Jim{\'e}nez-Vicente}, J.,
  {et~al.} 2006, \mnras, 371, 703

\bibitem[{{Shapley} {et~al.}(2003){Shapley}, {Steidel}, {Pettini}, \&
  {Adelberger}}]{Shap03}
{Shapley}, A.~E., {Steidel}, C.~C., {Pettini}, M., \& {Adelberger}, K.~L. 2003,
  \apj, 588, 65

\bibitem[{{Stark} {et~al.}(2009){Stark}, {Ellis}, {Bunker}, {Bundy}, {Targett},
  {Benson}, \& {Lacy}}]{Star09}
{Stark}, D.~P., {Ellis}, R.~S., {Bunker}, A., {et~al.} 2009, \apj, 697, 1493

\bibitem[{{Stark} {et~al.}(2010){Stark}, {Ellis}, {Chiu}, {Ouchi}, \&
  {Bunker}}]{Star10}
{Stark}, D.~P., {Ellis}, R.~S., {Chiu}, K., {Ouchi}, M., \& {Bunker}, A. 2010,
  \mnras, 408, 1628

\bibitem[{{Steidel} {et~al.}(2010){Steidel}, {Erb}, {Shapley}, {Pettini},
  {Reddy}, {Bogosavljevi{\'c}}, {Rudie}, \& {Rakic}}]{Stei10}
{Steidel}, C.~C., {Erb}, D.~K., {Shapley}, A.~E., {et~al.} 2010, \apj, 717, 289

\bibitem[{{Straughn} {et~al.}(2008){Straughn}, {Meurer}, {Pirzkal}, {Cohen},
  {Malhotra}, {Rhoads}, {Windhorst}, {Gardner}, {Hathi}, {Xu}, {Gronwall},
  {Koekemoer}, {Walsh}, \& {di Serego Alighieri}}]{Stra08}
{Straughn}, A.~N., {Meurer}, G.~R., {Pirzkal}, N., {et~al.} 2008, \aj, 135,
  1624

\bibitem[{{Szokoly} {et~al.}(2004){Szokoly}, {Bergeron}, {Hasinger}, {Lehmann},
  {Kewley}, {Mainieri}, {Nonino}, {Rosati}, {Giacconi}, {Gilli}, {Gilmozzi},
  {Norman}, {Romaniello}, {Schreier}, {Tozzi}, {Wang}, {Zheng}, \&
  {Zirm}}]{Szok04}
{Szokoly}, G.~P., {Bergeron}, J., {Hasinger}, G., {et~al.} 2004, \apjs, 155,
  271

\bibitem[{{Tasca} {et~al.}(2017){Tasca}, {Le F{\`e}vre}, {Ribeiro}, {Thomas},
  {Moreau}, {Cassata}, {Garilli}, {Le Brun}, {Lemaux}, {Maccagni},
  {Pentericci}, {Schaerer}, {Vanzella}, {Zamorani}, {Zucca}, {Amorin},
  {Bardelli}, {Cassar{\`a}}, {Castellano}, {Cimatti}, {Cucciati}, {Durkalec},
  {Fontana}, {Giavalisco}, {Grazian}, {Hathi}, {Ilbert}, {Paltani}, {Pforr},
  {Scodeggio}, {Sommariva}, {Talia}, {Tresse}, {Vergani}, {Capak}, {Charlot},
  {Contini}, {de la Torre}, {Dunlop}, {Fotopoulou}, {Guaita}, {Koekemoer},
  {L{\'o}pez-Sanjuan}, {Mellier}, {Salvato}, {Scoville}, {Taniguchi}, \&
  {Wang}}]{Tasc17}
{Tasca}, L.~A.~M., {Le F{\`e}vre}, O., {Ribeiro}, B., {et~al.} 2017, \aap, 600,
  A110

\bibitem[{{Teplitz} {et~al.}(2013){Teplitz}, {Rafelski}, {Kurczynski}, {Bond},
  {Grogin}, {Koekemoer}, {Atek}, {Brown}, {Coe}, {Colbert}, {Ferguson},
  {Finkelstein}, {Gardner}, {Gawiser}, {Giavalisco}, {Gronwall}, {Hanish},
  {Lee}, {de Mello}, {Ravindranath}, {Ryan}, {Siana}, {Scarlata}, {Soto},
  {Voyer}, \& {Wolfe}}]{Tepl13}
{Teplitz}, H.~I., {Rafelski}, M., {Kurczynski}, P., {et~al.} 2013, \aj, 146,
  159

\bibitem[{{Tremonti} {et~al.}(2004){Tremonti}, {Heckman}, {Kauffmann},
  {Brinchmann}, {Charlot}, {White}, {Seibert}, {Peng}, {Schlegel}, {Uomoto},
  {Fukugita}, \& {Brinkmann}}]{Trem04}
{Tremonti}, C.~A., {Heckman}, T.~M., {Kauffmann}, G., {et~al.} 2004, \apj, 613,
  898

\bibitem[{{van Dokkum} {et~al.}(2006){van Dokkum}, {Quadri}, {Marchesini},
  {Rudnick}, {Franx}, {Gawiser}, {Herrera}, {Wuyts}, {Lira}, {Labb{\'e}},
  {Maza}, {Illingworth}, {F{\"o}rster Schreiber}, {Kriek}, {Rix}, {Taylor},
  {Toft}, {Webb}, \& {Yi}}]{vanD06}
{van Dokkum}, P.~G., {Quadri}, R., {Marchesini}, D., {et~al.} 2006, \apjl, 638,
  L59

\bibitem[{{Vanzella} {et~al.}(2008){Vanzella}, {Cristiani}, {Dickinson},
  {Giavalisco}, {Kuntschner}, {Haase}, {Nonino}, {Rosati}, {Cesarsky},
  {Ferguson}, {Fosbury}, {Grazian}, {Moustakas}, {Rettura}, {Popesso},
  {Renzini}, {Stern}, \& {GOODS Team}}]{Vanz08}
{Vanzella}, E., {Cristiani}, S., {Dickinson}, M., {et~al.} 2008, \aap, 478, 83

\bibitem[{{Vanzella} {et~al.}(2005){Vanzella}, {Cristiani}, {Dickinson},
  {Kuntschner}, {Moustakas}, {Nonino}, {Rosati}, {Stern}, {Cesarsky}, {Ettori},
  {Ferguson}, {Fosbury}, {Giavalisco}, {Haase}, {Renzini}, {Rettura}, {Serra},
  \& {GOODS Team}}]{Vanz05}
{Vanzella}, E., {Cristiani}, S., {Dickinson}, M., {et~al.} 2005, \aap, 434, 53

\bibitem[{{Vanzella} {et~al.}(2006){Vanzella}, {Cristiani}, {Dickinson},
  {Kuntschner}, {Nonino}, {Rettura}, {Rosati}, {Vernet}, {Cesarsky},
  {Ferguson}, {Fosbury}, {Giavalisco}, {Grazian}, {Haase}, {Moustakas},
  {Popesso}, {Renzini}, {Stern}, \& {GOODS Team}}]{Vanz06}
{Vanzella}, E., {Cristiani}, S., {Dickinson}, M., {et~al.} 2006, \aap, 454, 423

\bibitem[{{Vanzella} {et~al.}(2009){Vanzella}, {Giavalisco}, {Dickinson},
  {Cristiani}, {Nonino}, {Kuntschner}, {Popesso}, {Rosati}, {Renzini}, {Stern},
  {Cesarsky}, {Ferguson}, \& {Fosbury}}]{Vanz09}
{Vanzella}, E., {Giavalisco}, M., {Dickinson}, M., {et~al.} 2009, \apj, 695,
  1163

\bibitem[{{Ventou} {et~al.}(2017){Ventou}, {Contini}, {Bouch\'e},
  {et~al.}}]{subm_Vent17}
{Ventou}, A., {Contini}, T., {Bouch\'e}, N., {et~al.} 2017, \aap, submitted
  (MUSE UDF SI paper IX)

\bibitem[{{Walter} {et~al.}(2016){Walter}, {Decarli}, {Aravena}, {Carilli},
  {Bouwens}, {da Cunha}, {Daddi}, {Ivison}, {Riechers}, {Smail}, {Swinbank},
  {Weiss}, {Anguita}, {Assef}, {Bacon}, {Bauer}, {Bell}, {Bertoldi}, {Chapman},
  {Colina}, {Cortes}, {Cox}, {Dickinson}, {Elbaz}, {G{\'o}nzalez-L{\'o}pez},
  {Ibar}, {Inami}, {Infante}, {Hodge}, {Karim}, {Le Fevre}, {Magnelli}, {Neri},
  {Oesch}, {Ota}, {Popping}, {Rix}, {Sargent}, {Sheth}, {van der Wel}, {van der
  Werf}, \& {Wagg}}]{Walt16}
{Walter}, F., {Decarli}, R., {Aravena}, M., {et~al.} 2016, \apj, 833, 67

\bibitem[{{Weilbacher} {et~al.}(2012){Weilbacher}, {Streicher}, {Urrutia},
  {Jarno}, {P{\'e}contal-Rousset}, {Bacon}, \& {B{\"o}hm}}]{Weil12}
{Weilbacher}, P.~M., {Streicher}, O., {Urrutia}, T., {et~al.} 2012, in
  \procspie, Vol. 8451, Software and Cyberinfrastructure for Astronomy II,
  84510B

\bibitem[{{Wisotzki} {et~al.}(2016){Wisotzki}, {Bacon}, {Blaizot},
  {Brinchmann}, {Herenz}, {Schaye}, {Bouch{\'e}}, {Cantalupo}, {Contini},
  {Carollo}, {Caruana}, {Courbot}, {Emsellem}, {Kamann}, {Kerutt}, {Leclercq},
  {Lilly}, {Patr{\'{\i}}cio}, {Sandin}, {Steinmetz}, {Straka}, {Urrutia},
  {Verhamme}, {Weilbacher}, \& {Wendt}}]{Wiso16}
{Wisotzki}, L., {Bacon}, R., {Blaizot}, J., {et~al.} 2016, \aap, 587, A98

\bibitem[{{York} {et~al.}(2000){York}, {Adelman}, {Anderson}, {Anderson},
  {Annis}, {Bahcall}, {Bakken}, {Barkhouser}, {Bastian}, {Berman}, {Boroski},
  {Bracker}, {Briegel}, {Briggs}, {Brinkmann}, {Brunner}, {Burles}, {Carey},
  {Carr}, {Castander}, {Chen}, {Colestock}, {Connolly}, {Crocker}, {Csabai},
  {Czarapata}, {Davis}, {Doi}, {Dombeck}, {Eisenstein}, {Ellman}, {Elms},
  {Evans}, {Fan}, {Federwitz}, {Fiscelli}, {Friedman}, {Frieman}, {Fukugita},
  {Gillespie}, {Gunn}, {Gurbani}, {de Haas}, {Haldeman}, {Harris}, {Hayes},
  {Heckman}, {Hennessy}, {Hindsley}, {Holm}, {Holmgren}, {Huang}, {Hull},
  {Husby}, {Ichikawa}, {Ichikawa}, {Ivezi{\'c}}, {Kent}, {Kim}, {Kinney},
  {Klaene}, {Kleinman}, {Kleinman}, {Knapp}, {Korienek}, {Kron}, {Kunszt},
  {Lamb}, {Lee}, {Leger}, {Limmongkol}, {Lindenmeyer}, {Long}, {Loomis},
  {Loveday}, {Lucinio}, {Lupton}, {MacKinnon}, {Mannery}, {Mantsch}, {Margon},
  {McGehee}, {McKay}, {Meiksin}, {Merelli}, {Monet}, {Munn}, {Narayanan},
  {Nash}, {Neilsen}, {Neswold}, {Newberg}, {Nichol}, {Nicinski}, {Nonino},
  {Okada}, {Okamura}, {Ostriker}, {Owen}, {Pauls}, {Peoples}, {Peterson},
  {Petravick}, {Pier}, {Pope}, {Pordes}, {Prosapio}, {Rechenmacher}, {Quinn},
  {Richards}, {Richmond}, {Rivetta}, {Rockosi}, {Ruthmansdorfer}, {Sandford},
  {Schlegel}, {Schneider}, {Sekiguchi}, {Sergey}, {Shimasaku}, {Siegmund},
  {Smee}, {Smith}, {Snedden}, {Stone}, {Stoughton}, {Strauss}, {Stubbs},
  {SubbaRao}, {Szalay}, {Szapudi}, {Szokoly}, {Thakar}, {Tremonti}, {Tucker},
  {Uomoto}, {Vanden Berk}, {Vogeley}, {Waddell}, {Wang}, {Watanabe},
  {Weinberg}, {Yanny}, {Yasuda}, \& {SDSS Collaboration}}]{York00}
{York}, D.~G., {Adelman}, J., {Anderson}, Jr., J.~E., {et~al.} 2000, \aj, 120,
  1579

\end{thebibliography}

\clearpage
\begin{appendix}

\section{MUSE UDF redshift catalogs}\label{app:cat}

Along with this paper, we release the MUSE ultra deep field (\udft)
and deep field (\mosaic) redshift and line flux catalogs. Because the
\udft field is within the \mosaic field, there is some overlap in the
object entries. The combined catalog, which only includes a single set
of redshift and the associated parameters (CONFID, TYPE, and DEFECT,
and measurements in detected lines) for each unique galaxy, is also
provided. For the cases with conflicting redshifts, we give the
redshift discussed in \S\ref{subsec:comp_udf10_mosaic}, otherwise the
\udft redshifts and the associated parameters are employed. In the
Table~\ref{tbl:cat}, we describe the contents in the catalogs.

The UVUDF ID numbers are stored as strings (CSV) instead of
integer numbers because some of the UVUDF objects are unresolved with
MUSE and are merged to be treated as a single object (see
Section~\ref{sec:analysis_cont}).  In order to find an object with the
object ID from the UVUDF catalog, users can use the following
commands in both Python 2 and 3:

\begin{verbatim}

from astropy.table import Table
import numpy as np

# Read catalog
>>> catalog = Table.read("catalog.fits", format="fits")

# Convert the catalog UVUDF_ID column to the string type
>>> UVUDF_IDs = catalog["UVUDF_ID"].astype(str)

# The UVUDF ID to search
>>> search_UVUDF_ID = "9525"

# Find the index number of the above UVUDF ID in the catalog
>>> idx = np.where(np.char.count(UVUDF_IDs, search_UVUDF_ID))[0][0]

# Print the above UVUDF ID and 
#  its corresponding MUSE ID and the index number in the catalog
>>> print_format = "UVUDF ID={0} is MUSE ID={1} (UVUDF ID {2}) at index={3}"
>>> print(print_format.format(search_UVUDF_ID, catalog["MUSE_ID"][idx], UVUDF_IDs[idx], idx))

UVUDF ID=9525 is MUSE ID=70 (UVUDF ID 9525,9515) at index=61

\end{verbatim}

\clearpage

\begin{table}[htp]
\begin{minipage}{\textwidth}
  \caption{Column description of the redshift catalogs for the MUSE
    Hubble Ultra Deep Field (\udft, \mosaic, and combined) }
\begin{center}
\begin{tabular}{cll}
\hline \hline
No. & Title & Description \\
\hline
  1 & MUSE\_ID    & MUSE object identification number~~\tablefootmark{a}\\
  2 & RA          & R.A. (J2000) in units of decimal degrees~~\tablefootmark{b}\\
  3 & DEC         & Decl. (J2000) in units of decimal degrees~~\tablefootmark{b} \\
  4 & MERGED      & Flag (boolean) indicating a merged object \\
  5 & UVUDF\_ID   & UVUDF object identification number \citep{Rafe15} \\
  6 & ORIG\_ID    & {\tt ORIGIN} object identification number~~\tablefootmark{c} \\
  7 & ORIG\_RA    & {\tt ORIGIN} R.A. (J2000) in units of decimal degrees \\
  8 & ORIG\_DEC   & {\tt ORIGIN} Decl. (J2000) in units of decimal degrees \\
  9 & MLET\_ID    & {\tt MUSELET} object identification number~~\tablefootmark{c,\,d} \\
 10 & MLET\_RA    & {\tt MUSELET} R.A. (J2000) in units of decimal degrees~~\tablefootmark{d} \\
 11 & MLET\_DEC   & {\tt MUSELET} Decl. (J2000) in units of decimal degrees~~\tablefootmark{d} \\
 12 & EML\_ONLY   & Flag (boolean) for the objects detected only by {\tt ORIGIN}
                    or {\tt MUSELET} (i.e., no UVUDF counterpart) \\
 13 & REF\_SPEC   & The main spectrum (PSF-weighted or MUSE white-light weighted) used to measure redshift \\
 14 & Z\_MUSE     & Spectroscopic redshift determined with MUSE \\
 15 & CONFID      & Confidence level of Z\_MUSE \\
 16 & TYPE        & Type of feature used to identified Z\_MUSE \\
 17 & DEFECT      & Flag (0 or 1) indicating defect in the data (0 for no defect and 1 for some issues with the data) \\
 18 & F775W\_MAG  & {\it HST} F775W magnitude (AB) from
                    \cite{Rafe15}~~\tablefootmark{e} or {\tt
                    NoiseChisel}~~\tablefootmark{f} \\
 19 & F775W\_MAG\_ERR   & The errors on the {\it HST} F775W magnitude (AB)~~\tablefootmark{e,\,f} \\
 \multirow{2}{*}{20-41} & [LINE]\_FLUX  & Line fluxes
                                          ($\rm 10^{-20} \, erg \, s^{-1} cm^{-2}$) summed over the entire
                                         mask aperture~~\tablefootmark{g} (see \S\ref{subsec:line_flux}) \\
    & [LINE]\_FLUX\_ERR & Errors on the line fluxes \\
\hline
\multicolumn{3}{l}{Columns that only exist in the combined catalog} \\
 45 & Z\_FROM     & The origin of the redshift (the \udft or the \mosaic catalog) \\
 46 & IN\_UDF10   & Flag (boolean) indicating an object in the \udft field \\
\hline
\end{tabular}
\end{center}
\tablefoot{ The full table is available at the CDS.
  \tablefoottext{a}{For the future release, if an object can be
    ``split'' into multiple objects in terms of their redshift
    identifications, then the original MUSE\_ID will be deleted and
    the new IDs of the split objects will be added to the catalog.}
  \tablefoottext{b}{When $\rm MERGED = True$, the coordinates are
    calculated as {\it HST} F775W flux weighted coordinates of all
    objects composing the new merged source (see
    \S\ref{sec:analysis_cont}). Otherwise, they are identical to the
    UVUDF catalog \citep{Rafe15}. } \tablefoottext{c}{When ORIG\_ID or
    MLET\_ID is empty, this indicates that the object is not detected
    by {\tt ORIGIN} or {\tt MUSELET}.  For the combined catalog, the
    column names of ORIG\_ID, ORIG\_RA, and ORIG\_DEC have a suffix of
    \_UDF10 and \_MOSAIC.  } \tablefoottext{d}{Emission line search
    with {\tt MUSELET} is only perfomed in the \mosaic.}
  \tablefoottext{e}{When $\rm MERGED = True$, this value is not
    provided.}  \tablefoottext{f}{When $\rm F775W\_MAG\_ERR = -1$,
    F775W\_MAG is a $3\sigma$ upper limit.} \tablefoottext{g}{ The
    emission lines include Balmer lines (\ha, \hb, \hg, and \hd),
    \oii$\lambda 3726$, \oii$\lambda 3729$, \oiii$\lambda 4959$,
    \oiii$\lambda 5007$, \ciii$\lambda 1907$, \ciii$\lambda 1909$, and
    \lya. The measurements are only provided when S/N~$> 3$.}  }
\label{tbl:cat}
\end{minipage}
\end{table}%

\clearpage

\begin{landscape}

\section{Spectral templates for {\tt MARZ}}\label{app:temples}

\begin{table}[htp]
\caption{Input spectral templates for {\tt MARZ}.}
\begin{center}
\begin{tabular}{cll}
\hline \hline
 Number & Template Name & Origin \\
\hline
1 & A star                       & Sloan Digital Sky Survey (ID~4) / {\tt AUTOZ} $\rm ^{(1)}$ \\
2 & G star                    & Sloan Digital Sky Survey (ID~9) / {\tt AUTOZ} $\rm ^{(1)}$ \\
3 & M3 star                   & Sloan Digital Sky Survey (ID~13) / {\tt AUTOZ} $\rm ^{(1)}$ \\
4 & White dwarf star          & Sloan Digital Sky Survey (ID~21) / {\tt AUTOZ} $\rm ^{(1)}$ \\
5 & Luminous red galaxy       & Sloan Digital Sky Survey (ID~29) / {\tt AUTOZ} $\rm ^{(1)}$ \\
6 & HDFS star-forming galaxies ($0.1<z<1.4$)     & MUSE HDFS $\rm CONFID=3$ $\rm ^{(2)}$ \\
7 & UDF star-forming galaxies ($0.21<z<1.0$)     & MUSE UDF IDs of 0, 1, 2, 6, 8, 10, 18, 21, 33, 36 $\rm ^{(3)}$ \\
8 & UDF absorption line galaxies ($0.62<z<0.95$) & MUSE UDF IDs of 4, 5, 7 $\rm ^{(3)}$ \\
9 & UDF intermediate-redshift galaxies ($1.8<z<2.5$) & MUSE UDF IDs of 16, 20, 22, 27, 31, 43, 45, 47, 53, 55, 90 $\rm ^{(3)}$ \\
10 & MUSE high-redshift galaxy ($z=3.5$) & MUSE smoothed and normalized spectrum of SMACSJ2031.8-4036 $\rm ^{(4)}$ \\
11 & HDFS z-bin 1 - \ha emitter                     & MUSE HDFS IDs of 1, 26, 28, 53, 61, 62, 63, 67, 70, 98 $\rm ^{(2)}$ \\
12 & HDFS z-bin 2 - \oii emitter - weak continuum   & MUSE HDFS IDs of 12, 32, 90, 101, 115, 120 $\rm ^{(2)}$ \\
13 & HDFS z-bin 2 - \oii emitter - strong continuum & MUSE HDFS IDs of 3, 4, 5, 6, 7, 8, 9, 11, 16, 20, 23, 29 $\rm ^{(2)}$ \\
14 & HDFS z-bin 3 - \oii emitter - weak continuum   & MUSE HDFS IDs of 37, 38, 39, 48, 49, 64, 68, 88, 96, 107, 160 $\rm ^{(2)}$ \\
15 & HDFS z-bin 3 - \oii emitter - strong continuum & MUSE HDFS IDs of 13, 24, 27, 35, 45 $\rm ^{(2)}$ \\
16 & HDFS z-bin 4 - Absorption line galaxy          & MUSE HDFS IDs of 50, 51, 55, 60, 65, 83 $\rm ^{(2)}$ \\
17 & HDFS z-bin 4 - \ciii emitter                   & MUSE HDFS IDs of 41, 78, 87, 97, 109 $\rm ^{(2)}$ \\
18 & HDFS z-bin 5 - \lya emitter - weak continuum   & MUSE HDFS IDs of
                                                      112, 139, 144,
                                                      159, 181, 200,
                                                      202, 216, 218,
                                                      225, 232, 246,
                                                      271, \\
   &  & 290, 294, 308, 311, 325, 334, 422, 430, 433, 437, 441, 449, 484, 489, 492, 499, 503,
        \\
   &  & 513, 546, 547, 549, 551, 552, 553, 560, 563, 577, 584 $\rm ^{(2)}$ \\
19 & HDFS z-bin 5 - \lya emitter - strong continuum & MUSE HDFS IDs of 40, 43, 56, 92, 95 $\rm ^{(2)}$ \\
\hline
\end{tabular}
\end{center}
\tablefoot{ References. (1) \cite{Bald14}; (2) \cite{Baco15}; (3) this
  work;  (4) \cite{Patr16}. See also
  \url{http://classic.sdss.org/dr5/algorithms/spectemplates/} for
  templates No. $1-5$. }
\label{tbl:tamplete}
\end{table}%

\end{landscape}

\end{appendix}

\end{document}